\documentclass[aip, amsmath,amssymb, reprint]{revtex4-2}
\usepackage{graphicx}           
\graphicspath{{figures/}}       
\usepackage{dcolumn}            
\usepackage{bm}                 

\usepackage[utf8]{inputenc}
\usepackage[T1]{fontenc}
\usepackage{mathptmx, etoolbox, dcolumn}
\usepackage{tabularray}
\usepackage{braket}
\usepackage{siunitx,  upgreek}

\makeatletter
\def\@email#1#2{%
 \endgroup
 \patchcmd{\titleblock@produce}
  {\frontmatter@RRAPformat}
  {\frontmatter@RRAPformat{\produce@RRAP{*#1\href{mailto:#2}{#2}}}\frontmatter@RRAPformat}
  {}{}
}%
\makeatother

\begin{document}
\title[]{A Cavity-Enhanced Spectroscopist’s Lens on Molecular Polaritons}
\author{Alexander M. McKillop and Marissa L. Weichman$^*$}
\email{weichman@princeton.edu}
\affiliation{Department of Chemistry, Princeton	University, Princeton, New Jersey 08544, United States}
\date{\today}

\begin{abstract}
Polariton chemistry has been hailed as a potential new route to direct molecular processes with electromagnetic fields. 
To make further strides, it is essential for the community to clarify which unusual polaritonic phenomena are true hallmarks of cavity quantum electrodynamics and which can be rationalized with classical optical physics.
Here, we provide a tutorial perspective on the formation, spectroscopy, and behavior of molecular polaritons using classical optics.
Where possible, we draw connections to cavity-enhanced spectroscopy and recast open questions in terms that may be more familiar to the broader community of physical chemists.
\end{abstract}

\maketitle


\section{\label{sec:intro} Introduction}

\begin{figure}[t]
    \includegraphics[width=0.75\linewidth]{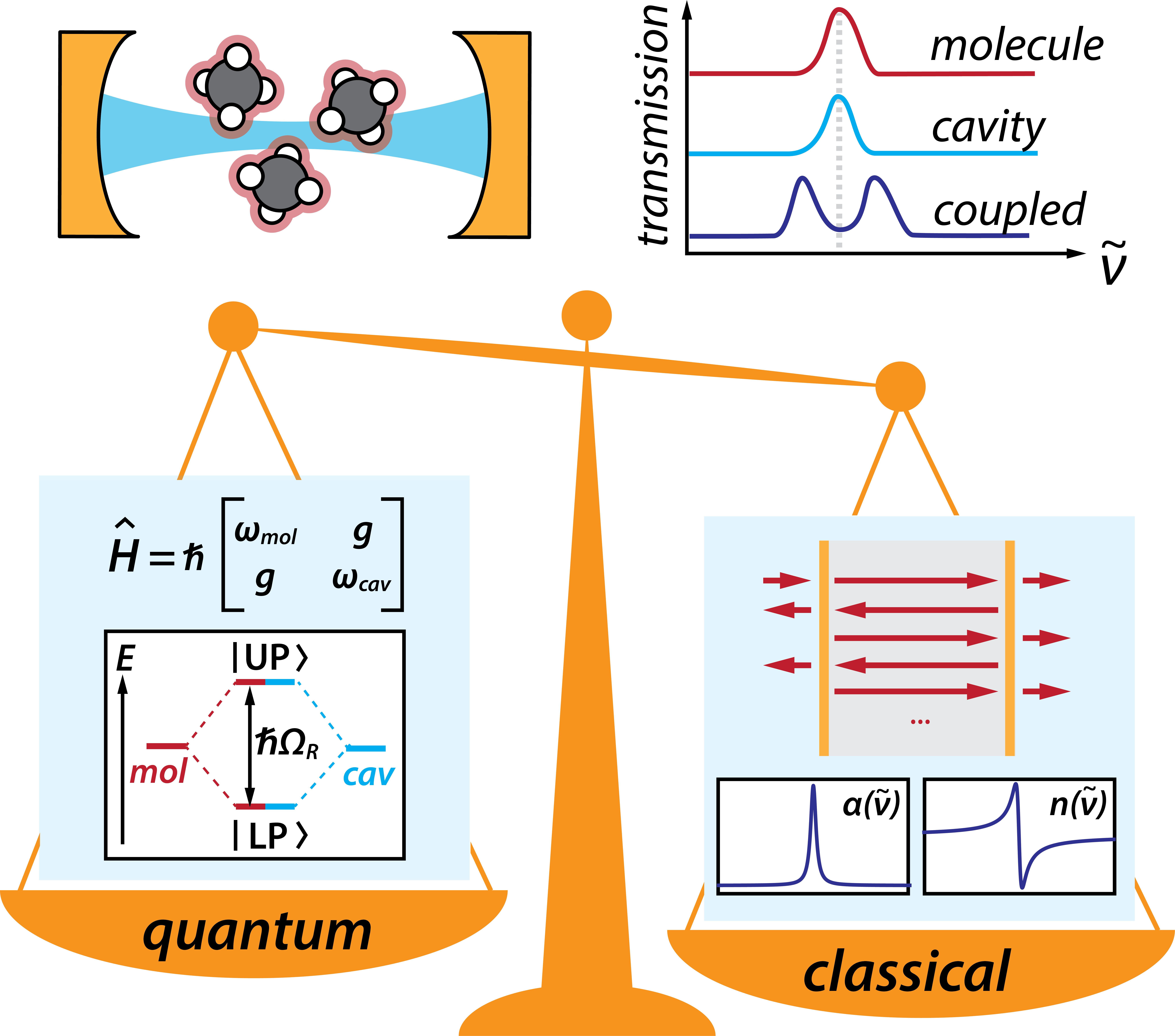}
    \caption{\label{fig:cartoon} 
    A bright optical transition of an ensemble of molecules can resonantly couple to a confined photonic mode of an optical cavity. 
    The emergence of the strong coupling regime coincides with the resolution of two distinct peaks in the cavity transmission spectrum. 
    This regime can be described using the language of either classical optical physics or cavity quantum electrodynamics. 
    }
\end{figure}

Molecular polaritons have become a subject of intense interest and debate among physical chemists.
Much of this excitement derives from the idea that polaritons may represent a new avenue for photonic control of chemistry and photophysics.
Polaritons are hybrid light-matter systems engineered by enclosing a molecular ensemble inside an optical cavity such that a confined photonic mode interacts strongly with a bright molecular optical transition. 
Intracavity molecules under such conditions have been reported to exhibit behavior distinct from the corresponding free-space systems under strong coupling of both vibrational\cite{thomas16,xiang20,hirai2020,sau2021,chen22,ahn23} and electronic\cite{hutchison12, munkhbat2018, zeng2023, lee2024controlling} transitions.
Of course, not all reported behavior has proved reproducible\cite{imperatore21,wiesehan21,michon24} and not all reactions exhibit polaritonic effects.\cite{fidler23,chen24,muller2024}
The community has not yet arrived at a unifying theory that can capture all observed phenomena, despite widespread efforts.\cite{herrera20,simpkins21,wang21,li22,campos23,mandal23,xiong23,schwennicke2024} 
Indeed, the field is still grappling with the choice of a unified framework with which to describe these systems (Fig.~\ref{fig:cartoon}).

Polaritons are often introduced using the language of cavity quantum electrodynamics (cQED).
The Jaynes-Cummings (JC) model treats the coupling of a two-level system to a single quantized cavity mode;\cite{shore93} the Tavis-Cummings (TC) model extends this description to the collective coupling of $N$ identical two-level systems to the same photonic mode.\cite{tavis68,garraway11}
In both the JC and TC models, polaritons arise as energy eigenstates of the coupled light-matter Hamiltonian: coherent quantum mechanical states that are a superposition of both molecular and photonic excitations.
These polaritonic states appear as new, optically bright features in the absorption spectrum of the coupled system, split in frequency by the vacuum Rabi splitting ($\Omega_R$), which scales as $\sqrt{N}$ in the TC collective coupling case. 
Some other subtleties arise in the TC model: the polariton states are delocalized over the entire ensemble of coupled molecules, and, in addition, a manifold of $N-1$ collective ``dark states'' are left at the original uncoupled molecular energy.
The dark states represent non-totally-symmetric linear combinations of molecular excitations which are formally forbidden from interacting with light and thus have no photonic contribution. 
The extent to which these dark states behave as uncoupled molecules remains an open question.

Most theoretical work in polariton chemistry is carried out within some variation of this cQED framework.
Ongoing efforts are updating cQED treatments to account for aspects of experimental complexity including structural and energetic disorder, detailed \textit{ab initio} molecular potential energy surfaces, treatment of large-$N$ intracavity ensembles,  multi-mode cavity geometries, and photonic losses.\cite{wang21, mandal23, campos23}

cQED is not the only way to understand strong light-matter coupling.
Classical optical cavity physics has been used to describe strongly-coupled systems since the early 1990s.\cite{zhu90, khitrova1999, rudin1999, novotny10, tanji2011, torma14, xiang18, herrera20, simpkins23}
Both explicit Fabry-P\'{e}rot cavity expressions and the transfer matrix method have proved popular tools; both approaches use Maxwell's equations to treat the classical interference of light inside a cavity structure.
Polaritons emerge as optical cavity modes that shift and split as the refractive index of the intracavity medium undergoes large changes near an optical resonance, reproducing the splitting of peaks in the cavity transmission spectrum with the same $\Omega_R$ that falls out of cQED.
This description applies so long as the ensemble of intracavity molecules is well-represented by a spatially homogeneous dielectric medium.
Throughout this perspective we will refer to the Maxwell description of the intracavity photonic field as the ``classical optics'' approach, in contrast to the cQED treatment that quantizes the field.
This ``classical'' approach is perhaps more accurately described as ``semiclassical'' in that the molecules are still quantum mechanical: their complex refractive index contains features that derive from dipole-allowed transitions between discrete quantum states.

In any event, the classical picture proves intuitive: just a case of coupled oscillators.\cite{novotny10} 
When two pendula with degenerate resonance frequencies are coupled, two new normal modes of the system result, corresponding to synchronous motion of the original oscillators either perfectly in-phase or $180^\textrm{o}$ out-of-phase.
These new normal modes oscillate at frequencies lying above and below those of the original uncoupled oscillators, and the frequency splitting between the new normal modes scales with the coupling strength.
The system is understood to be in the strong coupling regime when the splitting between the two new modes is experimentally resolvable, exceeding the resonance linewidths of the original oscillators.
In the case of polaritons, the original oscillators are an electromagnetic standing wave and a collection of intracavity molecular dipoles.
The upper and lower polariton modes capture the oscillation of the confined cavity field as it drags the molecular dipoles along, either in-phase or out-of-phase.

So, we have two seemingly appropriate frameworks -- the quantum optical and the classical -- with which to understand polaritons (Fig.~\ref{fig:cartoon}). 
As it stands, most theorists work within the cQED description, while experimentalists more often lean on classical optics to simulate linear cavity spectra.
It is not surprising that both approaches can be fruitful: recent work has shown that the cQED and classical optics frameworks produce formally equivalent linear cavity spectra in the large-$N$ limit.\cite{cwik2016, yuen2024, schwennicke2024} 
While these two approaches often reproduce the same observables, the intuition afforded by each is distinct.
The question therefore remains: which framework should we reach for to explain an interesting experimental result?

More than 25 years ago, Khitrova \textit{et al}.\cite{khitrova1999} argued that one need only invoke cQED when operating near the quantum statistical limit, defined as the regime where incrementing the number of intracavity molecules ($N$) or photons ($n$) by one quantum will detectably change the cavity transmission spectrum. 
The same authors contended that most semiconductor quantum well exciton-polariton systems operate nowhere near this regime.
Most experimental demonstrations of vibrational and electronic molecular polaritons lie similarly far from this stringent limit, featuring large-$N$ molecular ensembles and large-$n$ optical fields to perform spectroscopy.
Some seminal experiments fall within the quantum statistical regime, of course:\cite{tanji2011} 
Haroche and coworkers observed the reversible spontaneous emission of strongly-coupled Rydberg atoms;\cite{kaluzny1983,haroche1989} 
Kimble and coworkers strongly coupled a single cesium atom;\cite{thompson1992} 
Imamoğlu and coworkers observed anti-bunching of photons emitted from a single strongly-coupled quantum dot;\cite{hennessy2007} 
and several teams have now reached the single-emitter strong coupling regime in plasmonic nanocavities.\cite{chikkaraddy2016, santhosh2016, paoletta2024}
These examples aside, some suspension of disbelief is required to imagine polaritons as coherent quantum superposition states delocalized across millions of solution-phase, room temperature molecules.

Here, we examine how a range of polaritonic phenomena can be explained with classical optical cavity physics. 
Optical interference inside cavities can give rise to non-intuitive spectroscopic signals, and not all strange intracavity phenomena arise specifically from the formation of polariton states.\cite{renken21, simpkins23, simpkins23comment, thomas2024}
Our goal is to help distinguish behavior that requires us to invoke cQED from garden-variety optical cavity physics.
We hope to provide an entry point to these concepts for researchers embarking on work in polaritonics, for those more accustomed to working with cQED, and for the broader community of physical chemists interested in the spectroscopy of polaritonic systems.
For more comprehensive coverage of ongoing experimental and theoretical research in this field, we refer the reader to the vast, rapidly expanding library of excellent review articles.\cite{khitrova1999, torma14, ribeiro18, hertzog19, herrera20, garcia21, simpkins21, wang21, mandal23, li22, dunkelberger22, simpkins23, campos23, schwennicke2024}

We begin with a primer on the classical description of electromagnetic fields traveling in Fabry-P\'{e}rot optical cavities in Section \ref{sec:cavityprimer}. 
In Section \ref{sec:linear} we extend this treatment to the linear spectroscopy of strongly-coupled systems, and in Section \ref{sec:nonlinear} we turn our attention to nonlinear spectroscopy.
The examples we provide throughout are illustrated explicitly for micron-scale Fabry-P\'{e}rot cavities relevant to vibrational strong coupling, but are just as applicable to the nanoscale cavities typically used for electronic strong coupling.
We conclude in Section \ref{sec:conc} with our final thoughts and future outlook.

\section{\label{sec:cavityprimer} Classical Wave Interference in a Fabry-P\'{e}rot Optical Cavity}
We now introduce the description of a Fabry-P\'{e}rot (FP) optical cavity as two reflective interfaces sandwiching a dielectric medium.
We will derive closed-form expressions for the propagation of electromagnetic waves through this system.
Our derivation follows those found in various optics textbooks\cite{steck06,yariv07,born13,nagourney14} and resources on cavity-enhanced spectroscopy.\cite{lehmann96,gagliardi14}

\begin{figure}[b]
	\centering \includegraphics[width=3in]{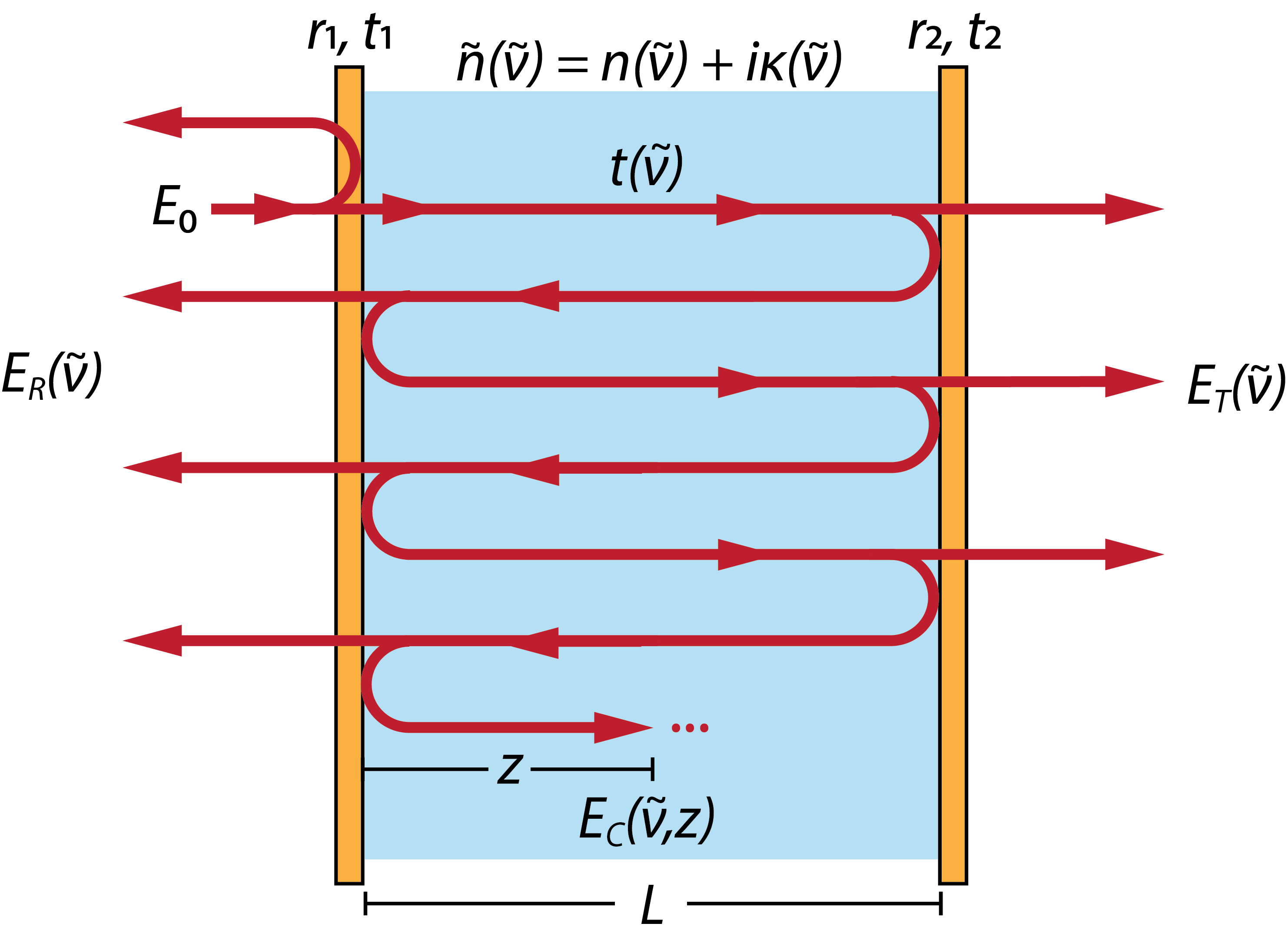}
        \caption{\label{fig:cavity} An incident electromagnetic wave of amplitude $E_0$ and wavenumber $\tilde\nu$ impinges on a Fabry-P\'{e}rot cavity of length $L$. The cavity is constructed from two mirrors with reflection and transmission amplitude coefficients $r_1$, $r_2$, $t_1$, and $t_2$, and is filled with a dielectric medium with complex refractive index $\tilde{n}(\tilde\nu)$. We calculate the transmitted [$E_T(\tilde\nu)$], reflected [$E_R(\tilde\nu)$], and circulating [$E_C(\tilde\nu,z)$] field amplitudes by considering the interference of partial waves transmitted and reflected by each cavity mirror.}
\end{figure}

\begin{table*}
        \caption{\label{tab:table1}\textbf{ Classical expressions for the field amplitudes and intensities of light transmitted, circulating, reflected, and absorbed by a Fabry-P\'{e}rot cavity filled with a dielectric medium.} We assume a cavity of length $L$ composed of two identical mirrors with intensity reflection coefficient $R = r_1^2 = r_2^2$ and intensity transmission coefficient $T = t_1^2 = t_2^2$. We take the intracavity material to have complex refractive index $\tilde{n}(\tilde\nu) = n(\tilde\nu) + i \kappa(\tilde\nu)$ and absorption coefficient $\alpha(\tilde\nu) = 4 \pi \kappa(\tilde\nu) \tilde\nu$. $\delta(\tilde\nu) = 4 \pi  L  n(\tilde\nu) \tilde\nu$ represents the round-trip phase accrued by light traveling in the cavity. The expressions for circulating amplitude and intensity represent interference from counter-propagating intracavity beams and are provided as a function of $z$, the distance along the longitudinal cavity axis past the input mirror.}
	\begin{tblr}{ colspec={ l  l  l }, rowsep = 5pt }
		\hline\hline
		& amplitude & intensity \\ \hline
		transmitted &  $\frac{E_T(\tilde\nu)}{E_0} = \frac{t_1 \, t_2 \, e^{-\alpha(\tilde\nu) L/2}} {1 - r_1 \, r_2 \, e^{-\alpha(\tilde\nu) L} \, e^{ i \delta(\tilde\nu)} }$ 
		&  $\frac{I_T(\tilde\nu)}{I_0} = \Big| \frac{E_T(\tilde\nu)}{E_0} \Big|^2 = \frac{T^2  e^{-\alpha(\tilde\nu) L}}{1 + R^2  e^{-2 \alpha(\tilde\nu) L} - 2 R e^{-\alpha(\tilde\nu) L} \cos \left[ \delta(\tilde\nu) \right]  }$ \\ \hline
	   circulating  &  
        $\frac{E_C(\tilde\nu, z)}{E_0} = \frac{t_1 \left[ e^{i \delta(\tilde\nu) z/2L} e^{-\alpha(\tilde\nu)z/2} - r_2 e^{i\delta(\tilde\nu)[1-z/2L]} e^{-\alpha(\tilde\nu) [L-z/2]} \right] }{1 - r_1 \, r_2 \, e^{-\alpha(\tilde\nu) L} \, e^{i \delta(\tilde\nu)} }$ 
		&  $\frac{I_C(\tilde\nu, z)}{I_0} = \Big| \frac{E_C(\tilde\nu, z)}{E_0} \Big|^2 = \frac{T \left[ e^{-\alpha(\tilde\nu)z} + Re^{-\alpha(\tilde\nu)[2L-z]} - 2\sqrt{R} e^{-\alpha(\tilde\nu)L} \cos\big[ \delta(\tilde\nu) [1 - z/L] \big] \right] }{1 + R^2  e^{-2 \alpha(\tilde\nu) L} - 2 R e^{-\alpha(\tilde\nu) L} \cos\left[ \delta(\tilde\nu) \right] }$ 
            \\ \hline
		reflected 	& $\frac{E_R (\tilde\nu)}{E_0} = \frac{ r_1 - r_2 \, (t_1^2 + r_1^2) \, e^{-\alpha(\tilde\nu) L} \, e^{i \delta(\tilde\nu)} }{1 - r_1 \, r_2 \, e^{-\alpha(\tilde\nu) L} \, e^{i \delta(\tilde\nu)} }$ 
		& $\frac{I_R(\tilde\nu)}{I_0} = \big| \frac{E_R(\tilde\nu)}{E_0} \big|^2 = \frac{R + R (R+T)^2 e^{-2 \alpha(\tilde\nu) L} - 2 R (R+T) e^{-\alpha(\tilde\nu) L} \cos \left[\delta(\tilde\nu) \right] }{1 + R^2  e^{-2 \alpha(\tilde\nu) L} - 2 R e^{-\alpha(\tilde\nu) L} \cos \left[ \delta(\tilde\nu) \right]}$ \\ \hline
		absorbed	&  &	$\frac{I_A(\tilde\nu)}{I_0} = 1 - \frac{I_T(\tilde\nu)}{I_0} - \frac{I_R(\tilde\nu)}{I_0}$\\
		\hline\hline
	\end{tblr}
\end{table*}

\subsection{Deriving the cavity transmission spectrum \label{subsec:derive}}

Consider an FP cavity of length $L$ composed of two parallel mirrors (Fig.~\ref{fig:cavity}). 
An electromagnetic wave of amplitude $E_0$ strikes the system from the left. 
For mathematical convenience, we take this incident light to be continuous-wave (cw) and monochromatic with wavenumber $\tilde\nu \equiv \nu/c$, where $\nu$ is frequency and $c$ is the speed of light.
Our strategy is to calculate the frequency-dependent amplitudes and intensities of light transmitted by, reflected by, and circulating within this structure by summing up the contributions of partial waves transmitted and reflected by each cavity mirror.
The final expressions are summarized in Table \ref{tab:table1}.

\textbf{\textit{Cavity mirrors.}}
We first consider what happens when light strikes a cavity mirror.
The reflected electromagnetic field amplitude is set by the mirror's reflection amplitude coefficient ($r_1, r_2$) while the transmitted amplitude is given by the transmission amplitude coefficient ($t_1, t_2)$.
We will eventually assume that the two FP cavity mirrors are identical (e.g.\ $r_1 = r_2$ and $t_1 = t_2$), but for the moment these indices are useful for bookkeeping.
We will not explicitly consider frequency-dependence of the amplitude coefficients, though this is straightforward to include if necessary.

These reflection and transmission amplitude coefficients are essentially Fresnel coefficients and therefore have some subtle properties.
The amplitude of light transmitted through an interface (at normal incidence) is equivalent when approached from either side of the interface while the amplitude of reflected light flips sign based on the direction of approach.\cite{nagourney14, yariv07}
In particular, a reflected wave undergoes a $\pi$ phase shift when striking an interface from the side with lower refractive index.
Here, we assume that the intracavity material has lower refractive index than the mirror material and use the following convention: if a wave of amplitude $E_0$ impinges on either mirror from \textit{outside} the cavity, the reflected wave will have amplitude $+r_{1,2} \, E_0 $, while the same wave approaching either mirror from \textit{inside} the cavity will have reflected amplitude $- r_{1,2} \, E_0$.
A transmitted wave will have amplitude $+ t_{1,2} \, E_0$ regardless of which side of the mirror it strikes.

The amplitude coefficients can be related to intensity coefficients defined as $R_{1,2} \equiv r_{1,2}^2$ and $T_{1,2} \equiv t_{1,2}^2$.
We will enforce $R_{1,2} + T_{1,2} + \ell_{1,2}= 1$, where $\ell$ represents the fraction of light intensity lost due to absorption or scattering upon a single mirror bounce.
We will often examine our results in the limit where $\ell_{1,2}=0$ and $R_{1,2} + T_{1,2} = 1$.

\textbf{\textit{Intracavity medium.}}
Now, we consider how an electromagnetic wave evolves as it travels between the cavity mirrors.
We assume that the cavity is uniformly filled with a slab of dielectric material with a complex refractive index given by $\tilde{n}(\tilde\nu) = n(\tilde\nu) + i \kappa(\tilde\nu)$, where $n(\tilde\nu)$ is the real (dispersive) component and $\kappa(\tilde\nu)$ is the imaginary (absorptive) component. 
The incident monochromatic light has vacuum wavelength $\lambda_0 = 1/\tilde\nu$, but as this light passes through the intracavity medium, the wavelength is given by $\lambda_n = \lambda_0 / \tilde{n}(\tilde\nu)$.

After a wave of amplitude $E_0$ travels pathlength $L$ through the intracavity medium, it emerges with amplitude $t(\tilde\nu) \, E_0$, where $t(\tilde\nu)$ is given by
\begin{align}
	t(\tilde\nu)	&= e^{2\pi i L / \lambda_n} = e^{2\pi i L \tilde{n}(\tilde\nu) / \lambda_0} \nonumber\\
			&= e^{2\pi i L \, \left[ n(\tilde\nu) + i \kappa(\tilde\nu) \right] \, \tilde\nu} \nonumber\\ 
			&\equiv e^{i \delta(\tilde\nu) / 2} \; e^{-\alpha(\tilde\nu) L / 2} \label{eq:t} 
\end{align}
In the above, we define $\delta(\tilde\nu)$ as the phase accrued by light as it makes one round trip pass through the cavity:
\begin{align}
    \delta(\tilde\nu) \equiv 4 \pi  L  n(\tilde\nu)  \tilde\nu \quad\quad [\textrm{rad}]\label{eq:delta}
\end{align}
and $\alpha(\tilde\nu)$ as the absorption coefficient of the intracavity material:
\begin{align}
	\alpha(\tilde\nu) = 4 \pi \kappa(\tilde\nu) \tilde\nu \quad\quad [\textrm{cm}^{-1}] \label{eq:alpha}
\end{align}
The $e^{i \delta(\tilde\nu) / 2}$ factor in Eq.~\ref{eq:t} captures the phase shift of the wave front caused by refraction in the intracavity medium, while the $e^{-\alpha(\tilde\nu) L / 2}$ term captures the exponential decay of the wave's amplitude due to absorption.
Examining the field intensity by taking $|t(\tilde\nu)|^2$ recovers the Beer-Lambert law.

\textbf{\textit{Cavity field amplitudes.}}
We are now in a position to consider the interference of partial waves traveling in the cavity.
Here, we work through an explicit derivation of $E_T(\tilde\nu)$, the frequency-dependent amplitude of light transmitted through the cavity structure.
$E_T(\tilde\nu)$ is constructed by summing the amplitudes of all partial waves that make any integer number of passes through the cavity and ultimately exit the structure through the back mirror:
\begin{align}
	E_T(\tilde\nu) &= E_0 \cdot t_1 \cdot t(\tilde\nu) \cdot t_2  & \textrm{[1 pass]}\nonumber\\
	&+ E_0 \cdot t_1 \cdot t(\tilde\nu) \cdot -r_2 \cdot t(\tilde\nu) \cdot -r_1  \cdot t(\tilde\nu) \cdot t_2 & \textrm{[2 passes]} \nonumber\\
	&+ \dots \nonumber \\
	&= \; E_0 \; t_1 \;  t_2 \; t(\tilde\nu) \sum_{k=0}^{\infty} \left[ r_1 r_2 \, t(\tilde\nu)^2 \right] ^k \label{eq:ET2} \\
	 &= \; E_0 \left[ \cfrac{t_1 \; t_2 \; t(\tilde\nu)}{1- r_1 r_2 \, t^2(\tilde\nu)^2} \right] \label{eq:ET}
\end{align}
To go from Eq.\ \ref{eq:ET2} to Eq.\ \ref{eq:ET} we make use of the expression for an infinite geometric series: $\sum_{k=0}^\infty a^k = (1-a)^{-1}$ for $|a|<1$. 

The field amplitude reflected from the cavity, $E_R(\tilde\nu)$, and circulating inside the cavity, $E_C(\tilde\nu)$, can be constructed using a similar process. 
The left half of Table \ref{tab:table1} summarizes the results, using the expression for $t(\tilde\nu)$ from Eq.~\ref{eq:t}.
Note that we drop all global phase factors from the amplitude expressions in Table \ref{tab:table1} for brevity, including factors of $e^{i\delta(\tilde\nu)/2}$, phases that result from interactions with the mirrors, and the Gouy phase.
These phase factors are important in some cases -- e.g. to obtain the correct frequencies of transverse cavity modes and the absolute phase of transmitted and reflected beams -- but are not crucial to our discussion here, where we focus on field intensities and neglect all but the lowest-order transverse cavity mode.

\textbf{\textit{Cavity field intensities.}}
Laboratory square-law detectors measure the intensity of light rather than its amplitude. 
We convert the reflected, circulating, and transmitted field amplitudes to fractional intensities according to $I(\tilde\nu)/I_0 = \left|E(\tilde\nu)/E_0 \right|^2 $.
Taking $|E_T(\tilde\nu)/E_0|^2$ with the results of Eq.~\ref{eq:ET}, we arrive at the expression for the frequency-dependent fractional intensity of light transmitted through an FP cavity:
\begin{align}
	 \frac{I_T(\tilde\nu)}{I_0} = \frac{T^2  e^{-\alpha(\tilde\nu) L}}{1 + R^2  e^{-2 \alpha(\tilde\nu) L} - 2 R e^{-\alpha(\tilde\nu) L} \cos \left[ \delta(\tilde\nu) \right]  }\label{eq:IT}
\end{align}
The right side of Table \ref{tab:table1} summarizes all expressions for transmitted, reflected, and circulating intensities of light.
The circulating field intensity $I_C(\tilde\nu,z)/I_0$ is expressed as a function of position along the longitudinal cavity $z$ axis; this will be discussed further in Sections \ref{sec:cavityprimer:empty} and \ref{sec:absorption} below.
The fraction of light absorbed, $I_A(\tilde\nu)/I_0$, can be constructed by assuming that $I_T(\tilde\nu)+I_R(\tilde\nu)+I_A(\tilde\nu)=1$, neglecting any mirror losses.

\subsection{Characterizing a Fabry-P\'{e}rot optical cavity}\label{sec:cavityprimer:empty}

\begin{figure}[t]
    \centering \includegraphics[width=0.95\linewidth]{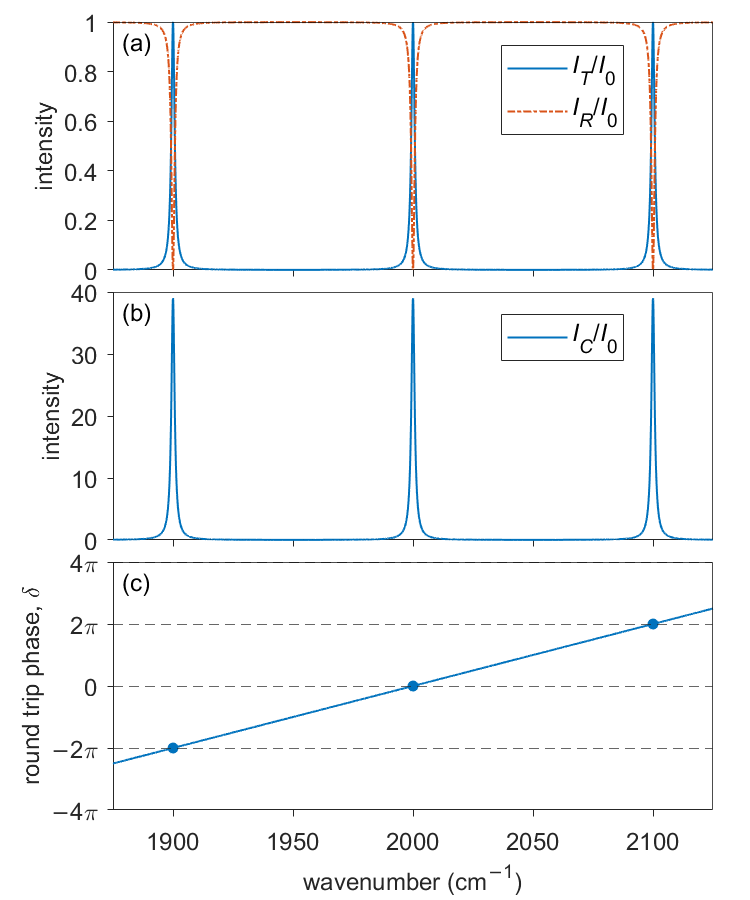}
    \caption{\label{fig:emptycavity} 
    \textbf{(a)} The transmission and reflection spectra of an empty Fabry-P\'{e}rot cavity composed of two identical mirrors, taking $R = 0.95$, $T=0.05$, $\ell = 0$, $L=50\,\upmu$m, $\alpha(\tilde\nu)=0$, and $n(\tilde\nu)=1$. 
    \textbf{(b)} The intensity of light circulating in the same cavity, spatially averaged along the cavity's longitudinal $z$ axis. 
    \textbf{(c)} Maxima in the transmitted and circulating spectra occur when the round-trip phase accrued by light traveling in the cavity, $\delta(\tilde\nu)= 4 \pi  L  n(\tilde\nu) \tilde\nu$, takes on a value equal to an integer multiple of $2\pi$, as marked with blue dots. 
    }
\end{figure}

We now use the FP expressions to derive useful figures of merit for optical cavities and examine their behavior in various limits.
We plot $I_T(\tilde\nu)/I_0$, $I_R(\tilde\nu)/I_0$, $I_C(\tilde\nu)/I_0$, and $\delta(\tilde\nu)$ for an empty cavity in Fig.~\ref{fig:emptycavity}. 

\textbf{\textit{Longitudinal cavity modes.}}
Cavity transmission maxima occur at special resonant wavenumber values $\tilde\nu_m$ that minimize the denominator of Eq.~\ref{eq:IT}. 
This occurs when $\cos\left[ \delta(\tilde\nu) \right] = 1$, and therefore when $\delta(\tilde\nu)$ is equal to an integer multiple of $2\pi$.
Using the expression for $\delta(\tilde\nu)$ from Eq.~\ref{eq:delta} and neglecting, for the moment, the frequency-dependence of the intracavity refractive index by taking $n(\tilde\nu)=n_0$, we find:
\begin{align}
	\delta(\tilde\nu_m) &= 4 \pi L n_0 \tilde\nu_{m} = 2 \pi m, \quad\quad m=1,2,\,\dots  \label{eq:deltam} \\[2mm]
	\rightarrow \quad  \tilde\nu_m &= \frac{1}{2 n_0 L}\cdot m \quad\quad\quad [\textrm{cm}^{-1}] \label{eq:vm}
\end{align}
Fig.~\ref{fig:emptycavity}c shows how these special values of the round-trip phase coincide with the appearance of sharp features in the cavity spectra in Figs.~\ref{fig:emptycavity}ab.
The resonance at $\tilde\nu_m$ corresponds to the $m^\textrm{th}$-order longitudinal cavity mode where the cavity length is commensurate with $m$ half-wavelengths of light, leading to constructive interference as that light travels in the cavity.

\textbf{\textit{Free spectral range.}}
The spacing between neighboring longitudinal modes is termed the free spectral range (\textit{FSR}) and is given by 
\begin{align}
	FSR = \tilde\nu_{m+1} - \tilde\nu_{m}  = \frac{1}{2 n_0 L} \quad\quad [\textrm{cm}^{-1}] \label{eq:fsr}
\end{align}
The \textit{FSR} can be considered a constant so long as the intracavity refractive index is independent of frequency; this approximation breaks down in the presence of significant intracavity dispersion.
We also note that the \textit{FSR} is set by the cavity length -- the longer the cavity, the denser the longitudinal mode spacing -- with no dependence on the mirror reflectivity.

\textbf{\textit{Resonant cavity transmission.}}
On-resonance, when $\cos\left[ \delta(\tilde\nu_m) \right] = 1$, Eq.~\ref{eq:IT} gives:
\begin{align}
    \frac{I_T(\tilde\nu_m)}{I_0} &= \frac{T^2 e^{-\alpha(\tilde\nu)L}}{\left( 1-Re^{-\alpha(\tilde\nu)L}\right)^2}\label{eq:peak} \\[5pt]
    &= \frac{T^2}{(1-R)^2} = \frac{T^2}{(T+\ell)^2} \quad \textrm{for} \quad \alpha(\tilde\nu)=0 \label{eq:peak2}
\end{align}
Eq.~\ref{eq:peak2} approaches 1 in the limit of negligible mirror losses, when $\ell \ll T$.
Wave interference can therefore lead to perfect resonant transmission of light through the cavity as evidenced in Fig.~\ref{fig:emptycavity}a. 
Perfect resonant transmission through the structure occurs despite the high reflectivity of each cavity mirror in isolation; such is the emergent (seemingly magical) behavior of optical cavities.

\textbf{\textit{Intracavity field enhancement.}}
Wave interference also leads to circulating intracavity fields that can be much higher in intensity than the incident field.
The circulating intracavity field for a resonant cavity mode, $I_C(\tilde\nu_m, z)/I_0$, has contributions from both forward- and backward-propagating traveling waves and therefore manifests as a standing wave along the longitudinal cavity $z$ axis.

Again, let's consider the resonant condition for an empty cavity with $\alpha(\tilde\nu)=0$. 
Taking the expression for $I_C(\tilde\nu, z)/I_0$ from Table \ref{tab:table1}, setting $\cos\left[ \delta(\tilde\nu_m) \right] = 1$, and spatially averaging along the longitudinal cavity axis from $z= 0$ to $L$, we find:
\begin{align}
    \frac{I_C(\tilde\nu_m)}{I_0} &= \frac{T(1+R)}{(1-R)^2} = \frac{T(1+R)}{(T+\ell)^2} \\[5pt]
                    &\simeq \frac{(1+R)}{T} = \frac{(1+R)}{(1-R)} \quad\quad \textrm{for} \quad  \ell \ll T \label{eq:IC}
\end{align}
A cavity with $T=0.05$ and $\ell=0$ therefore features a nearly 40-fold magnification of the resonant intracavity field intensity (Fig.~\ref{fig:emptycavity}b).
This phenomenon is capitalized upon in cavity-enhanced spectroscopy (\textit{vide infra}) and for applications in nonlinear optics.\cite{adler2009, pupeza2021}

\textbf{\textit{Cavity mode linewidths.}}
The full-width-at-half-maximum (fwhm) linewidth of a cavity mode, $\Delta \tilde\nu$ can be found by solving for $\delta_{1/2}$, the round-trip phase that when plugged into Eq.~\ref{eq:IT} yields half the peak cavity transmission from Eq.~\ref{eq:peak}:\cite{nagourney14,gagliardi14} 
\begin{align}
    \frac{I_T(\delta_{1/2})}{I_0} &= \frac{T^2 e^{-\alpha(\tilde\nu)L}}{1 + R^2 e^{-2\alpha(\tilde\nu)L}  - 2 R e^{-\alpha(\tilde\nu)L} \cos \left(\delta_{1/2} \right)  } \nonumber \\[5pt] 
    &\stackrel{?}{=} \frac{1}{2} \; \left[ \frac{T^2 e^{-\alpha(\tilde\nu)L}}{\left(1-Re^{-\alpha(\tilde\nu)L} \right)^2} \right]\label{eq:del12}
\end{align}
Making use of the half-angle relation $\cos(\delta_{1/2}) = 1 - 2 \sin^2(\delta_{1/2} / 2)$, we can rearrange Eq.~\ref{eq:del12} to find:
\begin{align}
    \sin( \delta_{1/2}/2) = \frac{1-Re^{-\alpha(\tilde\nu)L}}{2\sqrt{Re^{-\alpha(\tilde\nu)L}}} \label{eq:sind12}
\end{align}
We now take $\delta_{1/2} \ll 1$ so that we can safely replace the sine function in Eq.~\ref{eq:sind12} with its Taylor expansion.
Taking small $\delta_{1/2}$ is equivalent to stating that our cavity has sharp fringes relative to its \textit{FSR}.
We therefore have:
\begin{align}
    \delta_{1/2} \simeq \frac{1-Re^{-\alpha(\tilde\nu)L}}{\sqrt{Re^{-\alpha(\tilde\nu)L}}} \quad\quad [\textrm{rad}]
\end{align}
Converting from phase to wavenumbers and from half-width-at-half-maximum to fwhm, we arrive at the cavity linewidth:
\begin{align}
    \Delta \tilde\nu = 2\delta_{1/2} \cdot \frac{1}{4 \pi n_0 L} = \frac{1}{2\pi n_0 L} \frac{1-R e^{-\alpha(\tilde\nu)L}}{\sqrt{R e^{-\alpha(\tilde\nu)L}}}  \quad [\textrm{cm}^{-1}, \textrm{ fwhm}] \label{eq:deltanu}
\end{align}
$\Delta \tilde\nu$ has an inverse Fourier relationship to the $1/e$ cavity photon lifetime $\tau$ with:\cite{gagliardi14}
\begin{align}
    \tau = \frac{1}{ 2 \pi c \Delta \tilde\nu } \quad\quad [\textrm{s}] \label{eq:tau}
\end{align}

Inspection of Eq.~\ref{eq:deltanu} yields the important finding that intracavity absorption broadens $\Delta \tilde\nu$ compared to the empty cavity case. 
Eq.\ \ref{eq:tau} provides some intuition for this effect: intracavity absorption shortens the residence time of a photon in the cavity, which in turn broadens the frequency-domain cavity mode via the time-frequency Fourier relationship.
Both $\Delta \tilde\nu$ and $\tau$ can therefore be used as proxies for intracavity absorption; see Section \ref{subsec:CES} for further discussion.

Finally, in the empty-cavity limit with $\alpha(\tilde\nu)=0$, Eq.~\ref{eq:deltanu} becomes:
\begin{align}
    \Delta \tilde\nu = \frac{1}{2\pi n_0 L} \frac{1-R}{\sqrt{R}}  \quad\quad [\textrm{cm}^{-1}, \textrm{ fwhm}] \label{eq:deltanu2}
\end{align}
$\Delta \tilde\nu$ is therefore set by both the the cavity length and the mirror reflectivity. 
The longer the cavity or the higher reflectivity the mirrors, the narrower the resonances.

\textbf{\textit{Cavity finesse and quality factor.}}
The cavity finesse, $\mathcal{F}$, is another useful figure of merit. 
$\mathcal{F}$ is a dimensionless quantity defined as the ratio between the \textit{FSR} and $\Delta \tilde\nu$. 
Making use of Eqs.~\ref{eq:fsr} and Eq.~\ref{eq:deltanu}, we find:
\begin{align}
    \mathcal{F} &= \frac{\textit{FSR}}{\Delta \tilde\nu} = \frac{ \pi \sqrt{Re^{-\alpha(\tilde\nu)L}}}{1-Re^{-\alpha(\tilde\nu)L}} \label{eq:finesse1} \\[5pt]
     &\simeq \frac{\pi}{1-R} \quad \textrm{as} \quad R \rightarrow 1, \;\; \alpha(\tilde\nu) \rightarrow 0 \label{eq:finesse2}
\end{align}
In the absence of intracavity absorption, the cavity finesse is set solely by the mirror reflectivity.
As the cavity length grows, both the \textit{FSR} and $\Delta \tilde\nu$ shrink, but their ratio remains fixed, leaving $\mathcal{F}$ unchanged.
That said, significant intracavity absorption will spoil $\mathcal{F}$ according to Eq.~\ref{eq:finesse1}.
We also note that our expression for $\mathcal{F}$ is only as good as our approximation for $\Delta \tilde\nu$, and is therefore meaningful only in the limit that the cavity modes are narrow compared to the \textit{FSR}.

The $R = 0.95$ cavity illustrated in Fig.~\ref{fig:emptycavity} has a modest $\mathcal{F} \sim 60$; this is typical of structures used for condensed-phase strong coupling.
High-finesse cavities with $\mathcal{F}$ on the order of $10^3 - 10^5$ are in common use for cavity-enhanced spectroscopy.\cite{gagliardi14}

Another closely-related dimensionless quantity is the quality factor, $Q$, defined as 
\begin{align}
    Q \equiv  \frac{\tilde\nu_m}{\Delta \tilde\nu} = \frac{\tilde\nu_m \mathcal{F}}{\textit{FSR}} = 2 n_0 L \tilde\nu_m \mathcal{F}
\end{align} 
where $\tilde\nu_m$ is the resonant frequency of the optical mode of interest.
$Q$ and $\mathcal{F}$ are equivalent for the $m=1$ longitudinal mode where $\tilde\nu_m = FSR$.
$Q$ is often used as the figure of merit when considering this lowest-order cavity mode, while finesse is a more handy metric for longer cavities where higher-order longitudinal modes are of more practical interest.

\subsection{Resonant cavity transmission in the presence of intracavity molecules \label{subsec:intra}}
We now inspect how the cavity transmission spectrum behaves when intracavity absorbers are introduced.
We will find that optical cavities can dramatically increase absorption signals as compared to single-pass measurements; this fact is exploited in cavity-enhanced spectroscopy (see Section \ref{subsec:CES}).

Consider an ensemble of molecules with absorption cross section $\sigma(\tilde\nu)$ [cm$^2$/molecule]. 
The absorption cross section is related to the absorption coefficient (Eq.~\ref{eq:alpha}) according to
\begin{align}
        \alpha(\tilde\nu) = \rho \; \sigma(\tilde\nu)    \quad\quad [\textrm{cm}^{-1}]
\end{align}
where $\rho$ [molecules/cm$^3$] is the molecular number density. 

The fraction of light transmitted through a free-space sample of pathlength $L$ is given by the Beer-Lambert law:
\begin{align}
	\frac{I_T(\tilde\nu)}{I_0} &= e^{-\alpha(\tilde\nu) L}\label{eq:beer} \\[5pt]
        &\simeq 1 - \alpha(\tilde\nu) L \quad\quad \textrm{for} \quad \alpha(\tilde\nu) L \ll 1  \label{eq:beer2} 
\end{align}
Some molecules feature very small intrinsic absorption cross sections and some transient or highly reactive species can only be prepared with small $\rho$, in both cases leading to small $\alpha(\tilde\nu)$.
For these challenging systems, the attenuation of transmitted light may be increased by extending $L$, though tabletop laboratory experiments enforce practical constraints.
Optical cavities represent one compact means to scale up the effective pathlength of light through a sample by orders of magnitude.

Consider an optical cavity filled with a material with absorption coefficient $\alpha(\tilde\nu)$. 
We probe this structure using monochromatic cw light resonant with one longitudinal cavity mode such that $\cos\left[ \delta(\tilde\nu) \right] = 1$. 
From Eq.~\ref{eq:peak}, the intensity of light transmitted through the cavity is given by
\begin{align}
	\frac{I_T(\tilde\nu)}{I_0} &= \frac{T^2 e^{-\alpha(\tilde\nu) L}}{(1-Re^{-\alpha(\tilde\nu) L})^2} \label{eq:Leff1}\\[5pt]
                &\simeq \frac{T^2}{(1-R)^2} \left[ 1 - \alpha(\tilde\nu) L \, \frac{(1+R)}{(1-R)}\right] \label{eq:Leff2}\\[5pt]
	           &\equiv \frac{T^2}{(1-R)^2} \left[ 1 - \alpha(\tilde\nu) L_\textrm{eff} \right]	\label{eq:Leff3}
\end{align}
To go from Eq.~\ref{eq:Leff1} to Eq.~\ref{eq:Leff2} we take $\alpha(\tilde\nu) L \ll 1-R$ and apply according Taylor expansions following the derivation given in Chapter 1 of Gagliardi and Loock.\cite{gagliardi14}
In Eq.~\ref{eq:Leff3}, we introduce the effective intracavity pathlength: 
\begin{align}
L_\textrm{eff} &\equiv \frac{(1+R)}{(1-R)} \; L \label{eq:CEF2} \\[5pt] 
                &\simeq \frac{2 \mathcal{F}}{\pi} L \quad \textrm{as} \quad R \rightarrow 1  \label{eq:CEF}
\end{align}
We arrive at Eq.~\ref{eq:CEF} from Eq.~\ref{eq:CEF2} using the approximation for $\mathcal{F}$ from Eq.~\ref{eq:finesse2}. 
Per Eq.~\ref{eq:CEF}, the effective cavity-enhancement of pathlength with respect to free space is often cited as $\beta \mathcal{F}/\pi$.
$\beta=2$ for resonant cw light continuously transmitted  through the cavity, as assumed here.
Other experimental schemes -- including broadband spectroscopy with unresolved cavity modes and cavity ringdown spectroscopy -- feature a smaller effective enhancement factor with $\beta=1$.\cite{thorpe08,adler10, gagliardi14}

In any event, it is instructive to compare the peak cavity transmission intensity in the presence of absorbers (Eq.~\ref{eq:Leff3}) to that of the empty cavity (Eq.~\ref{eq:peak2}). 
The empty cavity transmits a peak fractional intensity of $T^2/(1-R)^2$, while the absorber-containing cavity transmits a factor $1-\alpha(\tilde\nu) L_\textrm{eff}$ less light. 
In contrast, the fraction of light transmitted in a single-pass free-space measurement is given by $1-\alpha(\tilde\nu) L$ (Eq.~\ref{eq:beer2}).
The effective cavity-enhanced absorption signal, $\alpha(\tilde\nu) L_\textrm{eff}$, exceeds the free-space quantity $\alpha(\tilde\nu) L$ by a factor proportional to $\mathcal{F}$, per Eq.~\ref{eq:CEF}.
For the cavity in Fig. \ref{fig:emptycavity} we find an enhancement factor of 39, which is exactly commensurate with the resonant enhancement of the intracavity circulating field given by Eq.~\ref{eq:IC}. 
This convergence is no accident; it is precisely the enhanced intracavity field that leads to increased absorption by the intracavity medium.

We close by re-emphasizing that absorption signals are cavity-enhanced by a factor proportional to $\mathcal{F}$ only in the limit that $\alpha(\tilde\nu) L \ll 1-R \ll 1$. 
If $\alpha(\tilde\nu) L$ is too large, $\mathcal{F}$ is spoiled and is no longer such a useful metric.
Eq.~\ref{eq:Leff1} is still valid in this case, and illustrates that the cavity-enhanced absorption factor falls off as $\alpha(\tilde\nu) L$ increases.
For this reason, cavity-enhanced spectroscopy is used to amplify tiny absorption signals, where you get the most ``bang'' for your finesse.

\subsection{Cavity-enhanced spectroscopy \label{subsec:CES}}

The attenuation, broadening, and shifting of cavity modes by intracavity molecules are the bread and butter of cavity-enhanced spectroscopy.

The typical implementation of cavity-enhanced absorption spectroscopy (CEAS)\cite{gagliardi14} relies on how intracavity molecules attenuate the resonant transmission of light through a cavity, as seen in Section \ref{subsec:intra}. 
In CEAS, cw laser light is resonantly coupled into a single longitudinal cavity mode.
The laser frequency is then stepped -- together with the cavity length -- to sample the molecular absorption spectrum.
Multiplexed CEAS can also be implemented by coupling a broadband light source into the spectrum of longitudinal cavity modes.\cite{thorpe08, adler10, cossel2016, weichman2019, gagliardi14}

Cavity-ringdown spectroscopy (CRDS) is an alternative technique that operates in the time domain.\cite{berden2000} 
In CRDS, a cavity is transiently excited with a short optical pulse or cw light is resonantly coupled into the cavity and rapidly switched off.
In either case, the subsequent exponential decay of light leaking out of the cavity is detected and fit to recover the cavity photon lifetime $\tau$.
As we found in Section \ref{sec:cavityprimer:empty}, intracavity absorption leads to fatter cavity modes (Eq.~\ref{eq:deltanu}) and by extension a shorter cavity lifetime (Eq.~\ref{eq:tau}).
The experimental measure of $\tau$ with CRDS can therefore be used to recover $\alpha(\tilde\nu)$.
The laser frequency and cavity length are stepped together to sample $\tau$ and thereby $\alpha(\tilde\nu)$ as a function of frequency. 
CRDS is extremely sensitive, as tiny changes in cavity lifetime can be measured with high accuracy, resulting in excellent detection limits for intracavity absorption.

As a final example, cavity mode dispersion spectroscopy (CMDS) uses the frequency shifts of cavity modes to measure the intracavity refractive index, $n(\tilde\nu)$.\cite{cygan15} 
So far, we have treated $n(\tilde\nu)$ as a constant $n_0$. 
However, as is clear from Eq.~\ref{eq:vm}, any frequency-dependence of $n(\tilde\nu)$ will shift the positions of longitudinal cavity resonances.
In CMDS, an optical heterodyne scheme is used to measure the absolute shifts of cavity modes to recover $n(\tilde\nu)$, from which the absorption spectrum can be constructed via the Kramers-Kronig relation, which connects the the real and imaginary parts of any complex analytic function.\cite{peiponen2009}
Sufficiently strong intracavity dispersion gives rise to the appearance of polaritonic cavity mode splittings; more on this shortly in Section \ref{sec:linear}.

\subsection{A note on the transfer matrix method}

The transfer matrix method (TMM) is also commonly used to treat the propagation of light through FP cavities \cite{burkhard10, macleod17} and polaritonic devices.\cite{hutchison12, george15, herrera20, xiang21, simpkins23, lee2024controlling} 
TMM is a numerical generalization of the analytical FP cavity treatment which can be applied to multi-layer dielectric stacks.
TMM is more accurate than the FP expressions when the thickness of the cavity mirror coatings is comparable to the cavity length $L$ and explicit treatment of all layers in the structure becomes necessary.
This caveat aside, the analytical FP expressions are attractive both for their low computational cost and for their explicit dependence on physical parameters which can provide useful intuition.
We attempt to leverage this intuition throughout this perspective.


\section{\label{sec:linear} Linear Cavity Spectroscopy Under Strong Light-Matter Coupling}

We have now motivated how heightened light-matter interactions mediated by an optical cavity can enhance the absorption of light.
Strongly-absorbing intracavity molecules can both attenuate the intensity of light transmitted through the cavity and shift the resonant frequencies of cavity modes. 
We will invoke these same phenomena to describe the strong light-matter coupling regime and the formation of polaritons.
In particular, sufficiently large peaks in $\alpha(\tilde\nu)$ result in highly dispersive derivative lineshapes in $n(\tilde\nu)$ that cause drastic changes to the cavity resonance conditions.
We can understand the new spectral features that emerge under these conditions as polaritons.
We now examine the linear spectroscopy of strongly-coupled systems in this context. 

\subsection{Cavity spectra under strong light-matter coupling \label{sec:polaritons}}

\begin{figure}[t]   \includegraphics[width=0.95\linewidth]{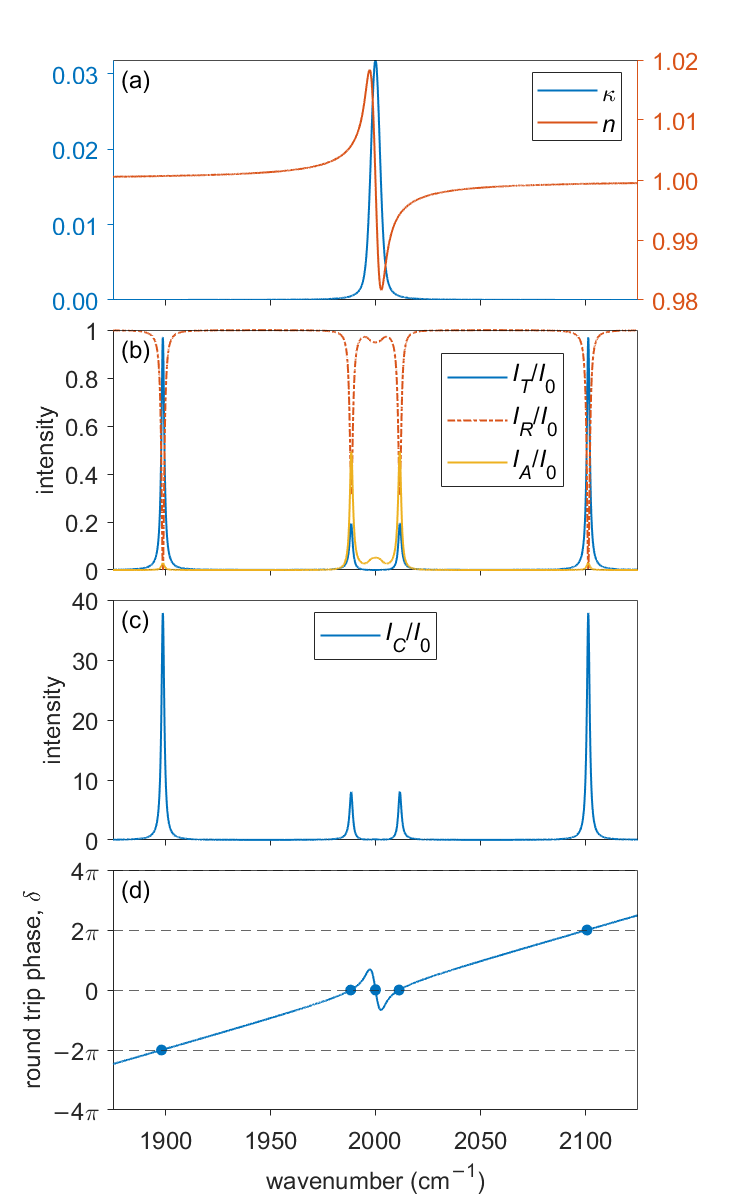}
    \caption{\label{fig:coupledcavity} Optical cavity spectra change drastically in the presence of a strongly-absorbing intracavity medium. 
    \textbf{(a)} The imaginary [$\kappa(\tilde\nu)$, blue] and real [$n(\tilde\nu)$, orange] components of the refractive index for a representative molecular absorber. The absorber has a Voigt lineshape centered at $\tilde\nu_0=$ \SI{2000}{cm^{-1}} with  Lorentzian broadening of $\gamma_L=\SI{2}{cm^{-1}}$ fwhm and  Gaussian broadening of $\gamma_G=\SI{4}{cm^{-1}}$ fwhm. The peak absorption coefficient is $\alpha(\tilde\nu_0) = \SI{800}{cm^{-1}}$. The background refractive index is taken to be $n_0=1$.
    \textbf{(b, c)}	The intensity spectra of light transmitted, reflected, absorbed, and circulating in an $L=50\, \upmu$m FP cavity composed of two identical mirrors, with $R = 0.95$ and $T = 0.05$. The cavity is filled with the material whose complex refractive index is shown in panel (a). Spectra are calculated using the relevant expressions from Table \ref{tab:table1}. The circulating field is spatially averaged along the cavity longitudinal $z$ axis. 
    \textbf{(d)} Maxima in the transmitted and circulating spectra occur when the round-trip phase accrued by light traveling in the cavity, $\delta(\tilde\nu)= 4 \pi  L  n(\tilde\nu) \tilde\nu$, coincides with an integer multiple of $2\pi$, as marked with blue dots.
    }
\end{figure}

Consider an optical cavity containing a material with a single strong absorption band. 
This medium has absorption coefficient $\alpha(\tilde\nu)$ from which we obtain the imaginary component of the refractive index, $\kappa(\tilde\nu)$ using Eq.~\ref{eq:alpha}.
The real component of the refractive index, $n(\tilde\nu)$ can be obtained from $\kappa(\tilde\nu)$ via the Kramers-Kronig relation.\cite{peiponen2009} 
Fig.~\ref{fig:coupledcavity}a plots $\kappa(\tilde\nu)$ and $n(\tilde\nu)$ for a representative absorber with a Voigt absorption lineshape centered at $\tilde\nu_0 = $ \si{2000}{cm$^{-1}$} and a peak absorption coefficient of  $\alpha(\tilde\nu_0) = \SI{800}{cm^{-1}}$.
The sharp peak in $\kappa(\tilde\nu)$ gives rise to a derivative-like feature in $n(\tilde\nu)$, superimposed on top of the background refractive index, here taken to be $n_0 = 1$.

We consider a cavity containing this medium by plugging $\alpha(\tilde\nu)$ and $n(\tilde\nu)$ into the equations in Table \ref{tab:table1} (Figs.~\ref{fig:coupledcavity}bc).
We understand this cavity to be resonant when $\cos\left[ \delta(\tilde\nu) \right] = 1$, where $\delta(\tilde\nu)$ is the round-trip phase accrued by light traveling in the cavity.
In an empty cavity with length $L =\SI{50}{\micro m}$ and intracavity refractive index $n_0 = 1$, the longitudinal mode with $m=20$ would meet the resonance condition for $\tilde\nu_0=\SI{2000}{cm^{-1}}$.
However, once when we place the Voigt absorber in this cavity, the frequency-dependent structure of $n(\tilde\nu)$ carries over into $\delta(\tilde\nu)$, and the cavity resonance conditions become more complicated (Fig.~\ref{fig:coupledcavity}d).
In particular, two additional zero-crossings in $\delta(\tilde\nu)$ arise on either side of $\tilde\nu_0$, representing new constructive interference conditions (see blue circles at $\SI{1988}{cm^{-1}}$ and $\SI{2012}{cm^{-1}}$ in Fig.~\ref{fig:coupledcavity}d).
Two new peaks in the cavity spectra result (Figs.~\ref{fig:coupledcavity}bc).
Note that $\delta(\tilde\nu)$ features a third constructive interference condition at $\tilde\nu_0=$ \SI{2000}{cm^{-1}}, yet no corresponding cavity mode appears in the transmission spectrum.
$\alpha(\tilde\nu)$ is simply too strong near $\tilde\nu_0$, killing the cavity finesse and preventing the transmission of light; see Section \ref{sec:absorption} for further discussion.

The appearance of two new spectral features symmetrically split about $\tilde\nu_0$ -- where only a single mode appears for the empty cavity (Fig.~\ref{fig:emptycavity}) -- is the classical manifestation of polariton formation.
The FP treatment provides an interpretation of these polaritonic modes as new cavity resonances that arise due to the strong dispersion of the intracavity medium.
While the FP treatment affords no simple closed-form expression for the splitting between the polariton modes, it is qualitatively clear that this splitting arises from intracavity dispersion and will therefore grow with the integrated line strength of $\alpha(\tilde\nu)$ and by extension the absorption cross section and number density of intracavity molecules.
The splitting between the polaritonic FP cavity modes is known to coincide empirically with what cQED predicts for the collective vacuum Rabi splitting, including scaling with $\sqrt{N}$.\cite{simpkins2015,herrera20,bala21,wright23}

\subsection{Cavity detuning and dispersion \label{sec:dispersion}}
\begin{figure}
\includegraphics[width=0.95\linewidth]{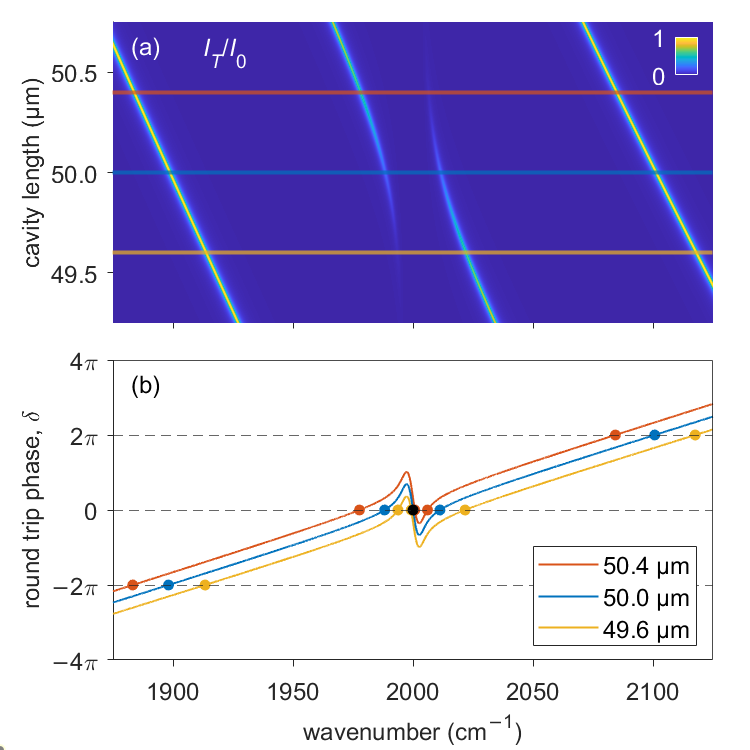}
   \caption{\label{fig:detuning}
   \textbf{(a)} Cavity transmission spectra simulated as a function of cavity length, $L$.  The cavity and intracavity medium are the same as those simulated in Fig.~\ref{fig:coupledcavity}, albeit with $L$ scanned from \SI{49.25}{\micro m} to \SI{50.75}{\micro m}. The polaritonic cavity modes exhibit an avoided crossing as they pass through resonance with the $\tilde\nu_0 = \SI{2000}{cm^{-1}}$ band center. 
   \textbf{(b)} The round-trip phase accrued by light traveling in the cavity, $\delta(\tilde\nu)= 4 \pi  L  n(\tilde\nu) \tilde\nu$, calculated for $L= 49.6$, $50.0$, and and \SI{50.4}{\micro m}, corresponding to the color-coded lines in panel (a). Resonant frequencies that yield constructive cavity interference are marked with colored dots. As $L$ increases, $\delta(\tilde\nu)$ is offset vertically and the resonant frequencies redshift. 
   }
\end{figure}

Systematic detuning of a cavity from resonance with a molecular transition is often used to characterize the dispersion of the polariton modes.
FP cavities may be detuned by scanning the cavity length $L$ or by tilting the angle of the mirrors with respect to the incident beam of light.
Fig.~\ref{fig:detuning} plots the transmission spectrum for the system introduced in Fig.~\ref{fig:coupledcavity} as a function of $L$.
The polaritonic cavity modes exhibit an avoided crossing as they pass through resonance with the absorption line at $\tilde\nu_0 = \SI{2000}{cm^{-1}}$.
The dependence of the transmission spectrum on cavity detuning can be understood by inspecting how the round-trip phase, $\delta(\tilde\nu)$ depends on $L$ (Eq.~\ref{eq:delta}).
As seen in Fig.~\ref{fig:detuning}b, sweeping $L$ vertically offsets the frequency-dependent structure in $\delta(\tilde\nu)$ and thereby changes the resonant frequencies where $\cos\left[ \delta(\tilde\nu) \right] = 1$.

The minimum gap between the two polariton modes across all detuning conditions yields the Rabi splitting.
This minimum-splitting condition occurs when the two polaritonic cavity modes feature equal transmission intensity, provided the absorption lineshape of the intracavity absorber is symmetric about $\tilde\nu_0$.
The dispersing polariton modes decrease in transmission intensity as they draw closer to $\tilde\nu_0$ and their spectral overlap with the intracavity molecular absorption band increases.
We will return to discuss how light is absorbed at the polariton frequencies in Section~\ref{sec:absorption}.

\subsection{Polariton linewidths\label{sec:linewidths}}

Broadening of the intracavity molecular absorption band can have interesting consequences for the linewidths of the resulting polaritonic resonances.
Polariton modes are known to exhibit linewidths consistent with the mean of the empty cavity linewidth ($\Delta\tilde\nu$) and the homogeneous molecular absorption linewidth ($\gamma_L$), while remaining seemingly insensitive to the inhomogeneous linewidth of intracavity molecules ($\gamma_G$).
This trend was first noted by Houdr\'{e} \textit{et al}.\cite{houdre1996} and has since been discussed by many others.\cite{ell1998,long2015,herrera20, schwennicke2024}
FP cavity physics provides some useful intuition for this phenomenon.
Throughout our treatment here, we make the standard assumption that homogeneous broadening processes convolute the molecular absorption coefficient with a Lorentzian lineshape while inhomogeneous broadening processes convolute with a Gaussian lineshape.

\begin{figure}
   \includegraphics[width=0.95\linewidth]{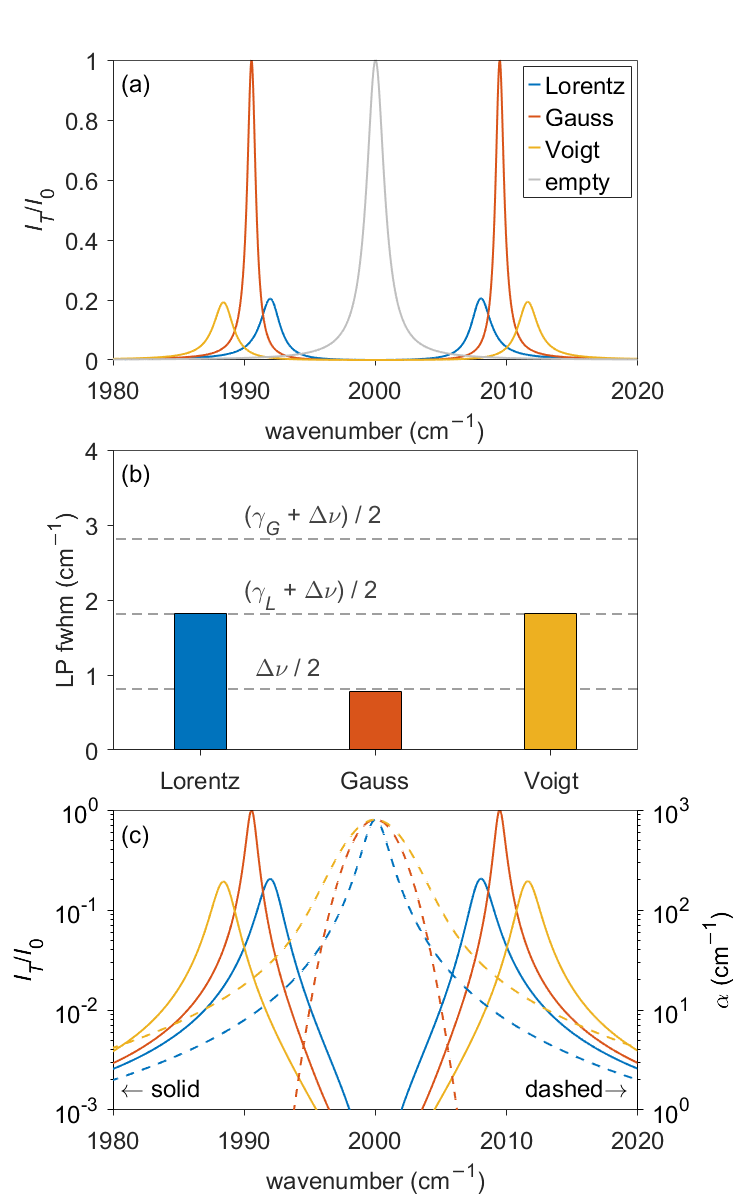}
    \caption{\label{fig:linewidths} 
    \textbf{(a)} Simulated transmission spectra for cavities containing molecular absorbers with Lorentzian (blue), Gaussian (orange), and Voigt (yellow) lineshapes centered at $\tilde\nu_0=\SI{2000}{cm^{-1}}$, plotted against the transmission spectrum of an empty cavity (grey). The Lorentzian and Gaussian linewidths are $\gamma_L=\SI{2}{cm^{-1}}$ fwhm and $\gamma_G=\SI{4}{cm^{-1}}$ fwhm, respectively, while the Voigt lineshape is a convolution of these two components. The peak absorption coefficient is $\alpha(\tilde\nu_0) = \SI{800}{cm^{-1}}$ for all three species though their integrated line strengths differ, leading to distinct polaritonic splittings. The cavity parameters are identical to those of Fig.~\ref{fig:coupledcavity} with $L=\SI{50}{\micro m}$, $R = 0.95$, and $T=0.05$. The cavity has a linewidth of $\Delta\tilde\nu =$ \SI{1.6}{cm^{-1}} fwhm when empty.
    \textbf{(b)} Extracted fwhm linewidth of the lower polariton (LP) for each transmission spectrum shown in panel (a). 
    \textbf{(c)} Cavity transmission spectra (solid lines) and absorption coefficients (dashed lines) for the three absorbing systems plotted on a log scale. The degree of intracavity absorption at the polariton mode frequency determines both the polariton linewidth and transmitted intensity.}
\end{figure}

Fig.~\ref{fig:linewidths}a shows simulated transmission spectra for cavities containing molecular absorbers with three distinct lineshapes:
a homogeneous Lorentzian lineshape ($\gamma_L = \SI{2}{cm^{-1}}$), 
an inhomogeneous Gaussian lineshape ($\gamma_G = \SI{4}{cm^{-1}}$), 
and a Voigt lineshape (convolution of $\gamma_L = \SI{2}{cm^{-1}}$ and $\gamma_G = \SI{4}{cm^{-1}}$).
All three lineshapes peak at $\tilde\nu_0 = \SI{2000}{cm^{-1}}$ with a maximum absorption coefficient of $\alpha(\tilde\nu_0) = \SI{800}{cm^{-1}}$, though the three bands differ in integrated line strength due to their different linewidths.
Each absorber is coupled to a cavity that features linewidth $\Delta\tilde\nu = \SI{1.6}{cm^{-1}}$ fwhm when empty.

The polariton modes in the three transmission spectra shown in Fig.~\ref{fig:linewidths}a display distinct linewidths, transmission intensities, and Rabi splittings.
The variation in Rabi splittings for the three absorbers derives from differences in their integrated line strengths, as we have kept only the peak absorption coefficients fixed across the three systems.
The variations in polariton linewidth and intensity are more interesting; we argue that both changes in both these quantities stem from the degree of intracavity absorption at the frequencies of the polariton modes.

We chart the fwhm linewidths of the lower polariton (LP) modes for each intracavity system in Fig.~\ref{fig:linewidths}b.
The LPs of the Lorentzian and Voigt systems both feature linewidths commensurate with the mean of the homogeneous absorption linewidth and empty cavity linewidth, $(\gamma_L + \Delta \tilde\nu)/2 = \SI{1.8}{cm^{-1}}$ fwhm.
Meanwhile, the Gaussian system's LP features a linewidth commensurate with just half the empty cavity linewidth, $\Delta \tilde\nu/2 = \SI{0.8}{cm^{-1}}$ fwhm.
In all cases, the polariton linewidths are seemingly blind to Gaussian broadening of intracavity molecules.
This is in some sense an amazing result: the linear cavity transmission spectrum reads out the Lorentzian linewidth of the intracavity molecules, cutting through Gaussian broadening of the ensemble.

We can draw upon the FP expression that relates cavity linewidth to intracavity absorption (Eq.~\ref{eq:deltanu}) to rationalize these findings.
Fig.~\ref{fig:linewidths}c plots $\alpha(\tilde\nu)$ for the Lorentzian, Gaussian, and Voigt absorbers (dashed lines) alongside the strongly-coupled cavity transmission spectrum of each system (solid lines).
The Lorentzian and Voigt lineshapes feature significant absorption at the polariton frequencies, while the wings of the Gaussian lineshape drop off more dramatically away from the band center. 
We can therefore understand the insensitivity of the polariton mode linewidths to inhomogeneous broadening as a result of the compact shape of the Gaussian distribution.
Meanwhile, both the Lorentzian and Voigt absorbers confer their homogeneous linewidths onto the polariton bands by virtue of their long absorbing tails.
With reference to Eq.~\ref{eq:peak}, the same considerations explain the nearly perfect transmission of light through the polariton bands for the cavity containing the Gaussian absorber, as compared to the weakly transmitting Lorentzian and Voigt systems.

Inhomogeneous broadening of the intracavity absorber may still influence polariton linewidths in certain cases.
For instance, when the Rabi splitting is only marginally larger than the Gaussian linewidth, the polaritons fall close enough to the peak molecular absorption to spoil their linewidths, as Houdr\'{e} \textit{et al.} also note.\cite{houdre1996}
In real-world experiments, imperfect parallelism between the two mirrors of a planar FP cavity can also lead to inhomogeneous spatial broadening of the polariton modes.
Imperfectly parallel cavities have been simulated using TMM;\cite{michon24} such effects could also be accounted for by averaging the FP expressions over a distribution of cavity lengths.

\subsection{Absorption of light by a strongly-coupled system \label{sec:absorption}} 

\begin{figure}
    \includegraphics[width=0.95\linewidth]{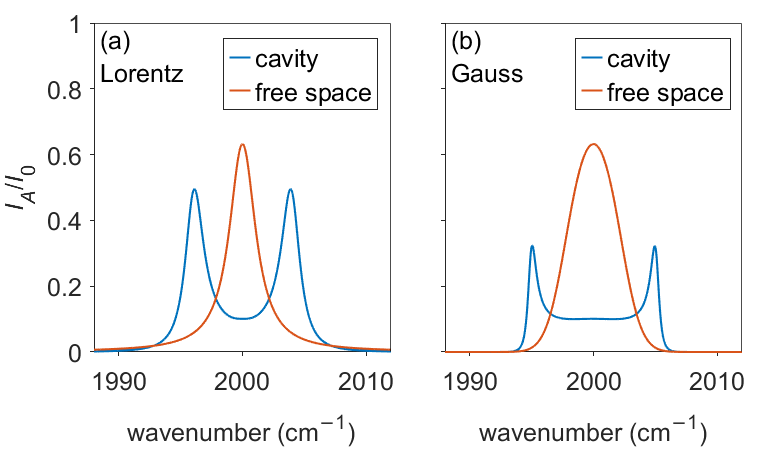}
    \caption{\label{fig:polabs} 
    Simulated absorption spectra for cavities containing \textbf{(a)} a Lorentzian absorber with $\gamma_L=\SI{2}{cm^{-1}}$ fwhm and \textbf{(b)} a Gaussian absorber with $\gamma_G=\SI{4}{cm^{-1}}$ fwhm. Both absorption lineshapes are centered at $\tilde\nu_0=\SI{2000}{cm^{-1}}$ with a maximum absorption coefficient of $\alpha(\tilde\nu_0) = \SI{200}{cm^{-1}}$. Traces in blue represent the fraction of light absorbed by species embedded in an $L=\SI{50}{\micro m}$, $R = 0.95$, $T=0.05$ FP cavity. Traces in orange represent the Beer-Lambert absorption of the same species over a \SI{50}{\micro m} free-space pathlength. 
    These systems lie closer to the onset of strong coupling than those in Fig.~\ref{fig:linewidths} to emphasize absorption at the polariton frequencies. 
    }
\end{figure}

We have described polaritons as optical cavity modes that arise from extreme dispersion of light by the intracavity medium. 
What, then, does it mean for these modes to absorb light?
Within the FP framework, only the intracavity molecules can absorb light (neglecting mirror losses).
Fig.~\ref{fig:polabs} shows fractional absorption spectra for molecules with Lorentzian and Gaussian lineshapes, both when enclosed in a resonant strongly-coupled $L=\SI{50}{\micro m}$ cavity (blue traces) and over a \SI{50}{\micro m} free-space pathlength (orange traces).
For both molecular absorbers, strong cavity coupling leads to enhanced absorption at the polariton frequencies and reduced absorption at the original molecular band center.
We now explore these observations further.

\textit{\textbf{Absorption of light at the polariton frequencies.}} 
While the intracavity absorption spectra in Fig.~\ref{fig:polabs} feature differences in spectral shape for the Lorentzian and Gaussian absorbers, both systems feature strong absorption at the polariton modes that dwarfs the free-space single-pass absorption at the same frequencies.
We can understand this as cavity-enhanced absorption (see Section \ref{subsec:intra}) in the wings of the intracavity absorption lineshape.

Incident light resonant with a polaritonic cavity mode will make several passes through the intracavity medium, increasing the effective pathlength for absorption.
In the examples shown in Fig.~\ref{fig:polabs} for an FP cavity with $R=0.95$, we would naively expect, per Eq.~\ref{eq:CEF2}, a cavity enhancement factor of $(1+R)/(1-R) = 39$.
This rule of thumb will overestimate the polariton absorption, as the large values of $\alpha(\tilde\nu)$ required to achieve strong coupling do not permit the use of the Taylor expansions necessary to recover Eq.~\ref{eq:CEF2}.
In the Lorentzian case (Fig.~\ref{fig:polabs}a), absorption at the polariton modes is enhanced by 8-fold compared to free space, while the Gaussian case (Fig.~\ref{fig:polabs}b) achieves a 25-fold absorption enhancement.

Cavity-enhanced absorption under strong coupling therefore has the interesting consequence of favoring excitation in the wings of the absorption lineshape.
This phenomenon has distinct consequences for homogeneously and inhomogeneously broadened intracavity molecules, as others have also argued.\cite{pyles2024}
For molecules featuring a single homogeneously-broadened transition, incident light of any frequency can be absorbed by any member of the ensemble.
For an inhomogeneously-broadened system, on the other hand, different subclasses of the ensemble absorb preferentially at different frequencies.
In this latter case, light traveling through the cavity at the polariton frequencies will be selectively absorbed by the fringe sub-populations whose transitions lie far from the band center.
We will return to this topic in Section \ref{sec:nonlinear} in the context of nonlinear spectroscopy of strongly-coupled systems.

\begin{figure}
	\includegraphics[width=0.95\linewidth]{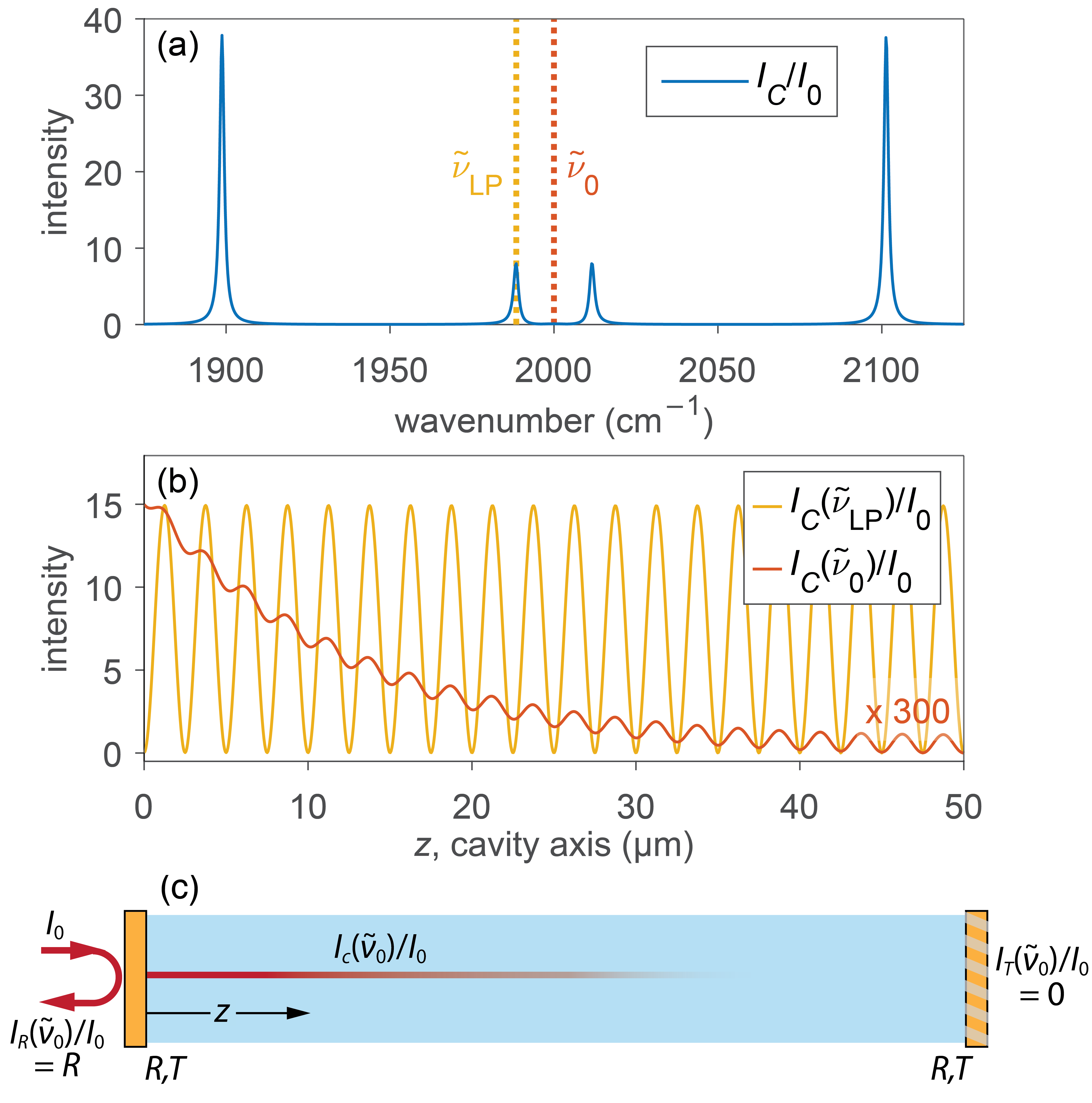}
    \caption{\label{fig:resabs} 
    Cavity interference conditions break down in the face of strong intracavity absorption. 
    \textbf{(a)} Circulating intracavity field intensity spectrum reproduced from Fig.~\ref{fig:coupledcavity}c for an $L=50\, \upmu$m, $R = 0.95$, $T = 0.05$ cavity containing a Voigt absorber with $\tilde\nu_0=$ \SI{2000}{cm^{-1}}, $\gamma_L=\SI{2}{cm^{-1}}$ fwhm, $\gamma_G=\SI{4}{cm^{-1}}$ fwhm, and $\alpha(\tilde\nu_0) = \SI{800}{cm^{-1}}$. The circulating field is spatially averaged along the cavity longitudinal axis from $z = 0$ to $L$. The vertical yellow and orange dashed lines indicate the positions of the lower polariton ($\tilde\nu_\textrm{LP}$) and absorption band center ($\tilde\nu_0$), respectively. 
    \textbf{(b)} The intracavity field intensity plotted as a function of the longitudinal cavity \textit{z} axis for incident light of wavenumber $\tilde\nu_\textrm{LP}$ (yellow) and $\tilde\nu_0$ (orange). 
    \textbf{(c)} When an electromagnetic wave of intensity $I_0$ and wavenumber $\tilde\nu_0$ impinges on a strongly-coupled FP cavity, 
    the intracavity intensity decays exponentially with $z$ due to strong absorption at $\tilde\nu_0$. Most of the incident light never reaches the back cavity mirror, resulting in negligible transmission through the structure and minimal back-reflection to interfere with the incident beam.
    }
\end{figure}

\textit{\textbf{Absorption of light at the band center.}} 
The strongly-coupled systems explored so far in Figs.~\ref{fig:coupledcavity}, \ref{fig:linewidths}, and \ref{fig:polabs}  all feature non-negligible absorption at the absorption band center, $\tilde\nu_0$, despite minimal transmission of light through the cavity at $\tilde\nu_0$.
To understand this, we consider the spatial structure of the circulating intracavity field. 
Fig.~\ref{fig:resabs}a plots the average intracavity intensity spectrum for a representative strongly-coupled FP cavity, while Fig.~\ref{fig:resabs}b plots spatial profiles of the intracavity field for light resonant with the lower polariton mode ($\tilde\nu_\textrm{LP}$) and the absorption band center ($\tilde\nu_0$).
Light coupled into the lower polariton forms a standing wave along $z$, as expected for a typical longitudinal cavity mode.   
Incident light at $\tilde\nu_0$ has very different behavior, decaying exponentially along the $z$ axis with only slight oscillatory character. 

Fig.~\ref{fig:resabs}c illustrates how we can understand this phenomenon.
When light of intensity $I_0$ and wavenumber $\tilde\nu_0$ is incident on the input cavity mirror, a fraction $R$ of that light is immediately reflected, while the remaining fraction $1-R$ is transmitted into the cavity.
The intensity of this transmitted beam decays exponentially with $e^{-\alpha(\tilde\nu_0) z}$ as it travels along the longitudinal cavity axis.
As $\alpha(\tilde\nu_0)$ must be large to facilitate strong coupling, only a very small intensity of light survives to reach the back cavity mirror.
As a result, minimal light is reflected from the back mirror, so the intracavity wave interference depicted in Fig.~\ref{fig:cavity} cannot occur. 
We can therefore neglect partial reflections and trace only the first pass of light through the structure:
$I_T(\tilde\nu_0)/I_0$ tends towards zero as no light makes it to the back mirror;
$I_R(\tilde\nu_0)/I_0$ tends towards $R$, deriving only from the first bounce of incident light off the input mirror;
and the $1-R$ fraction of light initially transmitted past the input mirror is entirely absorbed in its first pass through the cavity, so $I_A(\tilde\nu_0)/I_0$ tends towards $1-R$.

We have now motivated how strongly-coupled molecules directly absorb a fraction $1-R$ of incident light at the free-space band center $\tilde\nu_0$.
This phenomenon has been observed experimentally, particularly along the excitation axis in two-dimensional spectra of strongly coupled systems.\cite{xiang18} 
Absorption of light at $\tilde\nu_0$ is sometimes ascribed to a signature of the dark states which arise in the collective cQED description as delocalized molecular excitations that gain slight photonic character only in the presence of disorder.\cite{Botzung2020}
We are now in a position to motivate a classical analog for these dark states.
In the FP picture, the dark states represent the ensemble of intracavity molecules which are ``protected'' from incident light by the input cavity mirror.
These molecules are not perfectly dark because the input cavity mirror is not perfectly reflective; in the limit that $R\rightarrow 1$, the system would indeed absorb no light at $\tilde\nu_0$.

\subsection{Multi-mode strong coupling}

Experimental realizations of strong light-matter coupling are rarely so simple as the interaction of a single molecular transition with a single cavity mode.
Here, we touch on two situations with more complex cavity-coupling conditions that give rise to additional polaritonic resonances.

\begin{figure}
   \includegraphics[width=0.95\linewidth]{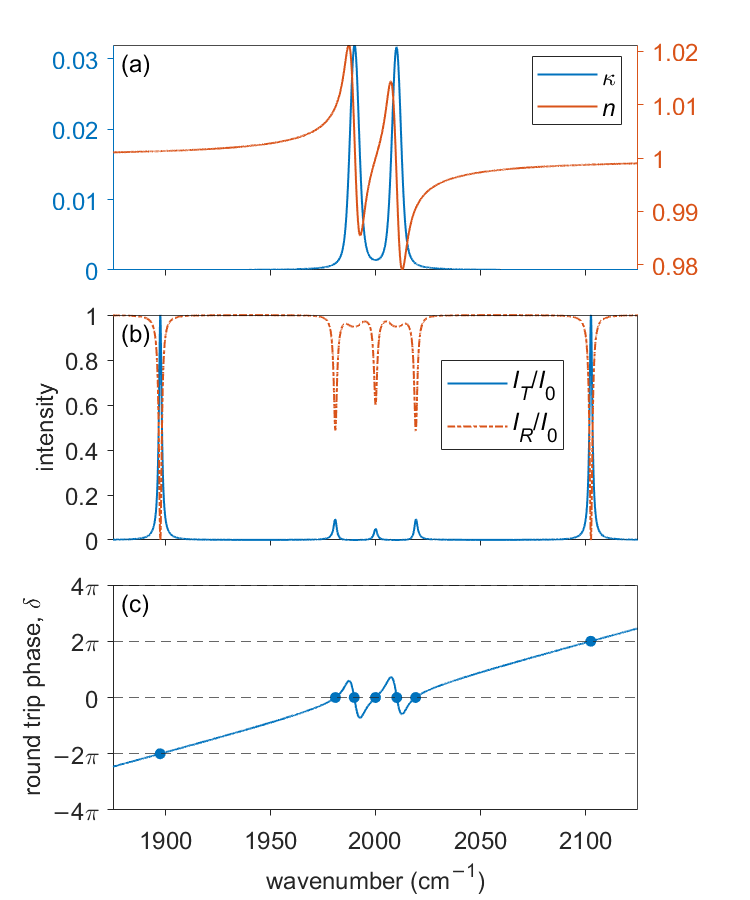}
    \caption{\label{fig:middle_polariton} Middle polariton formation. 
    \textbf{(a)} The imaginary [$\kappa(\tilde\nu)$, blue] and real [$n(\tilde\nu)$, orange] components of the refractive index for an  intracavity molecular absorber with two close-lying transitions. The absorption lines have Voigt lineshapes centered at $\tilde\nu_A=$ \SI{1990}{cm^{-1}}  and $\tilde\nu_B=$ \SI{2010}{cm^{-1}} with $\gamma_L=\SI{2}{cm^{-1}}$ fwhm Lorentzian broadening and $\gamma_G=\SI{4}{cm^{-1}}$ fwhm Gaussian broadening. The peak absorption coefficient of each line is $\alpha(\tilde\nu_A) = \alpha(\tilde\nu_B) = \SI{800}{cm^{-1}}$. The background refractive index is taken to be $n_0=1$.
    \textbf{(b)}	The intensity spectra of light transmitted and reflected by an $L=50\, \upmu$m FP cavity composed of two identical mirrors, with $R = 0.95$ and $T = 0.05$. The cavity is filled with the material whose complex refractive index is shown in panel (a). 
    \textbf{(c)} Maxima in the transmitted spectrum occur when the round-trip phase accrued by light traveling in the cavity, $\delta(\tilde\nu)= 4 \pi  L  n(\tilde\nu) \tilde\nu$, takes on a value equal to an integer multiple of $2\pi$, as marked with blue dots. 
    }
\end{figure}

\textbf{\textit{Multiple molecular transitions coupled to a single cavity mode.}}
When intracavity molecules feature two close-lying absorption lines and a single cavity mode is tuned between them, three peaks -- the upper, middle, and lower polaritons -- can be observed in the cavity transmission spectrum.
This has been demonstrated experimentally in various systems.\cite{xiang20, georgiou2021, chen22, son2022}
To capture this phenomenon, we simulate an intracavity medium with two absorption lines at $\tilde\nu_A=\SI{1990}{cm^{-1}}$ and $\tilde\nu_B=\SI{2010}{cm^{-1}}$ (Fig.~\ref{fig:middle_polariton}a) embedded in an $L=\SI{50}{\micro m}$ FP cavity.
The transmission and absorption spectra of this system indeed feature three polaritonic resonances (Fig.~\ref{fig:middle_polariton}b).
In Fig.~\ref{fig:middle_polariton}c, we plot $\delta(\tilde\nu)$, the round-trip phase accrued by light traveling in this structure, which can be compared to that of the simpler system in Fig.~\ref{fig:coupledcavity}d.
Here, additional constructive interference conditions arise for $\delta(\tilde\nu)$ caused by the two absorption features which yield two corresponding derivative lineshapes in $n(\tilde\nu)$.
A middle polariton must arise between the two absorption features because $\delta(\tilde\nu)$ must cross back through 0 to connect the two derivative lineshapes centered at $\tilde\nu_A$ and $\tilde\nu_B$ Fig.~\ref{fig:middle_polariton}c.

The middle polariton is described within cQED as a hybrid light-matter state delocalized across two molecular transitions.
The classical analogy is evident here: the middle polariton can be understood as a cavity mode that overlaps with the tails of two distinct absorption lines. 
Were we to pump the middle polariton with resonant light, simultaneous cavity-enhanced excitation of both molecular transitions would result.

\textbf{\textit{Multiple cavity modes coupled to a single molecular transition.}}
We now consider intracavity molecules with a single strong absorption line coupled to a cavity.
When the Rabi splitting approaches the cavity \textit{FSR}, such a system can enter the ``multimode'' coupling regime, where multiple nested pairs of polaritons form, as has been observed experimentally.\cite{yu2009,sundaresan2015,george2016,hertzog2020,bala21,wright2023_2}

\begin{figure}
   \includegraphics[width=0.95\linewidth]{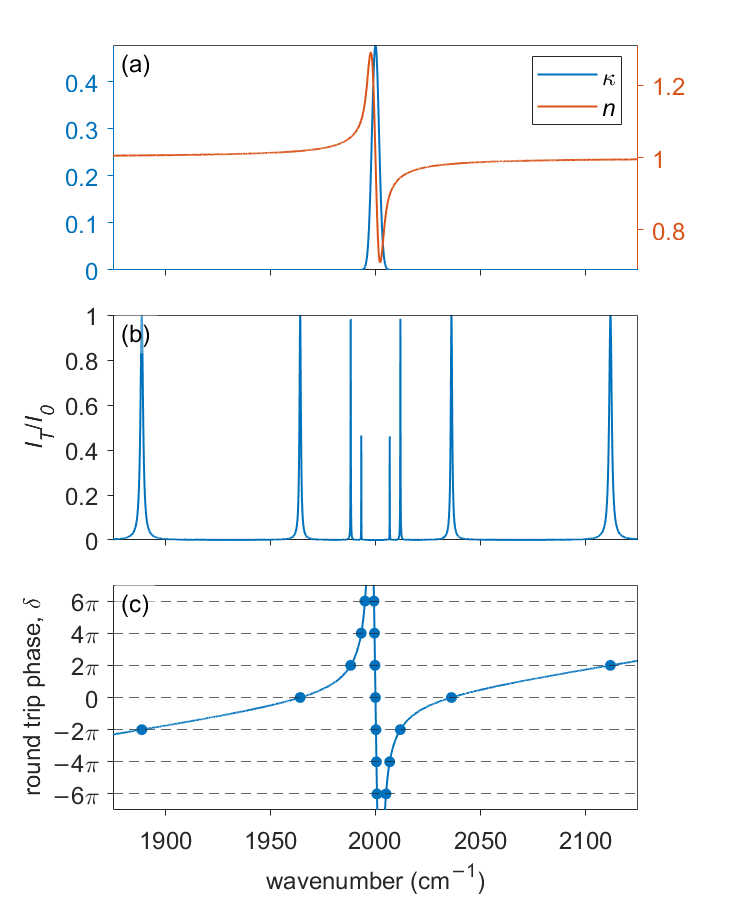}
    \caption{\label{fig:multimode} 
    \textbf{(a)} The imaginary [$\kappa(\tilde\nu)$, blue] and real [$n(\tilde\nu)$, orange] components of the refractive index for an intracavity absorber with a single strong transition. The absorption line is centered at $\tilde\nu_0=$ \SI{2000}{cm^{-1}} with $\gamma_G=\SI{4}{cm^{-1}}$ fwhm Gaussian broadening. The peak absorption coefficient is $\alpha(\tilde\nu_0) = \SI{12000}{cm^{-1}}$. The background refractive index is taken to be $n_0=1$.
    \textbf{(b)} The intensity spectrum of light transmitted by an $L=50\, \upmu$m FP cavity composed of two identical mirrors, with $R = 0.95$ and $T = 0.05$. The cavity is filled with the material whose complex refractive index is shown in panel (a).
    \textbf{(c)} Maxima in the transmitted spectrum occur when the round-trip phase accrued by light traveling in the cavity, $\delta(\tilde\nu)= 4 \pi  L  n(\tilde\nu) \tilde\nu$, takes on a value equal to an integer multiple of $2\pi$, as marked with blue dots. 
    }
\end{figure}

To illustrate this phenomenon, Fig.~\ref{fig:multimode}a plots the complex refractive index of a medium with a Gaussian absorption line at $\tilde\nu_0 = \SI{2000}{cm^{-1}}$ whose peak absorption coefficient of $\alpha(\tilde\nu_0) = \SI{12000}{cm^{-1}}$ is 15 times stronger than that of the medium considered in Fig.~\ref{fig:coupledcavity}.
Fig.~\ref{fig:multimode}b plots the transmission spectrum for this absorber embedded in an $L=\SI{50}{\micro m}$ cavity.
At least three nested pairs of polaritonic modes are visible in the transmission spectrum centered about $\tilde\nu_0$.
The cavity round-trip phase plotted in Fig.~\ref{fig:multimode}c provides an explanation: $n(\tilde\nu)$ is sufficiently dispersive that $\delta(\tilde\nu)$ crosses several integer multiples of 2$\pi$ near $\tilde\nu_0$, leading to several new constructive interference conditions.
Only a handful of these interference conditions lead to observable peaks in the cavity transmission spectrum; would-be modes lying too close to $\tilde\nu_0$ are damped by strong intracavity absorption.
Indeed, we use an absorber with a Gaussian lineshape to illustrate this effect, as long Lorentzian absorption tails would exacerbate the damping of these modes and make this phenomenon more difficult to observe. 
The formation of these nested polaritons can also be understood as off-resonant coupling of the one strong absorption line to several neighboring longitudinal cavity modes.

\section{\label{sec:nonlinear} Nonlinear Cavity Spectroscopy Under Strong Light-Matter Coupling}
So far we have considered the linear spectroscopy of strongly-coupled cavities, assuming interactions with weak incident fields that leave the vast majority of intracavity molecules in their ground state.
Linear spectroscopy is very useful as a characterization tool to confirm polariton formation or to monitor how slow thermal processes proceed under cavity coupling.
In nonlinear experiments, by contrast, interactions with multiple intense incident electromagnetic fields can be used to interrogate the dynamics of intracavity molecular processes with high time resolution.

A central question in molecular polaritonics is whether strongly-coupled intracavity species exhibit distinct behavior from their free-space counterparts.
It is appealing to use transient spectroscopy to examine, e.g., excited state relaxation pathways, energy transport, and reaction dynamics in strongly-coupled systems. 
Transient nonlinear spectroscopies are extremely powerful, but require nuance when applied to polaritonic systems which feature strong feedback between the intracavity molecules and the cavity field that can manifest in non-intuitive ways.
We must account for the classical optical interference effects that crop up in nonlinear experiments before we can confidently assess the behavior of intracavity molecules.

Here, we discuss some subtleties inherent in performing nonlinear spectroscopy experiments in cavities, and motivate how the FP cavity expressions prove useful for treating nonlinear spectroscopy.
We also refer the reader to other recent work covering related aspects of transient cavity spectroscopy.\cite{renken21, liu2021, schwennicke2024, pyles2024}

\subsection{Transient cavity spectroscopy with classical optics} \label{sec:workflow}

Let's begin by laying out how a simple transient pump-probe cavity transmission measurement proceeds.
Consider an intracavity molecular ensemble with an absorption transition centered at $\tilde\nu_0$ resonantly coupled to an FP cavity mode.
We excite this system along the longitudinal cavity axis at time $\tau_1$ with a broadband pump pulse with spectrum $I_0(\tilde\nu)$. 
We then record a transmission spectrum at some later time $\tau_2 > \tau_1$ with a probe pulse that features the same broadband spectrum as the pump.
We can break the light-matter feedback during this experiment up into the following steps:
\begin{enumerate}

    \item   \textbf{\textit{The pump light undergoes optical filtering.}}
    When incident broadband pump light impinges on the cavity at time $\tau_1$, it is optically filtered by the initial cavity transmission spectrum $I_T(\tilde\nu, \tau_1)/I_0(\tilde\nu)$, calculated using Eq.~\ref{eq:IT}.
    The pump pulse drives a corresponding intracavity field with spatially-averaged intensity $I_C(\tilde\nu, \tau_1)/I_0(\tilde\nu)$.
    Regions of the pump spectrum that are resonant with cavity transmission windows experience cavity enhancement, while off-resonant frequencies are rejected.

    \item   \textbf{\textit{The intracavity molecules absorb filtered pump light.}} 
    The intracavity molecules absorb pump light at time $\tau_1$ according to the initial cavity absorption spectrum $I_A(\tilde\nu, \tau_1)/I_0(\tilde\nu)$.
    The frequency-dependence of this absorption arises from the spectral overlap of the circulating field, $I_C(\tilde\nu, \tau_1)/I_0(\tilde\nu)$, with the initial absorption coefficient $\alpha(\tilde\nu, \tau_1)$ of the intracavity molecules. 
    These considerations lead to cavity-enhanced absorption of light at the polariton frequencies and reduced (though non-negligible) absorption at the band center $\tilde\nu_0$.
    Note that nonlinear effects like multiphoton absorption or saturation may be amplified by cavity-enhancement of the pump field at resonant frequencies.

    \item   \textbf{\textit{The intracavity absorption coefficient and refractive index are transiently modified.}} 
    Excitation with pump light transiently modifies the absorption coefficient and refractive index of the intracavity molecules for times $t>\tau_1$. 
    We define a time-dependent absorption coefficient $\alpha(\tilde\nu, t) \equiv \alpha(\tilde\nu, \tau_1) + \Delta\alpha(\tilde\nu, t)$. 
    Here $\Delta\alpha(\tilde\nu,t)$ captures transient changes that arise from pump-induced phenomena including depletion of the ground state and the appearance of excited state absorption features. 
    The corresponding $n(\tilde\nu,t)$ can be constructed from $\alpha(\tilde\nu,t)$ using Eq.~\ref{eq:alpha} and the Kramers-Kronig relation.

    \item \textbf{\textit{The transmitted, reflected, and circulating cavity fields are transiently modified.}}
    As $\alpha(\tilde\nu, t)$ and $n(\tilde\nu, t)$ evolve following pump excitation, the constructive interference conditions for light traveling through the structure are transiently altered and the cavity spectra will change in concert.
    The intense pump light may also induce transient changes in other quantities like cavity length, mirror reflectivity, and background refractive index.\cite{renken21, pyles2024}

    \item   \textit{\textbf{The probe light travels through a transiently modified cavity.}} 
    Finally, we record the time-dependent cavity transmission spectrum by sending in broadband probe light at time $\tau_2$.
    $I_T(\tilde\nu, \tau_2)/I_0(\tilde\nu)$ derives from Eq.~\ref{eq:IT} evaluated for an intracavity medium with appropriate $\alpha(\tilde\nu, \tau_2)$ and $n(\tilde\nu, \tau_2)$.

\end{enumerate}

Note that we have assumed here that the dynamics of the intracavity material are slow compared to the cavity photon lifetime $\tau$ (Eq.~\ref{eq:tau}).
$\tau$ is typically a few picoseconds for micron-scale FP cavities and tens of femtoseconds for nanoscale cavities.
Intracavity molecular dynamics that occur on timescales shorter than the photonic lifetime are difficult to observe, because the ultrafast pulses used to make the measurement self-interfere, distort, and temporally broaden inside the cavity.
This is a consequence of the frequency-time Fourier relationship: spectral filtering of a broadband pulse by the optical cavity results in commensurate changes to its time-domain pulse shape.
Indeed, one must go to great experimental lengths to preserve ultrafast pulse shapes inside cavities.\cite{reber2016}
The community has only recently developed careful treatments of the early time dynamics inside strongly-coupled cavities to interpret these complex signatures.\cite{reitz2024} 
We therefore focus only on the longer time dynamics here.

\begin{figure}
   \includegraphics[width=0.95\linewidth]{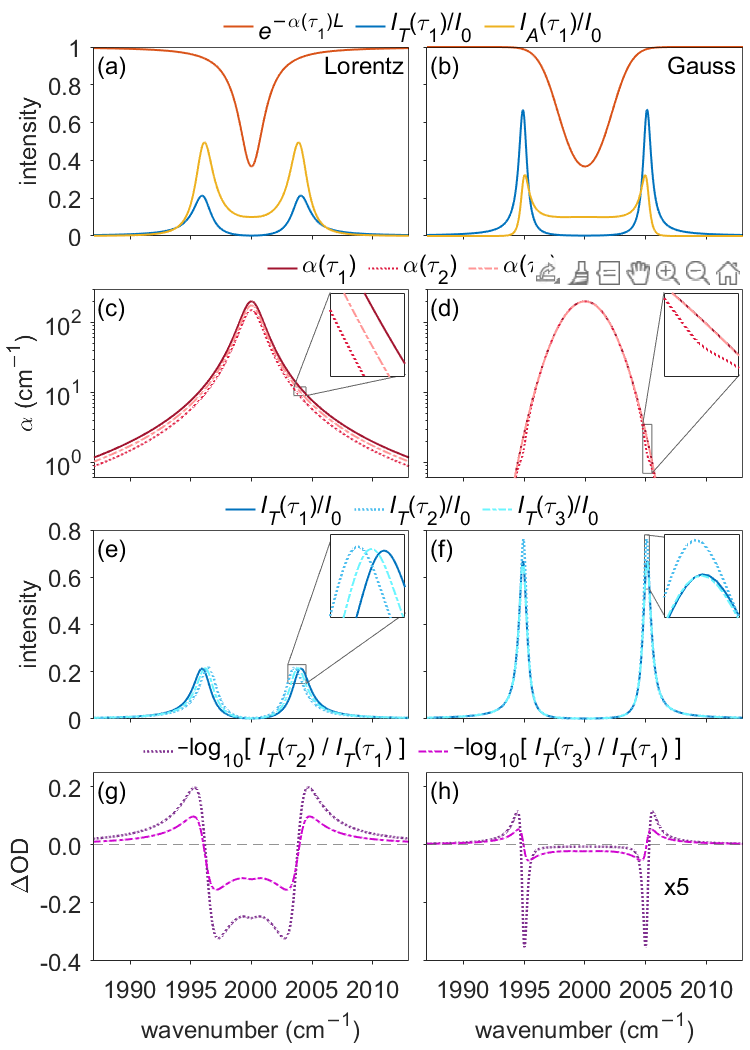}
    \caption{   \label{fig:nonlinear} 
    Transient pump-probe spectra for two strongly-coupled model systems. 
    \textbf{(a,b)} Linear free-space transmission spectra for molecular absorbers with Lorentzian and Gaussian broadening (orange), and fractional transmission (blue) and absorption (yellow) spectra for the same absorbers embedded in an optical cavity.
    Both systems have band center $\tilde\nu_0=$ \SI{2000}{cm^{-1}} and peak absorption coefficient $\alpha(\tilde\nu_0) = \SI{200}{cm^{-1}}$; the Lorentzian absorber has fwhm linewidth $\gamma_L=\SI{2}{cm^{-1}}$ while the Gaussian absorber has fwhm linewidth $\gamma_G=\SI{4}{cm^{-1}}$.
    Free-space transmission spectra are calculated over a $\SI{50}{\micro m}$ pathlength.
    Cavity spectra are simulated for an $L=\SI{50} {\micro m}$, $R = 0.95$, $T = 0.05$ FP cavity containing each absorber with background refractive index $n_0=1$. 
    These linear spectra are the reference point at time $\tau_1$ just before a pump-probe experiment starts.
    \textbf{(c,d)} Absorption coefficients plotted on a log scale for the systems in panels (a,b). $\alpha(\tilde\nu, \tau_1)$ is the intracavity absorption coefficient just before the pump arrives (solid lines) while $\alpha(\tilde\nu, \tau_2)$ (dotted lines) and $\alpha(\tilde\nu, \tau_3)$ (dash-dotted lines) represent transiently modulated intracavity absorption coefficients following broadband optical pumping, with $\tau_3 > \tau_2 > \tau_1$.
    We simulate uniform bleaching of the Lorentzian lineshape at time $\tau_2$ and partial recovery by time $\tau_3$.
    For the Gaussian lineshape, we initially simulate hole burning at the polariton frequencies at time $\tau_2$, followed by spectral diffusion to yield uniform bleaching by time $\tau_3$.
    \textbf{(e,f)}
    Transmission spectra of cavities containing the optically-modulated Lorentzian and Gaussian absorbers at times $\tau_2$ (dotted lines) and $\tau_3$ (dash-dotted lines) plotted against the reference cavity transmission spectra at time $\tau_1$ (solid lines) reproduced from panels (a,b).
    \textbf{(g,h)} 
    Differential cavity transmission spectra at times $\tau_2$ (dotted lines) and $\tau_3$ (dash-dotted lines) for both Lorentzian and Gaussian systems.
    } 
\end{figure}

\subsection{Simulating transient cavity spectroscopy} \label{sec:TA}

We now use the workflow introduced in Section \ref{sec:workflow} to model a transient pump-probe experiment and reproduce signals commonly seen in nonlinear cavity spectroscopy.
Here, we treat intracavity absorbers with either pure homogeneous Lorentzian or pure inhomogeneous Gaussian broadening.
In reality, most condensed-phase systems fall somewhere in between these edge cases, and an inhomogeneous absorption band represents an ensemble average over individual homogeneously-broadened transitions.
We direct the interested reader to the work of Pyles \textit{et al.},\cite{pyles2024} who present a careful treatment of this regime.
Here, we take a simpler approach with the goal of emphasizing the qualitatively distinct transient signals
afforded by intracavity molecules with homogeneous versus inhomogeneous absorption lineshapes.

Figs.~\ref{fig:nonlinear}ab plot static free-space transmission spectra for representative Lorentzian and Gaussian absorbers (orange) alongside cavity transmission (blue) and absorption (yellow) spectra for the same species strongly-coupled to a cavity. 
We take the cavity spectra in Figs.~\ref{fig:nonlinear}ab as our reference point at time $\tau_1$, just before the pump pulse arrives.
We assume that the intracavity medium absorbs broadband pump light according to the absorption spectrum at time $\tau_1$ (yellow traces in Figs.~\ref{fig:nonlinear}ab). 
This absorption then transiently modulates the intracavity absorption coefficient for later times $\tau_2$ and $\tau_3$. 
Here, we take time $\tau_2$ to occur very shortly after optical pumping, to treat the immediate response of the system before any relaxation  occurs.
We take $\tau_3 > \tau_2$ to represent a later time after the system has begun undergo population relaxation and spectral diffusion (\textit{vide infra}).
For simplicity, we assume that molecules that absorb pump light do not reabsorb probe light; in other words, we model only pump-induced bleaching and eventual recovery of intracavity absorption signals, neglecting any excited state absorption signals.

Let's first discuss how the Lorentzian system responds to optical pumping.
The homogeneously-broadened lineshape is preserved as it bleaches, effectively rescaling $\alpha(\tilde\nu, t)$ by a constant factor.
For the Lorentzian case, we therefore take $\alpha(\tilde\nu, \tau_2) = 0.75\times\alpha(\tilde\nu, \tau_1)$ to simulate the excitation of 25\% of intracavity molecules by the pump and commensurate bleaching of their ground state absorption signal, as plotted with a dotted line in Fig.~\ref{fig:nonlinear}c.
We take $\alpha(\tilde\nu, \tau_3) = 0.875\times\alpha(\tilde\nu, \tau_1)$ to illustrate partial population relaxation to the ground state at later times (dash-dotted lines in Fig.~\ref{fig:nonlinear}c).

By contrast, an inhomogeneously-broadened lineshape will experience pump-induced spectral hole burning of specific molecular sub-classes, which will transiently change the shape of $\alpha(\tilde\nu, t)$.
For the Gaussian case, we assume that molecules absorb cavity-filtered pump light according to the $\tau_1$ intracavity absorption spectrum (yellow trace in Fig.\ \ref{fig:nonlinear}b), transiently imprinting the shape of $I_A(\tilde\nu, \tau_1)$ as a bleach in $\alpha(\tilde\nu, t)$.
We therefore take $\alpha(\tilde\nu, \tau_2) = \alpha(\tilde\nu, \tau_1) - \alpha_0 \times I_A(\tilde\nu, \tau_1) / I_0(\tilde\nu)$ using a value of $\alpha_0$ that yields an excitation fraction of $\sim0.54\%$ of intracavity molecules.
$\alpha(\tilde\nu, \tau_2)$ is plotted in dotted lines in Fig.~\ref{fig:nonlinear}d.
This initial excitation is highly non-uniform across the inhomogeneously-broadened ensemble: $40\%$ of molecules that absorb near the polariton frequencies are excited while $<0.2\%$ of those that absorb near the $\tilde\nu_0$ band center are excited; this polariton-specific bleaching is evident in the inset of Fig.~\ref{fig:nonlinear}d. 
We simulate the effects of spectral diffusion a later time $\tau_3$ by assuming that the system has recovered its original Gaussian lineshape but $0.5\%$ of intracavity molecules remain excited.
We therefore assume a uniform bleach of the absorption spectrum at time $\tau_3$ with $\alpha(\tilde\nu, \tau_3) = 0.005\times\alpha(\tilde\nu, \tau_1)$, as plotted in dash-dotted lines in Fig.~\ref{fig:nonlinear}d.
$\alpha(\tilde\nu, \tau_3)$ is barely distinguishable by eye from $\alpha(\tilde\nu, \tau_1)$ due to the small excitation fraction.

Next, we calculate transient cavity transmission spectra using the modulated quantities for $\alpha(\tilde\nu, \tau_{2,3})$. 
Figs.~\ref{fig:nonlinear}ef show cavity transmission spectra with the original $\tau_1$ spectra reproduced as solid lines and the transient spectra plotted for $\tau_{2}$ with dotted lines and $\tau_{3}$ with dash-dotted lines.
In Figs.~\ref{fig:nonlinear}gh, we plot differential transmission spectra in terms of transient change in optical density given by $\Delta \textrm{OD} = - \log_{10} \left[ I_T(\tilde\nu, \tau_{2,3}) / I_T(\tilde\nu, \tau_1) \right]$, again with $\tau_{2}$ spectra plotted in dotted lines and $\tau_{3}$ spectra in dash-dotted lines.
With this convention, transiently increased transmission through the structure appears as a bleach ($\Delta\textrm{OD}<0$), while transiently reduced transmission looks like absorption ($\Delta\textrm{OD}>0$).
Note that transient cavity spectra are sometimes reported in terms of differential reflectivity, which flips these conventions.

Let's now inspect the signatures of these simulated intracavity transient absorption spectra.
The Lorentzian system experiences contraction of the Rabi splitting in response to optical pumping, which manifests as an inward shifting of the polaritonic transmission features in Fig.~\ref{fig:nonlinear}e.
Contraction occurs because the Rabi splitting scales with the square-root of the number of intracavity molecules available for coupling, and population is being transiently driven out of the lower state of the coupled transition.
The contraction of the Rabi splitting from \SI{8}{cm^{-1}} at $\tau_1$ to \SI{7}{cm^{-1}} at $\tau_2$ (solid versus dotted lines in Fig.~\ref{fig:nonlinear}e) is indeed commensurate with a 25\% reduction in $\alpha(\tilde\nu,t)$ from time $\tau_1$ to $\tau_2$.
%
Rabi contraction results in derivative lineshapes in the differential transmission spectra (Fig.~\ref{fig:nonlinear}g), as has been widely observed.\cite{xiang18, renken21, son2022, dunkelberger22, simpkins23comment, pyles2024}
The polariton transmission intensities increase only slightly from $\tau_1$ to $\tau_2$ as optical pumping does not drastically modulate the absorption coefficient at the polariton frequencies. 
The derivative lineshape in the differential $\tau_2$ spectrum (dotted lines in Fig.~\ref{fig:nonlinear}g) therefore feature similar excursion towards both positive and negative $\Delta$OD.
We expect these transient signatures to decay on the timescale with which the upper state population of the coupled transition relaxes and refills the ground state.
We illustrate this recovery process by plotting the transient transmission spectra at time $\tau_{3}$ in dash-dotted lines in Figs.\ \ref{fig:nonlinear}eg.
As the system recovers following optical pumping, the polariton modes move back apart and the derivative lineshapes in the differential signal decay -- largely uniformly -- in amplitude.

In the Gaussian case, optical pumping initially leads to hole-burning of the molecular sub-classes that absorb near the polaritonic frequencies. 
The most striking signature is the increase in polaritonic transmission at early time $\tau_2$ (dotted lines in Fig.~\ref{fig:nonlinear}f), which leads to significant excursions towards negative $\Delta$OD in the corresponding differential spectrum (Fig.~\ref{fig:nonlinear}h).
This phenomenon has been observed experimentally in transient spectra of inhomogeneously-broadened systems.\cite{grafton2021, pyles2024} 
The increase in transmission results from the intracavity absorption coefficient being drastically reduced at the polariton frequencies as is evident for $\alpha(\tilde\nu, \tau_2)$ in the inset of Fig.~\ref{fig:nonlinear}d.
We also note a minor Rabi contraction for $I_T(\tilde\nu, \tau_2)/I_0$ in Fig.~\ref{fig:nonlinear}f that leads to derivative-like features in the differential transmission spectrum.
The transient signatures of this system should evolve with two distinct timescales.
As in the homogeneous case, the Rabi contraction should recover on the timescale with which the upper state of the coupled transition relaxes and the lower state refills.
In addition, the peak polaritonic transmission intensity should drop on the timescale with which spectral diffusion refills holes burned in the inhomogeneously-broadened spectrum. 
We attempt to illustrate these effects by plotting the transient cavity transmission spectra at time $\tau_{3}$ (dash-dotted lines Figs.\ \ref{fig:nonlinear}fh).
We assume that by time $\tau_3$, spectral diffusion has healed the original Gaussian lineshape with just a small uniform bleach of the band as lingering evidence of optical pumping. 
The resulting $\tau_3$ cavity spectra look much like the homogeneous case.
The polariton bands exhibit a barely-detectable Rabi contraction and feature transmission intensities that nearly match those at $\tau_1$ (Fig.\ \ref{fig:nonlinear}f).
The differential spectrum at time $\tau_3$ (dash-dotted lines in Fig.\ \ref{fig:nonlinear}g) features derivative lineshapes with comparable excursion towards both positive and negative $\Delta$OD, in strong contrast to the dominant bleaching at the polariton frequencies in the $\tau_2$ differential spectrum (dotted lines). 
The key takeaway here is that inhomogeneously-broadened systems are expected to feature transient signatures whose shapes change with time.

The involvement of spectral diffusion certainly complicates the transient spectroscopy of inhomogeneously-broadened intracavity systems, as Pyles \textit{et al}.\cite{pyles2024} have also discussed. 
Spectral diffusion timescales depend sensitively on the specific molecular system and its environment, and molecules in the wings of the distribution may undergo spectral diffusion at different rates than those at the center.
If spectral diffusion outcompetes population relaxation, it may be challenging to detect the early-time spectral hole burning. 
Spectral diffusion and population relaxation timescales may also be similar in many systems and therefore difficult to disentangle.
In addition, if one considers more realistic systems where the homogeneous and inhomogeneous linewidths are comparable, the signatures of spectral hole burning will be smeared out. 
Thoughtful modeling of these effects will be necessary to determine best practices for analyzing the dynamics of inhomogeneously-broadened intracavity absorbers.

We close this section by noting that we have glossed over several other subtleties that crop up in real-world experiments.
We have neglected early-time dynamics within the cavity photonic lifetime, spectral congestion from excited state absorption signals, transient artifacts arising from the mirror coatings, and the response of the background refractive index and cavity length to optical pumping.
We have also assumed a uniform broadband pump spectrum, while in reality, transient signals will depend sensitively on the spectral shape of the pump.
For instance, narrowband pump pulses are sometimes used to separately excite each polariton band; for inhomogeneous species, such schemes result in smaller Rabi contraction and hole burning only in regions of overlap between the pump spectrum and cavity transmission spectrum.
We have also assumed here that the pump and probe beams strike the cavity at the same normal incidence angle.
In a non-collinear pump-probe geometry, the two beams would experience different optical filtering conditions due to the angle-dependence of the cavity spectrum.
These are all important factors to consider in simulations of experimental data.

\subsection{(De)localization of polariton excitations} \label{sec:delocal}

Our discussion of homogeneously and inhomogeneously broadened systems above dovetails into a broader conversation about the relative delocalization of polaritonic excitations.
The TC model describes polaritons and dark states as superpositions delocalized over all coupled intracavity molecules.
It is well known that heterogeneous disorder in the energy distribution of an intracavity ensemble leads to semi-localization of the dark states.\cite{Botzung2020, scholes2020, du2022} 
Liu \textit{et al.}\cite{liu2025} have also shown that the polaritons become semi-localized when the collective Rabi splitting does not greatly exceed the inhomogeneous linewidth of intracavity molecules.
In the presence of energetic disorder, a dark state or polariton lying at wavenumber $\tilde\nu$ will have the largest contributions from the sub-class of intracavity molecules whose transition frequencies also lie close to $\tilde\nu$.

We can now put forward a classical analog for these findings.
Consider how the absorption coefficients for the Lorentzian and Gaussian absorbers plotted in Fig.~\ref{fig:nonlinear}cd differ in their  modulation by optical pumping at time $\tau_2$.
In the Lorentzian system, absorption of light at the frequency of a polariton band can excite any member of the intracavity molecular ensemble. 
This captures the ``delocalized'' excitation of a homogeneous sample: any molecule in the ensemble may contribute equally to the nonlinear signals.
In the Gaussian system, on the other hand, excitation at the polariton frequencies selectively excites a small molecular sub-class whose transition frequencies are near-resonant with the polariton transmission window.
This captures the ``semi-localized'' excitation in the presence of inhomogeneous disorder: only a specific sub-class of the intracavity ensemble can be excited by the light that makes it into the cavity, and therefore only that specific sub-class will contribute to the nonlinear signal.
In reality, most molecules feature Voigt lineshapes with a convolution of both homogeneous and inhomogeneous broadening.
In this case, once the Rabi splitting significantly exceeds the inhomogeneous Gaussian linewidth, the polaritonic transmission windows will fall in the extended Lorentzian wings of the lineshape and one will recover a delocalized excitation upon optical pumping.
This picture is qualitatively consistent with the work of Liu \textit{et al.}\cite{liu2025}

\subsection{Some last words on transient polariton spectroscopy}
Nonlinear spectroscopy still awaits the adoption of a unified framework for simulating experimental data, particularly in systems that exhibit inhomogeneous broadening.
Transient cavity transmission spectra cannot be compared directly to extracavity data -- the observables are simply not the same.
Instead, as we lay out above and as others in the field are increasingly arguing,\cite{renken21,pyles2024} one must reconstruct the dynamics in the intracavity absorption coefficient $\alpha(\tilde\nu, t)$ that give rise to the transient cavity spectra.
Only once the intrinsic response of the intracavity molecules is isolated from the changing optical interference conditions can one determine if the molecules are behaving differently from their extracavity counterparts.
We have motivated here how the FP cavity expressions can help disentangle these dynamics.
Nonetheless, FP simulations of transient spectroscopy can be tricky to make quantitative.
We therefore close this section by noting some emerging strategies for transient spectroscopy in strongly-coupled systems.

One increasingly common approach is to construct the cavity from dichroic mirrors.
One can then form polaritons in a high-reflectivity spectral region, and make measurements in a high-transmission window far removed from the polaritonic bands.\cite{pang2020, wiesehan21, fidler23, chen24, michon24}
Another method to avoid spectral artifacts in the polaritonic region is to pump or probe a strongly-coupled structure orthogonal to the cavity axis, making use of creative cavity geometries\cite{ahn2020,nelson24} or open plasmonic platforms.\cite{wang2014, brawley2025}
There is no free lunch, of course, and these approaches come with caveats.
Measurements performed through a transmission window or along an orthogonal axis are sensitive to signals from uncoupled molecules, e.g.\ molecules that lie at the nodes of the cavity field or whose transition dipoles do not lie in the plane of the cavity mirrors.
These uncoupled species may dilute signals from the molecules that couple mostly strongly to the cavity, making potential cavity-modified behavior more difficult to detect.
Dichroic and orthogonal cavity schemes also enforce significant constraints on cavity geometry and mirror coatings. 
Nevertheless, we encourage the pursuit of more inventive cavity designs and unusual coupling schemes; complementary approaches may well yield new breakthroughs.


\section{\label{sec:conc}{Conclusions and Outlook}}

We have set out to provide a comprehensive picture of molecular polaritons with a classical framework that captures the regime in which many experiments operate.
We have tried, where possible, to translate between the classical optics and cQED descriptions of polaritonic systems.
We encourage initial interpretation of experimental data within the classical framework as a means to crosscheck that interesting signals cannot be simply explained by cavity interference effects.\cite{renken21, pyles2024, michon24}
Once classical signals are accounted for, any remaining behavior can be inspected and perhaps attributed to more exotic means of cavity-modification.
We hope that, in parallel, theoretical surveys of polaritonic phenomena will continue to explore both quantum and classical electromagnetic descriptions of the cavity field.

While the FP framework can explain much of linear and nonlinear cavity spectroscopy there are still major open questions and opportunities for future work. 
Vibrational polariton chemistry has not yet been fully rationalized within either cQED or classical optics\cite{simpkins21, wang21} 
and reports of altered photochemistry under electronic strong coupling remain poorly understood.\cite{hutchison12,thomas2024,schwartz2025, thomas2025}
There is also much work to be done to understand the fluorescence of intracavity molecules under strong coupling, where optical filtering, angle-dependent emission, Purcell enhancement, and higher-order processes like photon reabsorption may be in play.
Beyond molecular physics, strongly-coupled cavities may form a basis for next-generation optical technologies.
The possibilities for cavity-based nonlinear optical devices are self-evident, as enhanced intracavity light-matter interactions can augment higher-order and multiphoton processes.
New cavity-based approaches to nonlinear frequency generation,\cite{barachati2015} saturable absorbers,\cite{dunkelberger2019} and electro-optical modulators\cite{lee2024} have already been proposed or demonstrated.

We close by pointing out one final silver lining from our treatment here.
We have argued that distinct sub-classes of an inhomogeneous ensemble absorb light inside a cavity as compared to free space: molecules that absorb at the polariton frequencies are favored.
These molecules usually lie at the fringes of the inhomogeneous distribution and may feature conformations, solvation states, or other defects that can make their behavior distinct from the ensemble average. 
Cavity-mediated excitation of these unusual sub-populations may emphasize their distinct chemistry or photophysics, which would be washed out in excitation of the entire manifold.
This effect may be harnessed for optical control of chemistry via selective excitation of fringe populations or weakly absorbing states.
Such arguments are in fact in line with the early goals of polariton chemistry which sought to use a cavity as a tunable external knob to rationally steer molecular processes.\cite{simpkins21, garcia21}
The control afforded by this effect would stem not from intrinsic cavity-alteration of molecular behavior, but from the use of the cavity to selectively excite specific molecular populations.
These considerations are worth exploring in future work.

The field of molecular polaritonics has evolved rapidly over the past decade and shows no sign of slowing. 
We are entering a particularly productive time as the field matures and realistic prospects for cavity-based technologies come into focus.
In attempting to unify the language used to describe polaritons, we hope to provide deeper physical intuition, help bridge experiment and theory, and accelerate progress in the field.

\begin{acknowledgments}
\noindent 
This research was supported by the U.S. Department of Energy, Office of Science, Office of Basic Energy Sciences, CPIMS program, under Early Career Research Program award DE-SC0022948.

\end{acknowledgments}

\section*{Data Availability Statement}
\noindent The data that support the findings of this study are available from the corresponding author upon request.

\section*{References}
\nocite{*}
\bibliography{CPL_2025_v11}

\providecommand{\noopsort}[1]{}\providecommand{\singleletter}[1]{#1}%
\begin{thebibliography}{102}%
\makeatletter
\providecommand \@ifxundefined [1]{%
 \@ifx{#1\undefined}
}%
\providecommand \@ifnum [1]{%
 \ifnum #1\expandafter \@firstoftwo
 \else \expandafter \@secondoftwo
 \fi
}%
\providecommand \@ifx [1]{%
 \ifx #1\expandafter \@firstoftwo
 \else \expandafter \@secondoftwo
 \fi
}%
\providecommand \natexlab [1]{#1}%
\providecommand \enquote  [1]{``#1''}%
\providecommand \bibnamefont  [1]{#1}%
\providecommand \bibfnamefont [1]{#1}%
\providecommand \citenamefont [1]{#1}%
\providecommand \href@noop [0]{\@secondoftwo}%
\providecommand \href [0]{\begingroup \@sanitize@url \@href}%
\providecommand \@href[1]{\@@startlink{#1}\@@href}%
\providecommand \@@href[1]{\endgroup#1\@@endlink}%
\providecommand \@sanitize@url [0]{\catcode `\\12\catcode `\$12\catcode
  `\&12\catcode `\#12\catcode `\^12\catcode `\_12\catcode `\%12\relax}%
\providecommand \@@startlink[1]{}%
\providecommand \@@endlink[0]{}%
\providecommand \url  [0]{\begingroup\@sanitize@url \@url }%
\providecommand \@url [1]{\endgroup\@href {#1}{\urlprefix }}%
\providecommand \urlprefix  [0]{URL }%
\providecommand \Eprint [0]{\href }%
\providecommand \doibase [0]{https://doi.org/}%
\providecommand \selectlanguage [0]{\@gobble}%
\providecommand \bibinfo  [0]{\@secondoftwo}%
\providecommand \bibfield  [0]{\@secondoftwo}%
\providecommand \translation [1]{[#1]}%
\providecommand \BibitemOpen [0]{}%
\providecommand \bibitemStop [0]{}%
\providecommand \bibitemNoStop [0]{.\EOS\space}%
\providecommand \EOS [0]{\spacefactor3000\relax}%
\providecommand \BibitemShut  [1]{\csname bibitem#1\endcsname}%
\let\auto@bib@innerbib\@empty
\bibitem [{\citenamefont {Thomas}\ \emph {et~al.}(2016)\citenamefont {Thomas},
  \citenamefont {George}, \citenamefont {Shalabney}, \citenamefont {Dryzhakov},
  \citenamefont {Varma}, \citenamefont {Moran}, \citenamefont {Chervy},
  \citenamefont {Zhong}, \citenamefont {Devaux}, \citenamefont {Genet} \emph
  {et~al.}}]{thomas16}%
  \BibitemOpen
  \bibfield  {author} {\bibinfo {author} {\bibfnamefont {A.}~\bibnamefont
  {Thomas}}, \bibinfo {author} {\bibfnamefont {J.}~\bibnamefont {George}},
  \bibinfo {author} {\bibfnamefont {A.}~\bibnamefont {Shalabney}}, \bibinfo
  {author} {\bibfnamefont {M.}~\bibnamefont {Dryzhakov}}, \bibinfo {author}
  {\bibfnamefont {S.~J.}\ \bibnamefont {Varma}}, \bibinfo {author}
  {\bibfnamefont {J.}~\bibnamefont {Moran}}, \bibinfo {author} {\bibfnamefont
  {T.}~\bibnamefont {Chervy}}, \bibinfo {author} {\bibfnamefont
  {X.}~\bibnamefont {Zhong}}, \bibinfo {author} {\bibfnamefont
  {E.}~\bibnamefont {Devaux}}, \bibinfo {author} {\bibfnamefont
  {C.}~\bibnamefont {Genet}}, \emph {et~al.},\ }\bibfield  {title} {\enquote
  {\bibinfo {title} {Ground-state chemical reactivity under vibrational
  coupling to the vacuum electromagnetic field},}\ }\href@noop {} {\bibfield
  {journal} {\bibinfo  {journal} {Angew.\ Chem.\ Int.\ Ed.}\ }\textbf {\bibinfo
  {volume} {128}},\ \bibinfo {pages} {11634} (\bibinfo {year}
  {2016})}\BibitemShut {NoStop}%
\bibitem [{\citenamefont {Xiang}\ \emph {et~al.}(2020)\citenamefont {Xiang},
  \citenamefont {Ribeiro}, \citenamefont {Du}, \citenamefont {Chen},
  \citenamefont {Yang}, \citenamefont {Wang}, \citenamefont {Yuen-Zhou},\ and\
  \citenamefont {Xiong}}]{xiang20}%
  \BibitemOpen
  \bibfield  {author} {\bibinfo {author} {\bibfnamefont {B.}~\bibnamefont
  {Xiang}}, \bibinfo {author} {\bibfnamefont {R.~F.}\ \bibnamefont {Ribeiro}},
  \bibinfo {author} {\bibfnamefont {M.}~\bibnamefont {Du}}, \bibinfo {author}
  {\bibfnamefont {L.}~\bibnamefont {Chen}}, \bibinfo {author} {\bibfnamefont
  {Z.}~\bibnamefont {Yang}}, \bibinfo {author} {\bibfnamefont {J.}~\bibnamefont
  {Wang}}, \bibinfo {author} {\bibfnamefont {J.}~\bibnamefont {Yuen-Zhou}},\
  and\ \bibinfo {author} {\bibfnamefont {W.}~\bibnamefont {Xiong}},\ }\bibfield
   {title} {\enquote {\bibinfo {title} {Intermolecular vibrational energy
  transfer enabled by microcavity strong light--matter coupling},}\ }\href@noop
  {} {\bibfield  {journal} {\bibinfo  {journal} {Science}\ }\textbf {\bibinfo
  {volume} {368}},\ \bibinfo {pages} {665} (\bibinfo {year}
  {2020})}\BibitemShut {NoStop}%
\bibitem [{\citenamefont {Hirai}\ \emph {et~al.}(2020)\citenamefont {Hirai},
  \citenamefont {Takeda}, \citenamefont {Hutchison},\ and\ \citenamefont
  {Uji-i}}]{hirai2020}%
  \BibitemOpen
  \bibfield  {author} {\bibinfo {author} {\bibfnamefont {K.}~\bibnamefont
  {Hirai}}, \bibinfo {author} {\bibfnamefont {R.}~\bibnamefont {Takeda}},
  \bibinfo {author} {\bibfnamefont {J.~A.}\ \bibnamefont {Hutchison}},\ and\
  \bibinfo {author} {\bibfnamefont {H.}~\bibnamefont {Uji-i}},\ }\bibfield
  {title} {\enquote {\bibinfo {title} {Modulation of {P}rins cyclization by
  vibrational strong coupling},}\ }\href@noop {} {\bibfield  {journal}
  {\bibinfo  {journal} {Angew.\ Chem.\ In.\ Ed.}\ }\textbf {\bibinfo {volume}
  {132}},\ \bibinfo {pages} {5370} (\bibinfo {year} {2020})}\BibitemShut
  {NoStop}%
\bibitem [{\citenamefont {Sau}\ \emph {et~al.}(2021)\citenamefont {Sau},
  \citenamefont {Nagarajan}, \citenamefont {Patrahau}, \citenamefont
  {Lethuillier-Karl}, \citenamefont {Vergauwe}, \citenamefont {Thomas},
  \citenamefont {Moran}, \citenamefont {Genet},\ and\ \citenamefont
  {Ebbesen}}]{sau2021}%
  \BibitemOpen
  \bibfield  {author} {\bibinfo {author} {\bibfnamefont {A.}~\bibnamefont
  {Sau}}, \bibinfo {author} {\bibfnamefont {K.}~\bibnamefont {Nagarajan}},
  \bibinfo {author} {\bibfnamefont {B.}~\bibnamefont {Patrahau}}, \bibinfo
  {author} {\bibfnamefont {L.}~\bibnamefont {Lethuillier-Karl}}, \bibinfo
  {author} {\bibfnamefont {R.~M.}\ \bibnamefont {Vergauwe}}, \bibinfo {author}
  {\bibfnamefont {A.}~\bibnamefont {Thomas}}, \bibinfo {author} {\bibfnamefont
  {J.}~\bibnamefont {Moran}}, \bibinfo {author} {\bibfnamefont
  {C.}~\bibnamefont {Genet}},\ and\ \bibinfo {author} {\bibfnamefont {T.~W.}\
  \bibnamefont {Ebbesen}},\ }\bibfield  {title} {\enquote {\bibinfo {title}
  {Modifying {W}oodward--{H}offmann stereoselectivity under vibrational strong
  coupling},}\ }\href@noop {} {\bibfield  {journal} {\bibinfo  {journal}
  {Angew.\ Chem.\ Int.\ Ed.}\ }\textbf {\bibinfo {volume} {60}},\ \bibinfo
  {pages} {5712} (\bibinfo {year} {2021})}\BibitemShut {NoStop}%
\bibitem [{\citenamefont {Chen}\ \emph {et~al.}(2022)\citenamefont {Chen},
  \citenamefont {Du}, \citenamefont {Yang}, \citenamefont {Yuen-Zhou},\ and\
  \citenamefont {Xiong}}]{chen22}%
  \BibitemOpen
  \bibfield  {author} {\bibinfo {author} {\bibfnamefont {T.-T.}\ \bibnamefont
  {Chen}}, \bibinfo {author} {\bibfnamefont {M.}~\bibnamefont {Du}}, \bibinfo
  {author} {\bibfnamefont {Z.}~\bibnamefont {Yang}}, \bibinfo {author}
  {\bibfnamefont {J.}~\bibnamefont {Yuen-Zhou}},\ and\ \bibinfo {author}
  {\bibfnamefont {W.}~\bibnamefont {Xiong}},\ }\bibfield  {title} {\enquote
  {\bibinfo {title} {Cavity-enabled enhancement of ultrafast intramolecular
  vibrational redistribution over pseudorotation},}\ }\href@noop {} {\bibfield
  {journal} {\bibinfo  {journal} {Science}\ }\textbf {\bibinfo {volume}
  {378}},\ \bibinfo {pages} {790} (\bibinfo {year} {2022})}\BibitemShut
  {NoStop}%
\bibitem [{\citenamefont {Ahn}\ \emph {et~al.}(2023)\citenamefont {Ahn},
  \citenamefont {Triana}, \citenamefont {Recabal}, \citenamefont {Herrera},\
  and\ \citenamefont {Simpkins}}]{ahn23}%
  \BibitemOpen
  \bibfield  {author} {\bibinfo {author} {\bibfnamefont {W.}~\bibnamefont
  {Ahn}}, \bibinfo {author} {\bibfnamefont {J.~F.}\ \bibnamefont {Triana}},
  \bibinfo {author} {\bibfnamefont {F.}~\bibnamefont {Recabal}}, \bibinfo
  {author} {\bibfnamefont {F.}~\bibnamefont {Herrera}},\ and\ \bibinfo {author}
  {\bibfnamefont {B.~S.}\ \bibnamefont {Simpkins}},\ }\bibfield  {title}
  {\enquote {\bibinfo {title} {Modification of ground-state chemical reactivity
  via light-matter coherence in infrared cavities},}\ }\href@noop {} {\bibfield
   {journal} {\bibinfo  {journal} {Science}\ }\textbf {\bibinfo {volume}
  {380}},\ \bibinfo {pages} {1165} (\bibinfo {year} {2023})}\BibitemShut
  {NoStop}%
\bibitem [{\citenamefont {Hutchison}\ \emph {et~al.}(2012)\citenamefont
  {Hutchison}, \citenamefont {Schwartz}, \citenamefont {Genet}, \citenamefont
  {Devaux},\ and\ \citenamefont {Ebbesen}}]{hutchison12}%
  \BibitemOpen
  \bibfield  {author} {\bibinfo {author} {\bibfnamefont {J.~A.}\ \bibnamefont
  {Hutchison}}, \bibinfo {author} {\bibfnamefont {T.}~\bibnamefont {Schwartz}},
  \bibinfo {author} {\bibfnamefont {C.}~\bibnamefont {Genet}}, \bibinfo
  {author} {\bibfnamefont {E.}~\bibnamefont {Devaux}},\ and\ \bibinfo {author}
  {\bibfnamefont {T.~W.}\ \bibnamefont {Ebbesen}},\ }\bibfield  {title}
  {\enquote {\bibinfo {title} {Modifying chemical landscapes by coupling to
  vacuum fields},}\ }\href@noop {} {\bibfield  {journal} {\bibinfo  {journal}
  {Angew.\ Chem.\ Int.\ Ed.}\ }\textbf {\bibinfo {volume} {51}},\ \bibinfo
  {pages} {1592} (\bibinfo {year} {2012})}\BibitemShut {NoStop}%
\bibitem [{\citenamefont {Munkhbat}\ \emph {et~al.}(2018)\citenamefont
  {Munkhbat}, \citenamefont {Wers{\"a}ll}, \citenamefont {Baranov},
  \citenamefont {Antosiewicz},\ and\ \citenamefont {Shegai}}]{munkhbat2018}%
  \BibitemOpen
  \bibfield  {author} {\bibinfo {author} {\bibfnamefont {B.}~\bibnamefont
  {Munkhbat}}, \bibinfo {author} {\bibfnamefont {M.}~\bibnamefont
  {Wers{\"a}ll}}, \bibinfo {author} {\bibfnamefont {D.~G.}\ \bibnamefont
  {Baranov}}, \bibinfo {author} {\bibfnamefont {T.~J.}\ \bibnamefont
  {Antosiewicz}},\ and\ \bibinfo {author} {\bibfnamefont {T.}~\bibnamefont
  {Shegai}},\ }\bibfield  {title} {\enquote {\bibinfo {title} {Suppression of
  photo-oxidation of organic chromophores by strong coupling to plasmonic
  nanoantennas},}\ }\href@noop {} {\bibfield  {journal} {\bibinfo  {journal}
  {Sci.\ Adv.}\ }\textbf {\bibinfo {volume} {4}},\ \bibinfo {pages} {eaas9552}
  (\bibinfo {year} {2018})}\BibitemShut {NoStop}%
\bibitem [{\citenamefont {Zeng}\ \emph {et~al.}(2023)\citenamefont {Zeng},
  \citenamefont {P{\'e}rez-S{\'a}nchez}, \citenamefont {Eckdahl}, \citenamefont
  {Liu}, \citenamefont {Chang}, \citenamefont {Weiss}, \citenamefont {Kalow},
  \citenamefont {Yuen-Zhou},\ and\ \citenamefont {Stern}}]{zeng2023}%
  \BibitemOpen
  \bibfield  {author} {\bibinfo {author} {\bibfnamefont {H.}~\bibnamefont
  {Zeng}}, \bibinfo {author} {\bibfnamefont {J.~B.}\ \bibnamefont
  {P{\'e}rez-S{\'a}nchez}}, \bibinfo {author} {\bibfnamefont {C.~T.}\
  \bibnamefont {Eckdahl}}, \bibinfo {author} {\bibfnamefont {P.}~\bibnamefont
  {Liu}}, \bibinfo {author} {\bibfnamefont {W.~J.}\ \bibnamefont {Chang}},
  \bibinfo {author} {\bibfnamefont {E.~A.}\ \bibnamefont {Weiss}}, \bibinfo
  {author} {\bibfnamefont {J.~A.}\ \bibnamefont {Kalow}}, \bibinfo {author}
  {\bibfnamefont {J.}~\bibnamefont {Yuen-Zhou}},\ and\ \bibinfo {author}
  {\bibfnamefont {N.~P.}\ \bibnamefont {Stern}},\ }\bibfield  {title} {\enquote
  {\bibinfo {title} {Control of photoswitching kinetics with strong
  light--matter coupling in a cavity},}\ }\href@noop {} {\bibfield  {journal}
  {\bibinfo  {journal} {J.\ Am.\ Chem.\ Soc.}\ }\textbf {\bibinfo {volume}
  {145}},\ \bibinfo {pages} {19655} (\bibinfo {year} {2023})}\BibitemShut
  {NoStop}%
\bibitem [{\citenamefont {Lee}\ \emph {et~al.}(2024{\natexlab{a}})\citenamefont
  {Lee}, \citenamefont {Melton}, \citenamefont {Xu},\ and\ \citenamefont
  {Delor}}]{lee2024controlling}%
  \BibitemOpen
  \bibfield  {author} {\bibinfo {author} {\bibfnamefont {I.}~\bibnamefont
  {Lee}}, \bibinfo {author} {\bibfnamefont {S.~R.}\ \bibnamefont {Melton}},
  \bibinfo {author} {\bibfnamefont {D.}~\bibnamefont {Xu}},\ and\ \bibinfo
  {author} {\bibfnamefont {M.}~\bibnamefont {Delor}},\ }\bibfield  {title}
  {\enquote {\bibinfo {title} {Controlling molecular photoisomerization in
  photonic cavities through polariton funneling},}\ }\href@noop {} {\bibfield
  {journal} {\bibinfo  {journal} {J.\ Am.\ Chem.\ Soc.}\ }\textbf {\bibinfo
  {volume} {146}},\ \bibinfo {pages} {9544} (\bibinfo {year}
  {2024}{\natexlab{a}})}\BibitemShut {NoStop}%
\bibitem [{\citenamefont {Imperatore}, \citenamefont {Asbury},\ and\
  \citenamefont {Giebink}(2021)}]{imperatore21}%
  \BibitemOpen
  \bibfield  {author} {\bibinfo {author} {\bibfnamefont {M.~V.}\ \bibnamefont
  {Imperatore}}, \bibinfo {author} {\bibfnamefont {J.~B.}\ \bibnamefont
  {Asbury}},\ and\ \bibinfo {author} {\bibfnamefont {N.~C.}\ \bibnamefont
  {Giebink}},\ }\bibfield  {title} {\enquote {\bibinfo {title} {Reproducibility
  of cavity-enhanced chemical reaction rates in the vibrational strong coupling
  regime},}\ }\href@noop {} {\bibfield  {journal} {\bibinfo  {journal} {J.\
  Chem.\ Phys.}\ }\textbf {\bibinfo {volume} {154}},\ \bibinfo {pages} {191103}
  (\bibinfo {year} {2021})}\BibitemShut {NoStop}%
\bibitem [{\citenamefont {Wiesehan}\ and\ \citenamefont
  {Xiong}(2021)}]{wiesehan21}%
  \BibitemOpen
  \bibfield  {author} {\bibinfo {author} {\bibfnamefont {G.~D.}\ \bibnamefont
  {Wiesehan}}\ and\ \bibinfo {author} {\bibfnamefont {W.}~\bibnamefont
  {Xiong}},\ }\bibfield  {title} {\enquote {\bibinfo {title} {Negligible rate
  enhancement from reported cooperative vibrational strong coupling
  catalysis},}\ }\href@noop {} {\bibfield  {journal} {\bibinfo  {journal} {J.\
  Chem.\ Phys.}\ }\textbf {\bibinfo {volume} {155}},\ \bibinfo {pages} {241103}
  (\bibinfo {year} {2021})}\BibitemShut {NoStop}%
\bibitem [{\citenamefont {Michon}\ and\ \citenamefont
  {Simpkins}(2024)}]{michon24}%
  \BibitemOpen
  \bibfield  {author} {\bibinfo {author} {\bibfnamefont {M.~A.}\ \bibnamefont
  {Michon}}\ and\ \bibinfo {author} {\bibfnamefont {B.~S.}\ \bibnamefont
  {Simpkins}},\ }\bibfield  {title} {\enquote {\bibinfo {title} {Impact of
  cavity length non-uniformity on reaction rate extraction in strong coupling
  experiments},}\ }\href@noop {} {\bibfield  {journal} {\bibinfo  {journal}
  {J.\ Am.\ Chem.\ Soc.}\ }\textbf {\bibinfo {volume} {146}},\ \bibinfo {pages}
  {30596} (\bibinfo {year} {2024})}\BibitemShut {NoStop}%
\bibitem [{\citenamefont {Fidler}\ \emph {et~al.}(2023)\citenamefont {Fidler},
  \citenamefont {Chen}, \citenamefont {McKillop},\ and\ \citenamefont
  {Weichman}}]{fidler23}%
  \BibitemOpen
  \bibfield  {author} {\bibinfo {author} {\bibfnamefont {A.~P.}\ \bibnamefont
  {Fidler}}, \bibinfo {author} {\bibfnamefont {L.}~\bibnamefont {Chen}},
  \bibinfo {author} {\bibfnamefont {A.~M.}\ \bibnamefont {McKillop}},\ and\
  \bibinfo {author} {\bibfnamefont {M.~L.}\ \bibnamefont {Weichman}},\
  }\bibfield  {title} {\enquote {\bibinfo {title} {Ultrafast dynamics of {CN}
  radical reactions with chloroform solvent under vibrational strong
  coupling},}\ }\href@noop {} {\bibfield  {journal} {\bibinfo  {journal} {J.\
  Chem.\ Phys.}\ }\textbf {\bibinfo {volume} {159}},\ \bibinfo {pages} {164302}
  (\bibinfo {year} {2023})}\BibitemShut {NoStop}%
\bibitem [{\citenamefont {Chen}\ \emph {et~al.}(2024)\citenamefont {Chen},
  \citenamefont {Fidler}, \citenamefont {McKillop},\ and\ \citenamefont
  {Weichman}}]{chen24}%
  \BibitemOpen
  \bibfield  {author} {\bibinfo {author} {\bibfnamefont {L.}~\bibnamefont
  {Chen}}, \bibinfo {author} {\bibfnamefont {A.~P.}\ \bibnamefont {Fidler}},
  \bibinfo {author} {\bibfnamefont {A.~M.}\ \bibnamefont {McKillop}},\ and\
  \bibinfo {author} {\bibfnamefont {M.~L.}\ \bibnamefont {Weichman}},\
  }\bibfield  {title} {\enquote {\bibinfo {title} {Exploring the impact of
  vibrational cavity coupling strength on ultrafast {CN} +
  $c$-{C}$_6${H}$_{12}$ reaction dynamics},}\ }\href@noop {} {\bibfield
  {journal} {\bibinfo  {journal} {Nanophotonics}\ }\textbf {\bibinfo {volume}
  {13}},\ \bibinfo {pages} {2591} (\bibinfo {year} {2024})}\BibitemShut
  {NoStop}%
\bibitem [{\citenamefont {Muller}\ \emph {et~al.}(2024)\citenamefont {Muller},
  \citenamefont {Mayer}, \citenamefont {Piejko}, \citenamefont {Patrahau},
  \citenamefont {Bauer},\ and\ \citenamefont {Moran}}]{muller2024}%
  \BibitemOpen
  \bibfield  {author} {\bibinfo {author} {\bibfnamefont {C.}~\bibnamefont
  {Muller}}, \bibinfo {author} {\bibfnamefont {R.~J.}\ \bibnamefont {Mayer}},
  \bibinfo {author} {\bibfnamefont {M.}~\bibnamefont {Piejko}}, \bibinfo
  {author} {\bibfnamefont {B.}~\bibnamefont {Patrahau}}, \bibinfo {author}
  {\bibfnamefont {V.}~\bibnamefont {Bauer}},\ and\ \bibinfo {author}
  {\bibfnamefont {J.}~\bibnamefont {Moran}},\ }\bibfield  {title} {\enquote
  {\bibinfo {title} {Measuring kinetics under vibrational strong coupling:
  Testing for a change in the nucleophilicity of water and alcohols},}\
  }\href@noop {} {\bibfield  {journal} {\bibinfo  {journal} {Angew.\ Chem.\
  Int.\ Ed.}\ }\textbf {\bibinfo {volume} {136}},\ \bibinfo {pages}
  {e202410770} (\bibinfo {year} {2024})}\BibitemShut {NoStop}%
\bibitem [{\citenamefont {Herrera}\ and\ \citenamefont
  {Owrutsky}(2020)}]{herrera20}%
  \BibitemOpen
  \bibfield  {author} {\bibinfo {author} {\bibfnamefont {F.}~\bibnamefont
  {Herrera}}\ and\ \bibinfo {author} {\bibfnamefont {J.}~\bibnamefont
  {Owrutsky}},\ }\bibfield  {title} {\enquote {\bibinfo {title} {Molecular
  polaritons for controlling chemistry with quantum optics},}\ }\href@noop {}
  {\bibfield  {journal} {\bibinfo  {journal} {J.\ Chem.\ Phys.}\ }\textbf
  {\bibinfo {volume} {152}},\ \bibinfo {pages} {100902} (\bibinfo {year}
  {2020})}\BibitemShut {NoStop}%
\bibitem [{\citenamefont {Simpkins}, \citenamefont {Dunkelberger},\ and\
  \citenamefont {Owrutsky}(2021)}]{simpkins21}%
  \BibitemOpen
  \bibfield  {author} {\bibinfo {author} {\bibfnamefont {B.~S.}\ \bibnamefont
  {Simpkins}}, \bibinfo {author} {\bibfnamefont {A.~D.}\ \bibnamefont
  {Dunkelberger}},\ and\ \bibinfo {author} {\bibfnamefont {J.~C.}\ \bibnamefont
  {Owrutsky}},\ }\bibfield  {title} {\enquote {\bibinfo {title} {Mode-specific
  chemistry through vibrational strong coupling (or a wish come true)},}\
  }\href@noop {} {\bibfield  {journal} {\bibinfo  {journal} {J.\ Phys.\ Chem.\
  C}\ }\textbf {\bibinfo {volume} {125}},\ \bibinfo {pages} {19081} (\bibinfo
  {year} {2021})}\BibitemShut {NoStop}%
\bibitem [{\citenamefont {Wang}\ and\ \citenamefont {Yelin}(2021)}]{wang21}%
  \BibitemOpen
  \bibfield  {author} {\bibinfo {author} {\bibfnamefont {D.~S.}\ \bibnamefont
  {Wang}}\ and\ \bibinfo {author} {\bibfnamefont {S.~F.}\ \bibnamefont
  {Yelin}},\ }\bibfield  {title} {\enquote {\bibinfo {title} {A roadmap toward
  the theory of vibrational polariton chemistry},}\ }\href@noop {} {\bibfield
  {journal} {\bibinfo  {journal} {ACS Photonics}\ }\textbf {\bibinfo {volume}
  {8}},\ \bibinfo {pages} {2818} (\bibinfo {year} {2021})}\BibitemShut
  {NoStop}%
\bibitem [{\citenamefont {Li}\ \emph {et~al.}(2022)\citenamefont {Li},
  \citenamefont {Cui}, \citenamefont {Subotnik},\ and\ \citenamefont
  {Nitzan}}]{li22}%
  \BibitemOpen
  \bibfield  {author} {\bibinfo {author} {\bibfnamefont {T.~E.}\ \bibnamefont
  {Li}}, \bibinfo {author} {\bibfnamefont {B.}~\bibnamefont {Cui}}, \bibinfo
  {author} {\bibfnamefont {J.~E.}\ \bibnamefont {Subotnik}},\ and\ \bibinfo
  {author} {\bibfnamefont {A.}~\bibnamefont {Nitzan}},\ }\bibfield  {title}
  {\enquote {\bibinfo {title} {Molecular polaritonics: {C}hemical dynamics
  under strong light--matter coupling},}\ }\href@noop {} {\bibfield  {journal}
  {\bibinfo  {journal} {Annu.\ Rev.\ Phys.\ Chem.}\ }\textbf {\bibinfo {volume}
  {73}},\ \bibinfo {pages} {43} (\bibinfo {year} {2022})}\BibitemShut {NoStop}%
\bibitem [{\citenamefont {Campos-Gonzalez-Angulo}\ \emph
  {et~al.}(2023)\citenamefont {Campos-Gonzalez-Angulo}, \citenamefont {Poh},
  \citenamefont {Du},\ and\ \citenamefont {Yuen-Zhou}}]{campos23}%
  \BibitemOpen
  \bibfield  {author} {\bibinfo {author} {\bibfnamefont {J.}~\bibnamefont
  {Campos-Gonzalez-Angulo}}, \bibinfo {author} {\bibfnamefont {Y.}~\bibnamefont
  {Poh}}, \bibinfo {author} {\bibfnamefont {M.}~\bibnamefont {Du}},\ and\
  \bibinfo {author} {\bibfnamefont {J.}~\bibnamefont {Yuen-Zhou}},\ }\bibfield
  {title} {\enquote {\bibinfo {title} {Swinging between shine and shadow:
  {T}heoretical advances on thermally activated vibropolaritonic chemistry},}\
  }\href@noop {} {\bibfield  {journal} {\bibinfo  {journal} {J.\ Chem.\ Phys.}\
  }\textbf {\bibinfo {volume} {158}},\ \bibinfo {pages} {230901} (\bibinfo
  {year} {2023})}\BibitemShut {NoStop}%
\bibitem [{\citenamefont {Mandal}\ \emph {et~al.}(2023)\citenamefont {Mandal},
  \citenamefont {Taylor}, \citenamefont {Weight}, \citenamefont {Koessler},
  \citenamefont {Li},\ and\ \citenamefont {Huo}}]{mandal23}%
  \BibitemOpen
  \bibfield  {author} {\bibinfo {author} {\bibfnamefont {A.}~\bibnamefont
  {Mandal}}, \bibinfo {author} {\bibfnamefont {M.~A.}\ \bibnamefont {Taylor}},
  \bibinfo {author} {\bibfnamefont {B.~M.}\ \bibnamefont {Weight}}, \bibinfo
  {author} {\bibfnamefont {E.~R.}\ \bibnamefont {Koessler}}, \bibinfo {author}
  {\bibfnamefont {X.}~\bibnamefont {Li}},\ and\ \bibinfo {author}
  {\bibfnamefont {P.}~\bibnamefont {Huo}},\ }\bibfield  {title} {\enquote
  {\bibinfo {title} {Theoretical advances in polariton chemistry and molecular
  cavity quantum electrodynamics},}\ }\href@noop {} {\bibfield  {journal}
  {\bibinfo  {journal} {Chem.\ Rev.}\ }\textbf {\bibinfo {volume} {123}},\
  \bibinfo {pages} {9786} (\bibinfo {year} {2023})}\BibitemShut {NoStop}%
\bibitem [{\citenamefont {Xiong}(2023)}]{xiong23}%
  \BibitemOpen
  \bibfield  {author} {\bibinfo {author} {\bibfnamefont {W.}~\bibnamefont
  {Xiong}},\ }\bibfield  {title} {\enquote {\bibinfo {title} {Molecular
  vibrational polariton dynamics: {W}hat can polaritons do?}}\ }\href@noop {}
  {\bibfield  {journal} {\bibinfo  {journal} {Acc.\ Chem.\ Res.}\ }\textbf
  {\bibinfo {volume} {56}},\ \bibinfo {pages} {776} (\bibinfo {year}
  {2023})}\BibitemShut {NoStop}%
\bibitem [{\citenamefont {Schwennicke}\ \emph {et~al.}(2024)\citenamefont
  {Schwennicke}, \citenamefont {Koner}, \citenamefont {P{\'e}rez-S{\'a}nchez},
  \citenamefont {Xiong}, \citenamefont {Giebink}, \citenamefont {Weichman},\
  and\ \citenamefont {Yuen-Zhou}}]{schwennicke2024}%
  \BibitemOpen
  \bibfield  {author} {\bibinfo {author} {\bibfnamefont {K.}~\bibnamefont
  {Schwennicke}}, \bibinfo {author} {\bibfnamefont {A.}~\bibnamefont {Koner}},
  \bibinfo {author} {\bibfnamefont {J.~B.}\ \bibnamefont
  {P{\'e}rez-S{\'a}nchez}}, \bibinfo {author} {\bibfnamefont {W.}~\bibnamefont
  {Xiong}}, \bibinfo {author} {\bibfnamefont {N.~C.}\ \bibnamefont {Giebink}},
  \bibinfo {author} {\bibfnamefont {M.~L.}\ \bibnamefont {Weichman}},\ and\
  \bibinfo {author} {\bibfnamefont {J.}~\bibnamefont {Yuen-Zhou}},\ }\bibfield
  {title} {\enquote {\bibinfo {title} {When do molecular polaritons behave like
  optical filters?}}\ }\href@noop {} {\bibfield  {journal} {\bibinfo  {journal}
  {arXiv:2408.05036}\ } (\bibinfo {year} {2024})}\BibitemShut {NoStop}%
\bibitem [{\citenamefont {Shore}\ and\ \citenamefont {Knight}(1993)}]{shore93}%
  \BibitemOpen
  \bibfield  {author} {\bibinfo {author} {\bibfnamefont {B.~W.}\ \bibnamefont
  {Shore}}\ and\ \bibinfo {author} {\bibfnamefont {P.~L.}\ \bibnamefont
  {Knight}},\ }\bibfield  {title} {\enquote {\bibinfo {title} {The
  {J}aynes-{C}ummings model},}\ }\href@noop {} {\bibfield  {journal} {\bibinfo
  {journal} {J.\ Mod.\ Optics}\ }\textbf {\bibinfo {volume} {40}},\ \bibinfo
  {pages} {1195} (\bibinfo {year} {1993})}\BibitemShut {NoStop}%
\bibitem [{\citenamefont {Tavis}\ and\ \citenamefont
  {Cummings}(1968)}]{tavis68}%
  \BibitemOpen
  \bibfield  {author} {\bibinfo {author} {\bibfnamefont {M.}~\bibnamefont
  {Tavis}}\ and\ \bibinfo {author} {\bibfnamefont {F.~W.}\ \bibnamefont
  {Cummings}},\ }\bibfield  {title} {\enquote {\bibinfo {title} {Exact solution
  for an {N}-molecule—radiation-field {H}amiltonian},}\ }\href@noop {}
  {\bibfield  {journal} {\bibinfo  {journal} {Phys.\ Rev.}\ }\textbf {\bibinfo
  {volume} {170}},\ \bibinfo {pages} {379} (\bibinfo {year}
  {1968})}\BibitemShut {NoStop}%
\bibitem [{\citenamefont {Garraway}(2011)}]{garraway11}%
  \BibitemOpen
  \bibfield  {author} {\bibinfo {author} {\bibfnamefont {B.~M.}\ \bibnamefont
  {Garraway}},\ }\bibfield  {title} {\enquote {\bibinfo {title} {The {D}icke
  model in quantum optics: {D}icke model revisited},}\ }\href@noop {}
  {\bibfield  {journal} {\bibinfo  {journal} {Philos.\ Trans.\ R.\ Soc.\ A}\
  }\textbf {\bibinfo {volume} {369}},\ \bibinfo {pages} {1137} (\bibinfo {year}
  {2011})}\BibitemShut {NoStop}%
\bibitem [{\citenamefont {Zhu}\ \emph {et~al.}(1990)\citenamefont {Zhu},
  \citenamefont {Gauthier}, \citenamefont {Morin}, \citenamefont {Wu},
  \citenamefont {Carmichael},\ and\ \citenamefont {Mossberg}}]{zhu90}%
  \BibitemOpen
  \bibfield  {author} {\bibinfo {author} {\bibfnamefont {Y.}~\bibnamefont
  {Zhu}}, \bibinfo {author} {\bibfnamefont {D.~J.}\ \bibnamefont {Gauthier}},
  \bibinfo {author} {\bibfnamefont {S.}~\bibnamefont {Morin}}, \bibinfo
  {author} {\bibfnamefont {Q.}~\bibnamefont {Wu}}, \bibinfo {author}
  {\bibfnamefont {H.}~\bibnamefont {Carmichael}},\ and\ \bibinfo {author}
  {\bibfnamefont {T.}~\bibnamefont {Mossberg}},\ }\bibfield  {title} {\enquote
  {\bibinfo {title} {Vacuum {R}abi splitting as a feature of linear-dispersion
  theory: {A}nalysis and experimental observations},}\ }\href@noop {}
  {\bibfield  {journal} {\bibinfo  {journal} {Phys.\ Rev.\ Lett.}\ }\textbf
  {\bibinfo {volume} {64}},\ \bibinfo {pages} {2499} (\bibinfo {year}
  {1990})}\BibitemShut {NoStop}%
\bibitem [{\citenamefont {Khitrova}\ \emph {et~al.}(1999)\citenamefont
  {Khitrova}, \citenamefont {Gibbs}, \citenamefont {Jahnke}, \citenamefont
  {Kira},\ and\ \citenamefont {Koch}}]{khitrova1999}%
  \BibitemOpen
  \bibfield  {author} {\bibinfo {author} {\bibfnamefont {G.}~\bibnamefont
  {Khitrova}}, \bibinfo {author} {\bibfnamefont {H.}~\bibnamefont {Gibbs}},
  \bibinfo {author} {\bibfnamefont {F.}~\bibnamefont {Jahnke}}, \bibinfo
  {author} {\bibfnamefont {M.}~\bibnamefont {Kira}},\ and\ \bibinfo {author}
  {\bibfnamefont {S.~W.}\ \bibnamefont {Koch}},\ }\bibfield  {title} {\enquote
  {\bibinfo {title} {Nonlinear optics of normal-mode-coupling semiconductor
  microcavities},}\ }\href@noop {} {\bibfield  {journal} {\bibinfo  {journal}
  {Rev.\ Mod.\ Phys.}\ }\textbf {\bibinfo {volume} {71}},\ \bibinfo {pages}
  {1591} (\bibinfo {year} {1999})}\BibitemShut {NoStop}%
\bibitem [{\citenamefont {Rudin}\ and\ \citenamefont
  {Reinecke}(1999)}]{rudin1999}%
  \BibitemOpen
  \bibfield  {author} {\bibinfo {author} {\bibfnamefont {S.}~\bibnamefont
  {Rudin}}\ and\ \bibinfo {author} {\bibfnamefont {T.}~\bibnamefont
  {Reinecke}},\ }\bibfield  {title} {\enquote {\bibinfo {title} {Oscillator
  model for vacuum {R}abi splitting in microcavities},}\ }\href@noop {}
  {\bibfield  {journal} {\bibinfo  {journal} {Phys.\ Rev.\ B}\ }\textbf
  {\bibinfo {volume} {59}},\ \bibinfo {pages} {10227} (\bibinfo {year}
  {1999})}\BibitemShut {NoStop}%
\bibitem [{\citenamefont {Novotny}(2010)}]{novotny10}%
  \BibitemOpen
  \bibfield  {author} {\bibinfo {author} {\bibfnamefont {L.}~\bibnamefont
  {Novotny}},\ }\bibfield  {title} {\enquote {\bibinfo {title} {Strong
  coupling, energy splitting, and level crossings: A classical perspective},}\
  }\href@noop {} {\bibfield  {journal} {\bibinfo  {journal} {Am.\ J.\ Phys.}\
  }\textbf {\bibinfo {volume} {78}},\ \bibinfo {pages} {1199} (\bibinfo {year}
  {2010})}\BibitemShut {NoStop}%
\bibitem [{\citenamefont {Tanji-Suzuki}\ \emph {et~al.}(2011)\citenamefont
  {Tanji-Suzuki}, \citenamefont {Leroux}, \citenamefont {Schleier-Smith},
  \citenamefont {Cetina}, \citenamefont {Grier}, \citenamefont {Simon},\ and\
  \citenamefont {Vuletić}}]{tanji2011}%
  \BibitemOpen
  \bibfield  {author} {\bibinfo {author} {\bibfnamefont {H.}~\bibnamefont
  {Tanji-Suzuki}}, \bibinfo {author} {\bibfnamefont {I.~D.}\ \bibnamefont
  {Leroux}}, \bibinfo {author} {\bibfnamefont {M.~H.}\ \bibnamefont
  {Schleier-Smith}}, \bibinfo {author} {\bibfnamefont {M.}~\bibnamefont
  {Cetina}}, \bibinfo {author} {\bibfnamefont {A.~T.}\ \bibnamefont {Grier}},
  \bibinfo {author} {\bibfnamefont {J.}~\bibnamefont {Simon}},\ and\ \bibinfo
  {author} {\bibfnamefont {V.}~\bibnamefont {Vuletić}},\ }\bibfield  {title}
  {\enquote {\bibinfo {title} {Interaction between {A}tomic {E}nsembles and
  {O}ptical {R}esonators: {C}lassical {D}escription},}\ }in\ \href
  {https://doi.org/https://doi.org/10.1016/B978-0-12-385508-4.00004-8} {\emph
  {\bibinfo {booktitle} {Advances in Atomic, Molecular, and Optical
  Physics}}},\ \bibinfo {series} {Advances In Atomic, Molecular, and Optical
  Physics}, Vol.~\bibinfo {volume} {60},\ \bibinfo {editor} {edited by\
  \bibinfo {editor} {\bibfnamefont {E.}~\bibnamefont {Arimondo}}, \bibinfo
  {editor} {\bibfnamefont {P.}~\bibnamefont {Berman}},\ and\ \bibinfo {editor}
  {\bibfnamefont {C.}~\bibnamefont {Lin}}}\ (\bibinfo  {publisher} {Academic
  Press},\ \bibinfo {year} {2011})\ pp.\ \bibinfo {pages}
  {201--237}\BibitemShut {NoStop}%
\bibitem [{\citenamefont {T{\"o}rm{\"a}}\ and\ \citenamefont
  {Barnes}(2015)}]{torma14}%
  \BibitemOpen
  \bibfield  {author} {\bibinfo {author} {\bibfnamefont {P.}~\bibnamefont
  {T{\"o}rm{\"a}}}\ and\ \bibinfo {author} {\bibfnamefont {W.~L.}\ \bibnamefont
  {Barnes}},\ }\bibfield  {title} {\enquote {\bibinfo {title} {Strong coupling
  between surface plasmon polaritons and emitters: {A} review},}\ }\href@noop
  {} {\bibfield  {journal} {\bibinfo  {journal} {Rep.\ Prog.\ Phys.}\ }\textbf
  {\bibinfo {volume} {78}},\ \bibinfo {pages} {013901} (\bibinfo {year}
  {2015})}\BibitemShut {NoStop}%
\bibitem [{\citenamefont {Xiang}\ \emph {et~al.}(2018)\citenamefont {Xiang},
  \citenamefont {Ribeiro}, \citenamefont {Dunkelberger}, \citenamefont {Wang},
  \citenamefont {Li}, \citenamefont {Simpkins}, \citenamefont {Owrutsky},
  \citenamefont {Yuen-Zhou},\ and\ \citenamefont {Xiong}}]{xiang18}%
  \BibitemOpen
  \bibfield  {author} {\bibinfo {author} {\bibfnamefont {B.}~\bibnamefont
  {Xiang}}, \bibinfo {author} {\bibfnamefont {R.~F.}\ \bibnamefont {Ribeiro}},
  \bibinfo {author} {\bibfnamefont {A.~D.}\ \bibnamefont {Dunkelberger}},
  \bibinfo {author} {\bibfnamefont {J.}~\bibnamefont {Wang}}, \bibinfo {author}
  {\bibfnamefont {Y.}~\bibnamefont {Li}}, \bibinfo {author} {\bibfnamefont
  {B.~S.}\ \bibnamefont {Simpkins}}, \bibinfo {author} {\bibfnamefont {J.~C.}\
  \bibnamefont {Owrutsky}}, \bibinfo {author} {\bibfnamefont {J.}~\bibnamefont
  {Yuen-Zhou}},\ and\ \bibinfo {author} {\bibfnamefont {W.}~\bibnamefont
  {Xiong}},\ }\bibfield  {title} {\enquote {\bibinfo {title} {Two-dimensional
  infrared spectroscopy of vibrational polaritons},}\ }\href@noop {} {\bibfield
   {journal} {\bibinfo  {journal} {Proc.\ Natl.\ Acad.\ Sci.\ U.S.A.}\ }\textbf
  {\bibinfo {volume} {115}},\ \bibinfo {pages} {4845} (\bibinfo {year}
  {2018})}\BibitemShut {NoStop}%
\bibitem [{\citenamefont {Simpkins}, \citenamefont {Dunkelberger},\ and\
  \citenamefont {Vurgaftman}(2023)}]{simpkins23}%
  \BibitemOpen
  \bibfield  {author} {\bibinfo {author} {\bibfnamefont {B.~S.}\ \bibnamefont
  {Simpkins}}, \bibinfo {author} {\bibfnamefont {A.~D.}\ \bibnamefont
  {Dunkelberger}},\ and\ \bibinfo {author} {\bibfnamefont {I.}~\bibnamefont
  {Vurgaftman}},\ }\bibfield  {title} {\enquote {\bibinfo {title} {Control,
  modulation, and analytical descriptions of vibrational strong coupling},}\
  }\href@noop {} {\bibfield  {journal} {\bibinfo  {journal} {Chem.\ Rev.}\
  }\textbf {\bibinfo {volume} {123}},\ \bibinfo {pages} {5020} (\bibinfo {year}
  {2023})}\BibitemShut {NoStop}%
\bibitem [{\citenamefont {{\'C}wik}\ \emph {et~al.}(2016)\citenamefont
  {{\'C}wik}, \citenamefont {Kirton}, \citenamefont {De~Liberato},\ and\
  \citenamefont {Keeling}}]{cwik2016}%
  \BibitemOpen
  \bibfield  {author} {\bibinfo {author} {\bibfnamefont {J.~A.}\ \bibnamefont
  {{\'C}wik}}, \bibinfo {author} {\bibfnamefont {P.}~\bibnamefont {Kirton}},
  \bibinfo {author} {\bibfnamefont {S.}~\bibnamefont {De~Liberato}},\ and\
  \bibinfo {author} {\bibfnamefont {J.}~\bibnamefont {Keeling}},\ }\bibfield
  {title} {\enquote {\bibinfo {title} {Excitonic spectral features in strongly
  coupled organic polaritons},}\ }\href@noop {} {\bibfield  {journal} {\bibinfo
   {journal} {Phys.\ Rev.\ A}\ }\textbf {\bibinfo {volume} {93}},\ \bibinfo
  {pages} {033840} (\bibinfo {year} {2016})}\BibitemShut {NoStop}%
\bibitem [{\citenamefont {Yuen-Zhou}\ and\ \citenamefont
  {Koner}(2024)}]{yuen2024}%
  \BibitemOpen
  \bibfield  {author} {\bibinfo {author} {\bibfnamefont {J.}~\bibnamefont
  {Yuen-Zhou}}\ and\ \bibinfo {author} {\bibfnamefont {A.}~\bibnamefont
  {Koner}},\ }\bibfield  {title} {\enquote {\bibinfo {title} {Linear response
  of molecular polaritons},}\ }\href@noop {} {\bibfield  {journal} {\bibinfo
  {journal} {J.\ Chem.\ Phys.}\ }\textbf {\bibinfo {volume} {160}},\ \bibinfo
  {pages} {154107} (\bibinfo {year} {2024})}\BibitemShut {NoStop}%
\bibitem [{\citenamefont {Kaluzny}\ \emph {et~al.}(1983)\citenamefont
  {Kaluzny}, \citenamefont {Goy}, \citenamefont {Gross}, \citenamefont
  {Raimond},\ and\ \citenamefont {Haroche}}]{kaluzny1983}%
  \BibitemOpen
  \bibfield  {author} {\bibinfo {author} {\bibfnamefont {Y.}~\bibnamefont
  {Kaluzny}}, \bibinfo {author} {\bibfnamefont {P.}~\bibnamefont {Goy}},
  \bibinfo {author} {\bibfnamefont {M.}~\bibnamefont {Gross}}, \bibinfo
  {author} {\bibfnamefont {J.}~\bibnamefont {Raimond}},\ and\ \bibinfo {author}
  {\bibfnamefont {S.}~\bibnamefont {Haroche}},\ }\bibfield  {title} {\enquote
  {\bibinfo {title} {Observation of self-induced {R}abi oscillations in
  two-level atoms excited inside a resonant cavity: {T}he ringing regime of
  superradiance},}\ }\href@noop {} {\bibfield  {journal} {\bibinfo  {journal}
  {Phys.\ Rev.\ Lett.}\ }\textbf {\bibinfo {volume} {51}},\ \bibinfo {pages}
  {1175} (\bibinfo {year} {1983})}\BibitemShut {NoStop}%
\bibitem [{\citenamefont {Haroche}\ and\ \citenamefont
  {Kleppner}(1989)}]{haroche1989}%
  \BibitemOpen
  \bibfield  {author} {\bibinfo {author} {\bibfnamefont {S.}~\bibnamefont
  {Haroche}}\ and\ \bibinfo {author} {\bibfnamefont {D.}~\bibnamefont
  {Kleppner}},\ }\bibfield  {title} {\enquote {\bibinfo {title} {Cavity quantum
  electrodynamics},}\ }\href@noop {} {\bibfield  {journal} {\bibinfo  {journal}
  {Phys.\ Today}\ }\textbf {\bibinfo {volume} {42}},\ \bibinfo {pages} {24}
  (\bibinfo {year} {1989})}\BibitemShut {NoStop}%
\bibitem [{\citenamefont {Thompson}, \citenamefont {Rempe},\ and\ \citenamefont
  {Kimble}(1992)}]{thompson1992}%
  \BibitemOpen
  \bibfield  {author} {\bibinfo {author} {\bibfnamefont {R.}~\bibnamefont
  {Thompson}}, \bibinfo {author} {\bibfnamefont {G.}~\bibnamefont {Rempe}},\
  and\ \bibinfo {author} {\bibfnamefont {H.}~\bibnamefont {Kimble}},\
  }\bibfield  {title} {\enquote {\bibinfo {title} {Observation of normal-mode
  splitting for an atom in an optical cavity},}\ }\href@noop {} {\bibfield
  {journal} {\bibinfo  {journal} {Phys.\ Rev.\ Lett.}\ }\textbf {\bibinfo
  {volume} {68}},\ \bibinfo {pages} {1132} (\bibinfo {year}
  {1992})}\BibitemShut {NoStop}%
\bibitem [{\citenamefont {Hennessy}\ \emph {et~al.}(2007)\citenamefont
  {Hennessy}, \citenamefont {Badolato}, \citenamefont {Winger}, \citenamefont
  {Gerace}, \citenamefont {Atat{\"u}re}, \citenamefont {Gulde}, \citenamefont
  {F{\"a}lt}, \citenamefont {Hu},\ and\ \citenamefont
  {Imamo{\u{g}}lu}}]{hennessy2007}%
  \BibitemOpen
  \bibfield  {author} {\bibinfo {author} {\bibfnamefont {K.}~\bibnamefont
  {Hennessy}}, \bibinfo {author} {\bibfnamefont {A.}~\bibnamefont {Badolato}},
  \bibinfo {author} {\bibfnamefont {M.}~\bibnamefont {Winger}}, \bibinfo
  {author} {\bibfnamefont {D.}~\bibnamefont {Gerace}}, \bibinfo {author}
  {\bibfnamefont {M.}~\bibnamefont {Atat{\"u}re}}, \bibinfo {author}
  {\bibfnamefont {S.}~\bibnamefont {Gulde}}, \bibinfo {author} {\bibfnamefont
  {S.}~\bibnamefont {F{\"a}lt}}, \bibinfo {author} {\bibfnamefont {E.~L.}\
  \bibnamefont {Hu}},\ and\ \bibinfo {author} {\bibfnamefont {A.}~\bibnamefont
  {Imamo{\u{g}}lu}},\ }\bibfield  {title} {\enquote {\bibinfo {title} {Quantum
  nature of a strongly coupled single quantum dot--cavity system},}\
  }\href@noop {} {\bibfield  {journal} {\bibinfo  {journal} {Nature}\ }\textbf
  {\bibinfo {volume} {445}},\ \bibinfo {pages} {896} (\bibinfo {year}
  {2007})}\BibitemShut {NoStop}%
\bibitem [{\citenamefont {Chikkaraddy}\ \emph {et~al.}(2016)\citenamefont
  {Chikkaraddy}, \citenamefont {De~Nijs}, \citenamefont {Benz}, \citenamefont
  {Barrow}, \citenamefont {Scherman}, \citenamefont {Rosta}, \citenamefont
  {Demetriadou}, \citenamefont {Fox}, \citenamefont {Hess},\ and\ \citenamefont
  {Baumberg}}]{chikkaraddy2016}%
  \BibitemOpen
  \bibfield  {author} {\bibinfo {author} {\bibfnamefont {R.}~\bibnamefont
  {Chikkaraddy}}, \bibinfo {author} {\bibfnamefont {B.}~\bibnamefont
  {De~Nijs}}, \bibinfo {author} {\bibfnamefont {F.}~\bibnamefont {Benz}},
  \bibinfo {author} {\bibfnamefont {S.~J.}\ \bibnamefont {Barrow}}, \bibinfo
  {author} {\bibfnamefont {O.~A.}\ \bibnamefont {Scherman}}, \bibinfo {author}
  {\bibfnamefont {E.}~\bibnamefont {Rosta}}, \bibinfo {author} {\bibfnamefont
  {A.}~\bibnamefont {Demetriadou}}, \bibinfo {author} {\bibfnamefont
  {P.}~\bibnamefont {Fox}}, \bibinfo {author} {\bibfnamefont {O.}~\bibnamefont
  {Hess}},\ and\ \bibinfo {author} {\bibfnamefont {J.~J.}\ \bibnamefont
  {Baumberg}},\ }\bibfield  {title} {\enquote {\bibinfo {title}
  {Single-molecule strong coupling at room temperature in plasmonic
  nanocavities},}\ }\href@noop {} {\bibfield  {journal} {\bibinfo  {journal}
  {Nature}\ }\textbf {\bibinfo {volume} {535}},\ \bibinfo {pages} {127}
  (\bibinfo {year} {2016})}\BibitemShut {NoStop}%
\bibitem [{\citenamefont {Santhosh}\ \emph {et~al.}(2016)\citenamefont
  {Santhosh}, \citenamefont {Bitton}, \citenamefont {Chuntonov},\ and\
  \citenamefont {Haran}}]{santhosh2016}%
  \BibitemOpen
  \bibfield  {author} {\bibinfo {author} {\bibfnamefont {K.}~\bibnamefont
  {Santhosh}}, \bibinfo {author} {\bibfnamefont {O.}~\bibnamefont {Bitton}},
  \bibinfo {author} {\bibfnamefont {L.}~\bibnamefont {Chuntonov}},\ and\
  \bibinfo {author} {\bibfnamefont {G.}~\bibnamefont {Haran}},\ }\bibfield
  {title} {\enquote {\bibinfo {title} {Vacuum {R}abi splitting in a plasmonic
  cavity at the single quantum emitter limit},}\ }\href@noop {} {\bibfield
  {journal} {\bibinfo  {journal} {Nat.\ Comm.}\ }\textbf {\bibinfo {volume}
  {7}},\ \bibinfo {pages} {11823} (\bibinfo {year} {2016})}\BibitemShut
  {NoStop}%
\bibitem [{\citenamefont {Paoletta}\ \emph {et~al.}(2024)\citenamefont
  {Paoletta}, \citenamefont {Hoffmann}, \citenamefont {Cheng}, \citenamefont
  {York}, \citenamefont {Xu}, \citenamefont {Zhang}, \citenamefont {Delor},
  \citenamefont {Berkelbach},\ and\ \citenamefont
  {Venkataraman}}]{paoletta2024}%
  \BibitemOpen
  \bibfield  {author} {\bibinfo {author} {\bibfnamefont {A.~L.}\ \bibnamefont
  {Paoletta}}, \bibinfo {author} {\bibfnamefont {N.~M.}\ \bibnamefont
  {Hoffmann}}, \bibinfo {author} {\bibfnamefont {D.~W.}\ \bibnamefont {Cheng}},
  \bibinfo {author} {\bibfnamefont {E.}~\bibnamefont {York}}, \bibinfo {author}
  {\bibfnamefont {D.}~\bibnamefont {Xu}}, \bibinfo {author} {\bibfnamefont
  {B.}~\bibnamefont {Zhang}}, \bibinfo {author} {\bibfnamefont
  {M.}~\bibnamefont {Delor}}, \bibinfo {author} {\bibfnamefont {T.~C.}\
  \bibnamefont {Berkelbach}},\ and\ \bibinfo {author} {\bibfnamefont
  {L.}~\bibnamefont {Venkataraman}},\ }\bibfield  {title} {\enquote {\bibinfo
  {title} {Plasmon-exciton strong coupling in single-molecule junction
  electroluminescence},}\ }\href@noop {} {\bibfield  {journal} {\bibinfo
  {journal} {J.\ Am.\ Chem.\ Soc.}\ }\textbf {\bibinfo {volume} {146}},\
  \bibinfo {pages} {34394} (\bibinfo {year} {2024})}\BibitemShut {NoStop}%
\bibitem [{\citenamefont {Renken}\ \emph {et~al.}(2021)\citenamefont {Renken},
  \citenamefont {Pandya}, \citenamefont {Georgiou}, \citenamefont
  {Jayaprakash}, \citenamefont {Gai}, \citenamefont {Shen}, \citenamefont
  {Lidzey}, \citenamefont {Rao},\ and\ \citenamefont {Musser}}]{renken21}%
  \BibitemOpen
  \bibfield  {author} {\bibinfo {author} {\bibfnamefont {S.}~\bibnamefont
  {Renken}}, \bibinfo {author} {\bibfnamefont {R.}~\bibnamefont {Pandya}},
  \bibinfo {author} {\bibfnamefont {K.}~\bibnamefont {Georgiou}}, \bibinfo
  {author} {\bibfnamefont {R.}~\bibnamefont {Jayaprakash}}, \bibinfo {author}
  {\bibfnamefont {L.}~\bibnamefont {Gai}}, \bibinfo {author} {\bibfnamefont
  {Z.}~\bibnamefont {Shen}}, \bibinfo {author} {\bibfnamefont {D.~G.}\
  \bibnamefont {Lidzey}}, \bibinfo {author} {\bibfnamefont {A.}~\bibnamefont
  {Rao}},\ and\ \bibinfo {author} {\bibfnamefont {A.~J.}\ \bibnamefont
  {Musser}},\ }\bibfield  {title} {\enquote {\bibinfo {title} {Untargeted
  effects in organic exciton-polariton transient spectroscopy: {A} cautionary
  tale},}\ }\href@noop {} {\bibfield  {journal} {\bibinfo  {journal} {J.\
  Chem.\ Phys.}\ }\textbf {\bibinfo {volume} {155}},\ \bibinfo {pages} {154701}
  (\bibinfo {year} {2021})}\BibitemShut {NoStop}%
\bibitem [{\citenamefont {Simpkins}\ \emph {et~al.}(2023)\citenamefont
  {Simpkins}, \citenamefont {Yang}, \citenamefont {Dunkelberger}, \citenamefont
  {Vurgaftman}, \citenamefont {Owrutsky},\ and\ \citenamefont
  {Xiong}}]{simpkins23comment}%
  \BibitemOpen
  \bibfield  {author} {\bibinfo {author} {\bibfnamefont {B.~S.}\ \bibnamefont
  {Simpkins}}, \bibinfo {author} {\bibfnamefont {Z.}~\bibnamefont {Yang}},
  \bibinfo {author} {\bibfnamefont {A.~D.}\ \bibnamefont {Dunkelberger}},
  \bibinfo {author} {\bibfnamefont {I.}~\bibnamefont {Vurgaftman}}, \bibinfo
  {author} {\bibfnamefont {J.~C.}\ \bibnamefont {Owrutsky}},\ and\ \bibinfo
  {author} {\bibfnamefont {W.}~\bibnamefont {Xiong}},\ }\bibfield  {title}
  {\enquote {\bibinfo {title} {Comment on “{I}solating polaritonic {2D}-{IR}
  transmission spectra”},}\ }\href@noop {} {\bibfield  {journal} {\bibinfo
  {journal} {J.\ Phys.\ Chem.\ Lett.}\ }\textbf {\bibinfo {volume} {14}},\
  \bibinfo {pages} {983} (\bibinfo {year} {2023})}\BibitemShut {NoStop}%
\bibitem [{\citenamefont {Thomas}\ \emph {et~al.}(2024)\citenamefont {Thomas},
  \citenamefont {Tan}, \citenamefont {Kravets}, \citenamefont {Grigorenko},\
  and\ \citenamefont {Barnes}}]{thomas2024}%
  \BibitemOpen
  \bibfield  {author} {\bibinfo {author} {\bibfnamefont {P.~A.}\ \bibnamefont
  {Thomas}}, \bibinfo {author} {\bibfnamefont {W.~J.}\ \bibnamefont {Tan}},
  \bibinfo {author} {\bibfnamefont {V.~G.}\ \bibnamefont {Kravets}}, \bibinfo
  {author} {\bibfnamefont {A.~N.}\ \bibnamefont {Grigorenko}},\ and\ \bibinfo
  {author} {\bibfnamefont {W.~L.}\ \bibnamefont {Barnes}},\ }\bibfield  {title}
  {\enquote {\bibinfo {title} {Non-polaritonic effects in cavity-modified
  photochemistry},}\ }\href@noop {} {\bibfield  {journal} {\bibinfo  {journal}
  {Adv.\ Mater.}\ }\textbf {\bibinfo {volume} {36}},\ \bibinfo {pages}
  {2309393} (\bibinfo {year} {2024})}\BibitemShut {NoStop}%
\bibitem [{\citenamefont {Ribeiro}\ \emph {et~al.}(2018)\citenamefont
  {Ribeiro}, \citenamefont {Mart{\'\i}nez-Mart{\'\i}nez}, \citenamefont {Du},
  \citenamefont {Campos-Gonzalez-Angulo},\ and\ \citenamefont
  {Yuen-Zhou}}]{ribeiro18}%
  \BibitemOpen
  \bibfield  {author} {\bibinfo {author} {\bibfnamefont {R.~F.}\ \bibnamefont
  {Ribeiro}}, \bibinfo {author} {\bibfnamefont {L.~A.}\ \bibnamefont
  {Mart{\'\i}nez-Mart{\'\i}nez}}, \bibinfo {author} {\bibfnamefont
  {M.}~\bibnamefont {Du}}, \bibinfo {author} {\bibfnamefont {J.}~\bibnamefont
  {Campos-Gonzalez-Angulo}},\ and\ \bibinfo {author} {\bibfnamefont
  {J.}~\bibnamefont {Yuen-Zhou}},\ }\bibfield  {title} {\enquote {\bibinfo
  {title} {Polariton chemistry: {C}ontrolling molecular dynamics with optical
  cavities},}\ }\href@noop {} {\bibfield  {journal} {\bibinfo  {journal}
  {Chem.\ Sci.}\ }\textbf {\bibinfo {volume} {9}},\ \bibinfo {pages} {6325}
  (\bibinfo {year} {2018})}\BibitemShut {NoStop}%
\bibitem [{\citenamefont {Hertzog}\ \emph {et~al.}(2019)\citenamefont
  {Hertzog}, \citenamefont {Wang}, \citenamefont {Mony},\ and\ \citenamefont
  {B{\"o}rjesson}}]{hertzog19}%
  \BibitemOpen
  \bibfield  {author} {\bibinfo {author} {\bibfnamefont {M.}~\bibnamefont
  {Hertzog}}, \bibinfo {author} {\bibfnamefont {M.}~\bibnamefont {Wang}},
  \bibinfo {author} {\bibfnamefont {J.}~\bibnamefont {Mony}},\ and\ \bibinfo
  {author} {\bibfnamefont {K.}~\bibnamefont {B{\"o}rjesson}},\ }\bibfield
  {title} {\enquote {\bibinfo {title} {Strong light--matter interactions: {A}
  new direction within chemistry},}\ }\href@noop {} {\bibfield  {journal}
  {\bibinfo  {journal} {Chem.\ Soc.\ Rev.}\ }\textbf {\bibinfo {volume} {48}},\
  \bibinfo {pages} {937} (\bibinfo {year} {2019})}\BibitemShut {NoStop}%
\bibitem [{\citenamefont {Garcia-Vidal}, \citenamefont {Ciuti},\ and\
  \citenamefont {Ebbesen}(2021)}]{garcia21}%
  \BibitemOpen
  \bibfield  {author} {\bibinfo {author} {\bibfnamefont {F.~J.}\ \bibnamefont
  {Garcia-Vidal}}, \bibinfo {author} {\bibfnamefont {C.}~\bibnamefont
  {Ciuti}},\ and\ \bibinfo {author} {\bibfnamefont {T.~W.}\ \bibnamefont
  {Ebbesen}},\ }\bibfield  {title} {\enquote {\bibinfo {title} {Manipulating
  matter by strong coupling to vacuum fields},}\ }\href@noop {} {\bibfield
  {journal} {\bibinfo  {journal} {Science}\ }\textbf {\bibinfo {volume}
  {373}},\ \bibinfo {pages} {eabd0336} (\bibinfo {year} {2021})}\BibitemShut
  {NoStop}%
\bibitem [{\citenamefont {Dunkelberger}\ \emph {et~al.}(2022)\citenamefont
  {Dunkelberger}, \citenamefont {Simpkins}, \citenamefont {Vurgaftman},\ and\
  \citenamefont {Owrutsky}}]{dunkelberger22}%
  \BibitemOpen
  \bibfield  {author} {\bibinfo {author} {\bibfnamefont {A.~D.}\ \bibnamefont
  {Dunkelberger}}, \bibinfo {author} {\bibfnamefont {B.~S.}\ \bibnamefont
  {Simpkins}}, \bibinfo {author} {\bibfnamefont {I.}~\bibnamefont
  {Vurgaftman}},\ and\ \bibinfo {author} {\bibfnamefont {J.~C.}\ \bibnamefont
  {Owrutsky}},\ }\bibfield  {title} {\enquote {\bibinfo {title}
  {Vibration-cavity polariton chemistry and dynamics},}\ }\href@noop {}
  {\bibfield  {journal} {\bibinfo  {journal} {Annu.\ Rev.\ Phys.\ Chem.}\
  }\textbf {\bibinfo {volume} {73}},\ \bibinfo {pages} {429} (\bibinfo {year}
  {2022})}\BibitemShut {NoStop}%
\bibitem [{\citenamefont {Steck}(2006)}]{steck06}%
  \BibitemOpen
  \bibfield  {author} {\bibinfo {author} {\bibfnamefont {D.~A.}\ \bibnamefont
  {Steck}},\ }\href@noop {} {\emph {\bibinfo {title} {Classical and Modern
  Optics}}}\ (\bibinfo  {publisher} {Oregon University,
  http://steck.us/teaching},\ \bibinfo {year} {2006})\BibitemShut {NoStop}%
\bibitem [{\citenamefont {Yariv}\ and\ \citenamefont {Yeh}(2007)}]{yariv07}%
  \BibitemOpen
  \bibfield  {author} {\bibinfo {author} {\bibfnamefont {A.}~\bibnamefont
  {Yariv}}\ and\ \bibinfo {author} {\bibfnamefont {P.}~\bibnamefont {Yeh}},\
  }\href@noop {} {\emph {\bibinfo {title} {Photonics: Optical Electronics in
  Modern Communications}}}\ (\bibinfo  {publisher} {Oxford University Press},\
  \bibinfo {year} {2007})\BibitemShut {NoStop}%
\bibitem [{\citenamefont {Born}\ and\ \citenamefont {Wolf}(2013)}]{born13}%
  \BibitemOpen
  \bibfield  {author} {\bibinfo {author} {\bibfnamefont {M.}~\bibnamefont
  {Born}}\ and\ \bibinfo {author} {\bibfnamefont {E.}~\bibnamefont {Wolf}},\
  }\href@noop {} {\emph {\bibinfo {title} {Principles of Optics:
  Electromagnetic Theory of Propagation, Interference and Diffraction of
  Light}}}\ (\bibinfo  {publisher} {Elsevier},\ \bibinfo {year}
  {2013})\BibitemShut {NoStop}%
\bibitem [{\citenamefont {Nagourney}(2014)}]{nagourney14}%
  \BibitemOpen
  \bibfield  {author} {\bibinfo {author} {\bibfnamefont {W.}~\bibnamefont
  {Nagourney}},\ }\href@noop {} {\emph {\bibinfo {title} {Quantum Electronics
  for Atomic Physics and Telecommunication}}}\ (\bibinfo  {publisher} {Oxford
  University Publishing},\ \bibinfo {year} {2014})\BibitemShut {NoStop}%
\bibitem [{\citenamefont {Lehmann}\ and\ \citenamefont
  {Romanini}(1996)}]{lehmann96}%
  \BibitemOpen
  \bibfield  {author} {\bibinfo {author} {\bibfnamefont {K.~K.}\ \bibnamefont
  {Lehmann}}\ and\ \bibinfo {author} {\bibfnamefont {D.}~\bibnamefont
  {Romanini}},\ }\bibfield  {title} {\enquote {\bibinfo {title} {The
  superposition principle and cavity ring-down spectroscopy},}\ }\href@noop {}
  {\bibfield  {journal} {\bibinfo  {journal} {J.\ Chem.\ Phys.}\ }\textbf
  {\bibinfo {volume} {105}},\ \bibinfo {pages} {10263} (\bibinfo {year}
  {1996})}\BibitemShut {NoStop}%
\bibitem [{\citenamefont {Gagliardi}\ and\ \citenamefont
  {Loock}(2014)}]{gagliardi14}%
  \BibitemOpen
  \bibinfo {editor} {\bibfnamefont {G.}~\bibnamefont {Gagliardi}}\ and\
  \bibinfo {editor} {\bibfnamefont {H.-P.}\ \bibnamefont {Loock}},\ eds.,\
  \href@noop {} {\emph {\bibinfo {title} {Cavity-Enhanced Spectroscopy and
  Sensing}}},\ \bibinfo {series} {Springer Series in Optical Sciences}, Vol.\
  \bibinfo {volume} {179}\ (\bibinfo  {publisher} {Springer},\ \bibinfo {year}
  {2014})\BibitemShut {NoStop}%
\bibitem [{\citenamefont {Adler}\ \emph {et~al.}(2009)\citenamefont {Adler},
  \citenamefont {Cossel}, \citenamefont {Thorpe}, \citenamefont {Hartl},
  \citenamefont {Fermann},\ and\ \citenamefont {Ye}}]{adler2009}%
  \BibitemOpen
  \bibfield  {author} {\bibinfo {author} {\bibfnamefont {F.}~\bibnamefont
  {Adler}}, \bibinfo {author} {\bibfnamefont {K.~C.}\ \bibnamefont {Cossel}},
  \bibinfo {author} {\bibfnamefont {M.~J.}\ \bibnamefont {Thorpe}}, \bibinfo
  {author} {\bibfnamefont {I.}~\bibnamefont {Hartl}}, \bibinfo {author}
  {\bibfnamefont {M.~E.}\ \bibnamefont {Fermann}},\ and\ \bibinfo {author}
  {\bibfnamefont {J.}~\bibnamefont {Ye}},\ }\bibfield  {title} {\enquote
  {\bibinfo {title} {Phase-stabilized, 1.5 {W} frequency comb at 2.8--4.8
  $\mu$m},}\ }\href@noop {} {\bibfield  {journal} {\bibinfo  {journal} {Opt.\
  Lett.}\ }\textbf {\bibinfo {volume} {34}},\ \bibinfo {pages} {1330} (\bibinfo
  {year} {2009})}\BibitemShut {NoStop}%
\bibitem [{\citenamefont {Pupeza}\ \emph {et~al.}(2021)\citenamefont {Pupeza},
  \citenamefont {Zhang}, \citenamefont {H{\"o}gner},\ and\ \citenamefont
  {Ye}}]{pupeza2021}%
  \BibitemOpen
  \bibfield  {author} {\bibinfo {author} {\bibfnamefont {I.}~\bibnamefont
  {Pupeza}}, \bibinfo {author} {\bibfnamefont {C.}~\bibnamefont {Zhang}},
  \bibinfo {author} {\bibfnamefont {M.}~\bibnamefont {H{\"o}gner}},\ and\
  \bibinfo {author} {\bibfnamefont {J.}~\bibnamefont {Ye}},\ }\bibfield
  {title} {\enquote {\bibinfo {title} {Extreme-ultraviolet frequency combs for
  precision metrology and attosecond science},}\ }\href@noop {} {\bibfield
  {journal} {\bibinfo  {journal} {Nat.\ Photon.}\ }\textbf {\bibinfo {volume}
  {15}},\ \bibinfo {pages} {175} (\bibinfo {year} {2021})}\BibitemShut
  {NoStop}%
\bibitem [{\citenamefont {Thorpe}\ and\ \citenamefont {Ye}(2008)}]{thorpe08}%
  \BibitemOpen
  \bibfield  {author} {\bibinfo {author} {\bibfnamefont {M.~J.}\ \bibnamefont
  {Thorpe}}\ and\ \bibinfo {author} {\bibfnamefont {J.}~\bibnamefont {Ye}},\
  }\bibfield  {title} {\enquote {\bibinfo {title} {Cavity-enhanced direct
  frequency comb spectroscopy},}\ }\href@noop {} {\bibfield  {journal}
  {\bibinfo  {journal} {Appl.\ Phys.\ B}\ }\textbf {\bibinfo {volume} {91}},\
  \bibinfo {pages} {397} (\bibinfo {year} {2008})}\BibitemShut {NoStop}%
\bibitem [{\citenamefont {Adler}\ \emph {et~al.}(2010)\citenamefont {Adler},
  \citenamefont {Thorpe}, \citenamefont {Cossel},\ and\ \citenamefont
  {Ye}}]{adler10}%
  \BibitemOpen
  \bibfield  {author} {\bibinfo {author} {\bibfnamefont {F.}~\bibnamefont
  {Adler}}, \bibinfo {author} {\bibfnamefont {M.~J.}\ \bibnamefont {Thorpe}},
  \bibinfo {author} {\bibfnamefont {K.~C.}\ \bibnamefont {Cossel}},\ and\
  \bibinfo {author} {\bibfnamefont {J.}~\bibnamefont {Ye}},\ }\bibfield
  {title} {\enquote {\bibinfo {title} {Cavity-enhanced direct frequency comb
  spectroscopy: {T}echnology and applications},}\ }\href@noop {} {\bibfield
  {journal} {\bibinfo  {journal} {Annu.\ Rev.\ Anal.\ Chem.}\ }\textbf
  {\bibinfo {volume} {3}},\ \bibinfo {pages} {175} (\bibinfo {year}
  {2010})}\BibitemShut {NoStop}%
\bibitem [{\citenamefont {Cossel}\ \emph {et~al.}(2016)\citenamefont {Cossel},
  \citenamefont {Waxman}, \citenamefont {Finneran}, \citenamefont {Blake},
  \citenamefont {Ye},\ and\ \citenamefont {Newbury}}]{cossel2016}%
  \BibitemOpen
  \bibfield  {author} {\bibinfo {author} {\bibfnamefont {K.~C.}\ \bibnamefont
  {Cossel}}, \bibinfo {author} {\bibfnamefont {E.~M.}\ \bibnamefont {Waxman}},
  \bibinfo {author} {\bibfnamefont {I.~A.}\ \bibnamefont {Finneran}}, \bibinfo
  {author} {\bibfnamefont {G.~A.}\ \bibnamefont {Blake}}, \bibinfo {author}
  {\bibfnamefont {J.}~\bibnamefont {Ye}},\ and\ \bibinfo {author}
  {\bibfnamefont {N.~R.}\ \bibnamefont {Newbury}},\ }\bibfield  {title}
  {\enquote {\bibinfo {title} {Gas-phase broadband spectroscopy using active
  sources: progress, status, and applications},}\ }\href@noop {} {\bibfield
  {journal} {\bibinfo  {journal} {J.\ Opt.\ Soc.\ Am.\ B}\ }\textbf {\bibinfo
  {volume} {34}},\ \bibinfo {pages} {104} (\bibinfo {year} {2016})}\BibitemShut
  {NoStop}%
\bibitem [{\citenamefont {Weichman}\ \emph {et~al.}(2019)\citenamefont
  {Weichman}, \citenamefont {Changala}, \citenamefont {Ye}, \citenamefont
  {Chen}, \citenamefont {Yan},\ and\ \citenamefont
  {Picqu{\'e}}}]{weichman2019}%
  \BibitemOpen
  \bibfield  {author} {\bibinfo {author} {\bibfnamefont {M.~L.}\ \bibnamefont
  {Weichman}}, \bibinfo {author} {\bibfnamefont {P.~B.}\ \bibnamefont
  {Changala}}, \bibinfo {author} {\bibfnamefont {J.}~\bibnamefont {Ye}},
  \bibinfo {author} {\bibfnamefont {Z.}~\bibnamefont {Chen}}, \bibinfo {author}
  {\bibfnamefont {M.}~\bibnamefont {Yan}},\ and\ \bibinfo {author}
  {\bibfnamefont {N.}~\bibnamefont {Picqu{\'e}}},\ }\bibfield  {title}
  {\enquote {\bibinfo {title} {Broadband molecular spectroscopy with optical
  frequency combs},}\ }\href@noop {} {\bibfield  {journal} {\bibinfo  {journal}
  {J.\ Mol.\ Spec.}\ }\textbf {\bibinfo {volume} {355}},\ \bibinfo {pages} {66}
  (\bibinfo {year} {2019})}\BibitemShut {NoStop}%
\bibitem [{\citenamefont {Berden}, \citenamefont {Peeters},\ and\ \citenamefont
  {Meijer}(2000)}]{berden2000}%
  \BibitemOpen
  \bibfield  {author} {\bibinfo {author} {\bibfnamefont {G.}~\bibnamefont
  {Berden}}, \bibinfo {author} {\bibfnamefont {R.}~\bibnamefont {Peeters}},\
  and\ \bibinfo {author} {\bibfnamefont {G.}~\bibnamefont {Meijer}},\
  }\bibfield  {title} {\enquote {\bibinfo {title} {Cavity ring-down
  spectroscopy: Experimental schemes and applications},}\ }\href@noop {}
  {\bibfield  {journal} {\bibinfo  {journal} {Int.\ Rev.\ Phys.\ Chem.}\
  }\textbf {\bibinfo {volume} {19}},\ \bibinfo {pages} {565} (\bibinfo {year}
  {2000})}\BibitemShut {NoStop}%
\bibitem [{\citenamefont {Cygan}\ \emph {et~al.}(2015)\citenamefont {Cygan},
  \citenamefont {Wcis{\l}o}, \citenamefont {W{\'o}jtewicz}, \citenamefont
  {Mas{\l}owski}, \citenamefont {Hodges}, \citenamefont {Ciury{\l}o},\ and\
  \citenamefont {Lisak}}]{cygan15}%
  \BibitemOpen
  \bibfield  {author} {\bibinfo {author} {\bibfnamefont {A.}~\bibnamefont
  {Cygan}}, \bibinfo {author} {\bibfnamefont {P.}~\bibnamefont {Wcis{\l}o}},
  \bibinfo {author} {\bibfnamefont {S.}~\bibnamefont {W{\'o}jtewicz}}, \bibinfo
  {author} {\bibfnamefont {P.}~\bibnamefont {Mas{\l}owski}}, \bibinfo {author}
  {\bibfnamefont {J.~T.}\ \bibnamefont {Hodges}}, \bibinfo {author}
  {\bibfnamefont {R.}~\bibnamefont {Ciury{\l}o}},\ and\ \bibinfo {author}
  {\bibfnamefont {D.}~\bibnamefont {Lisak}},\ }\bibfield  {title} {\enquote
  {\bibinfo {title} {One-dimensional frequency-based spectroscopy},}\
  }\href@noop {} {\bibfield  {journal} {\bibinfo  {journal} {Opt.\ Express}\
  }\textbf {\bibinfo {volume} {23}},\ \bibinfo {pages} {14472} (\bibinfo {year}
  {2015})}\BibitemShut {NoStop}%
\bibitem [{\citenamefont {Peiponen}\ and\ \citenamefont
  {Saarinen}(2009)}]{peiponen2009}%
  \BibitemOpen
  \bibfield  {author} {\bibinfo {author} {\bibfnamefont {K.}~\bibnamefont
  {Peiponen}}\ and\ \bibinfo {author} {\bibfnamefont {J.}~\bibnamefont
  {Saarinen}},\ }\bibfield  {title} {\enquote {\bibinfo {title} {Generalized
  {K}ramers--{K}ronig relations in nonlinear optical-and thz-spectroscopy},}\
  }\href@noop {} {\bibfield  {journal} {\bibinfo  {journal} {Rep.\ Prog.\
  Phys.}\ }\textbf {\bibinfo {volume} {72}},\ \bibinfo {pages} {056401}
  (\bibinfo {year} {2009})}\BibitemShut {NoStop}%
\bibitem [{\citenamefont {Burkhard}, \citenamefont {Hoke},\ and\ \citenamefont
  {McGehee}(2010)}]{burkhard10}%
  \BibitemOpen
  \bibfield  {author} {\bibinfo {author} {\bibfnamefont {G.~F.}\ \bibnamefont
  {Burkhard}}, \bibinfo {author} {\bibfnamefont {E.~T.}\ \bibnamefont {Hoke}},\
  and\ \bibinfo {author} {\bibfnamefont {M.~D.}\ \bibnamefont {McGehee}},\
  }\bibfield  {title} {\enquote {\bibinfo {title} {Accounting for interference,
  scattering, and electrode absorption to make accurate internal quantum
  efficiency measurements in organic and other thin solar cells},}\ }\href@noop
  {} {\bibfield  {journal} {\bibinfo  {journal} {Adv.\ Mater.}\ }\textbf
  {\bibinfo {volume} {22}},\ \bibinfo {pages} {3293} (\bibinfo {year}
  {2010})}\BibitemShut {NoStop}%
\bibitem [{\citenamefont {Macleod}(2017)}]{macleod17}%
  \BibitemOpen
  \bibfield  {author} {\bibinfo {author} {\bibfnamefont {H.~A.}\ \bibnamefont
  {Macleod}},\ }\href@noop {} {\emph {\bibinfo {title} {Thin-Film Optical
  Filters}}}\ (\bibinfo  {publisher} {CRC Press},\ \bibinfo {year}
  {2017})\BibitemShut {NoStop}%
\bibitem [{\citenamefont {George}\ \emph {et~al.}(2015)\citenamefont {George},
  \citenamefont {Shalabney}, \citenamefont {Hutchison}, \citenamefont {Genet},\
  and\ \citenamefont {Ebbesen}}]{george15}%
  \BibitemOpen
  \bibfield  {author} {\bibinfo {author} {\bibfnamefont {J.}~\bibnamefont
  {George}}, \bibinfo {author} {\bibfnamefont {A.}~\bibnamefont {Shalabney}},
  \bibinfo {author} {\bibfnamefont {J.~A.}\ \bibnamefont {Hutchison}}, \bibinfo
  {author} {\bibfnamefont {C.}~\bibnamefont {Genet}},\ and\ \bibinfo {author}
  {\bibfnamefont {T.~W.}\ \bibnamefont {Ebbesen}},\ }\bibfield  {title}
  {\enquote {\bibinfo {title} {Liquid-phase vibrational strong coupling},}\
  }\href@noop {} {\bibfield  {journal} {\bibinfo  {journal} {J.\ Phys.\ Chem.\
  Lett.}\ }\textbf {\bibinfo {volume} {6}},\ \bibinfo {pages} {1027} (\bibinfo
  {year} {2015})}\BibitemShut {NoStop}%
\bibitem [{\citenamefont {Xiang}\ \emph {et~al.}(2021)\citenamefont {Xiang},
  \citenamefont {Wang}, \citenamefont {Yang},\ and\ \citenamefont
  {Xiong}}]{xiang21}%
  \BibitemOpen
  \bibfield  {author} {\bibinfo {author} {\bibfnamefont {B.}~\bibnamefont
  {Xiang}}, \bibinfo {author} {\bibfnamefont {J.}~\bibnamefont {Wang}},
  \bibinfo {author} {\bibfnamefont {Z.}~\bibnamefont {Yang}},\ and\ \bibinfo
  {author} {\bibfnamefont {W.}~\bibnamefont {Xiong}},\ }\bibfield  {title}
  {\enquote {\bibinfo {title} {Nonlinear infrared polaritonic interaction
  between cavities mediated by molecular vibrations at ultrafast time scale},}\
  }\href@noop {} {\bibfield  {journal} {\bibinfo  {journal} {Sci.\ Adv.}\
  }\textbf {\bibinfo {volume} {7}},\ \bibinfo {pages} {eabf6397} (\bibinfo
  {year} {2021})}\BibitemShut {NoStop}%
\bibitem [{\citenamefont {Simpkins}\ \emph {et~al.}(2015)\citenamefont
  {Simpkins}, \citenamefont {Fears}, \citenamefont {Dressick}, \citenamefont
  {Spann}, \citenamefont {Dunkelberger},\ and\ \citenamefont
  {Owrutsky}}]{simpkins2015}%
  \BibitemOpen
  \bibfield  {author} {\bibinfo {author} {\bibfnamefont {B.}~\bibnamefont
  {Simpkins}}, \bibinfo {author} {\bibfnamefont {K.~P.}\ \bibnamefont {Fears}},
  \bibinfo {author} {\bibfnamefont {W.~J.}\ \bibnamefont {Dressick}}, \bibinfo
  {author} {\bibfnamefont {B.~T.}\ \bibnamefont {Spann}}, \bibinfo {author}
  {\bibfnamefont {A.~D.}\ \bibnamefont {Dunkelberger}},\ and\ \bibinfo {author}
  {\bibfnamefont {J.~C.}\ \bibnamefont {Owrutsky}},\ }\bibfield  {title}
  {\enquote {\bibinfo {title} {Spanning strong to weak normal mode coupling
  between vibrational and fabry--p{\'e}rot cavity modes through tuning of
  vibrational absorption strength},}\ }\href@noop {} {\bibfield  {journal}
  {\bibinfo  {journal} {ACS Photonics}\ }\textbf {\bibinfo {volume} {2}},\
  \bibinfo {pages} {1460} (\bibinfo {year} {2015})}\BibitemShut {NoStop}%
\bibitem [{\citenamefont {Balasubrahmaniyam}, \citenamefont {Genet},\ and\
  \citenamefont {Schwartz}(2021)}]{bala21}%
  \BibitemOpen
  \bibfield  {author} {\bibinfo {author} {\bibfnamefont {M.}~\bibnamefont
  {Balasubrahmaniyam}}, \bibinfo {author} {\bibfnamefont {C.}~\bibnamefont
  {Genet}},\ and\ \bibinfo {author} {\bibfnamefont {T.}~\bibnamefont
  {Schwartz}},\ }\bibfield  {title} {\enquote {\bibinfo {title} {Coupling and
  decoupling of polaritonic states in multimode cavities},}\ }\href@noop {}
  {\bibfield  {journal} {\bibinfo  {journal} {Phys.\ Rev.\ B}\ }\textbf
  {\bibinfo {volume} {103}},\ \bibinfo {pages} {L241407} (\bibinfo {year}
  {2021})}\BibitemShut {NoStop}%
\bibitem [{\citenamefont {Wright}, \citenamefont {Nelson},\ and\ \citenamefont
  {Weichman}(2023{\natexlab{a}})}]{wright23}%
  \BibitemOpen
  \bibfield  {author} {\bibinfo {author} {\bibfnamefont {A.~D.}\ \bibnamefont
  {Wright}}, \bibinfo {author} {\bibfnamefont {J.~C.}\ \bibnamefont {Nelson}},\
  and\ \bibinfo {author} {\bibfnamefont {M.~L.}\ \bibnamefont {Weichman}},\
  }\bibfield  {title} {\enquote {\bibinfo {title} {Rovibrational polaritons in
  gas-phase methane},}\ }\href@noop {} {\bibfield  {journal} {\bibinfo
  {journal} {J.\ Am.\ Chem.\ Soc.}\ }\textbf {\bibinfo {volume} {145}},\
  \bibinfo {pages} {5982} (\bibinfo {year} {2023}{\natexlab{a}})}\BibitemShut
  {NoStop}%
\bibitem [{\citenamefont {Houdr{\'e}}, \citenamefont {Stanley},\ and\
  \citenamefont {Ilegems}(1996)}]{houdre1996}%
  \BibitemOpen
  \bibfield  {author} {\bibinfo {author} {\bibfnamefont {R.}~\bibnamefont
  {Houdr{\'e}}}, \bibinfo {author} {\bibfnamefont {R.}~\bibnamefont
  {Stanley}},\ and\ \bibinfo {author} {\bibfnamefont {M.}~\bibnamefont
  {Ilegems}},\ }\bibfield  {title} {\enquote {\bibinfo {title} {Vacuum-field
  {R}abi splitting in the presence of inhomogeneous broadening: {R}esolution of
  a homogeneous linewidth in an inhomogeneously broadened system},}\
  }\href@noop {} {\bibfield  {journal} {\bibinfo  {journal} {Phys.\ Rev.\ A}\
  }\textbf {\bibinfo {volume} {53}},\ \bibinfo {pages} {2711} (\bibinfo {year}
  {1996})}\BibitemShut {NoStop}%
\bibitem [{\citenamefont {Ell}\ \emph {et~al.}(1998)\citenamefont {Ell},
  \citenamefont {Prineas}, \citenamefont {Nelson~Jr}, \citenamefont {Park},
  \citenamefont {Gibbs}, \citenamefont {Khitrova}, \citenamefont {Koch},\ and\
  \citenamefont {Houdr{\'e}}}]{ell1998}%
  \BibitemOpen
  \bibfield  {author} {\bibinfo {author} {\bibfnamefont {C.}~\bibnamefont
  {Ell}}, \bibinfo {author} {\bibfnamefont {J.}~\bibnamefont {Prineas}},
  \bibinfo {author} {\bibfnamefont {T.}~\bibnamefont {Nelson~Jr}}, \bibinfo
  {author} {\bibfnamefont {S.}~\bibnamefont {Park}}, \bibinfo {author}
  {\bibfnamefont {H.}~\bibnamefont {Gibbs}}, \bibinfo {author} {\bibfnamefont
  {G.}~\bibnamefont {Khitrova}}, \bibinfo {author} {\bibfnamefont
  {S.}~\bibnamefont {Koch}},\ and\ \bibinfo {author} {\bibfnamefont
  {R.}~\bibnamefont {Houdr{\'e}}},\ }\bibfield  {title} {\enquote {\bibinfo
  {title} {Influence of structural disorder and light coupling on the excitonic
  response of semiconductor microcavities},}\ }\href@noop {} {\bibfield
  {journal} {\bibinfo  {journal} {Phys. Rev. Lett.}\ }\textbf {\bibinfo
  {volume} {80}},\ \bibinfo {pages} {4795} (\bibinfo {year}
  {1998})}\BibitemShut {NoStop}%
\bibitem [{\citenamefont {Long}\ and\ \citenamefont
  {Simpkins}(2015)}]{long2015}%
  \BibitemOpen
  \bibfield  {author} {\bibinfo {author} {\bibfnamefont {J.~P.}\ \bibnamefont
  {Long}}\ and\ \bibinfo {author} {\bibfnamefont {B.}~\bibnamefont
  {Simpkins}},\ }\bibfield  {title} {\enquote {\bibinfo {title} {Coherent
  coupling between a molecular vibration and {F}abry--{P}\'{e}rot optical
  cavity to give hybridized states in the strong coupling limit},}\ }\href@noop
  {} {\bibfield  {journal} {\bibinfo  {journal} {ACS Photonics}\ }\textbf
  {\bibinfo {volume} {2}},\ \bibinfo {pages} {130} (\bibinfo {year}
  {2015})}\BibitemShut {NoStop}%
\bibitem [{\citenamefont {Pyles}\ \emph {et~al.}(2024)\citenamefont {Pyles},
  \citenamefont {Simpkins}, \citenamefont {Vurgaftman}, \citenamefont
  {Owrutsky},\ and\ \citenamefont {Dunkelberger}}]{pyles2024}%
  \BibitemOpen
  \bibfield  {author} {\bibinfo {author} {\bibfnamefont {C.~G.}\ \bibnamefont
  {Pyles}}, \bibinfo {author} {\bibfnamefont {B.~S.}\ \bibnamefont {Simpkins}},
  \bibinfo {author} {\bibfnamefont {I.}~\bibnamefont {Vurgaftman}}, \bibinfo
  {author} {\bibfnamefont {J.~C.}\ \bibnamefont {Owrutsky}},\ and\ \bibinfo
  {author} {\bibfnamefont {A.~D.}\ \bibnamefont {Dunkelberger}},\ }\bibfield
  {title} {\enquote {\bibinfo {title} {Revisiting cavity-coupled {2DIR}: {A}
  classical approach implicates reservoir modes},}\ }\href@noop {} {\bibfield
  {journal} {\bibinfo  {journal} {J.\ Chem.\ Phys.}\ }\textbf {\bibinfo
  {volume} {161}},\ \bibinfo {pages} {234202} (\bibinfo {year}
  {2024})}\BibitemShut {NoStop}%
\bibitem [{\citenamefont {Botzung}\ \emph {et~al.}(2020)\citenamefont
  {Botzung}, \citenamefont {Hagenm\"uller}, \citenamefont {Sch\"utz},
  \citenamefont {Dubail}, \citenamefont {Pupillo},\ and\ \citenamefont
  {Schachenmayer}}]{Botzung2020}%
  \BibitemOpen
  \bibfield  {author} {\bibinfo {author} {\bibfnamefont {T.}~\bibnamefont
  {Botzung}}, \bibinfo {author} {\bibfnamefont {D.}~\bibnamefont
  {Hagenm\"uller}}, \bibinfo {author} {\bibfnamefont {S.}~\bibnamefont
  {Sch\"utz}}, \bibinfo {author} {\bibfnamefont {J.}~\bibnamefont {Dubail}},
  \bibinfo {author} {\bibfnamefont {G.}~\bibnamefont {Pupillo}},\ and\ \bibinfo
  {author} {\bibfnamefont {J.}~\bibnamefont {Schachenmayer}},\ }\bibfield
  {title} {\enquote {\bibinfo {title} {Dark state semilocalization of quantum
  emitters in a cavity},}\ }\href {https://doi.org/10.1103/PhysRevB.102.144202}
  {\bibfield  {journal} {\bibinfo  {journal} {Phys.\ Rev.\ B}\ }\textbf
  {\bibinfo {volume} {102}},\ \bibinfo {pages} {144202} (\bibinfo {year}
  {2020})}\BibitemShut {NoStop}%
\bibitem [{\citenamefont {Georgiou}\ \emph {et~al.}(2021)\citenamefont
  {Georgiou}, \citenamefont {Jayaprakash}, \citenamefont {Othonos},\ and\
  \citenamefont {Lidzey}}]{georgiou2021}%
  \BibitemOpen
  \bibfield  {author} {\bibinfo {author} {\bibfnamefont {K.}~\bibnamefont
  {Georgiou}}, \bibinfo {author} {\bibfnamefont {R.}~\bibnamefont
  {Jayaprakash}}, \bibinfo {author} {\bibfnamefont {A.}~\bibnamefont
  {Othonos}},\ and\ \bibinfo {author} {\bibfnamefont {D.~G.}\ \bibnamefont
  {Lidzey}},\ }\bibfield  {title} {\enquote {\bibinfo {title} {Ultralong-range
  polariton-assisted energy transfer in organic microcavities},}\ }\href@noop
  {} {\bibfield  {journal} {\bibinfo  {journal} {Angew.\ Chem.\ Int.\ Ed.}\
  }\textbf {\bibinfo {volume} {133}},\ \bibinfo {pages} {16797} (\bibinfo
  {year} {2021})}\BibitemShut {NoStop}%
\bibitem [{\citenamefont {Son}\ \emph {et~al.}(2022)\citenamefont {Son},
  \citenamefont {Armstrong}, \citenamefont {Allen}, \citenamefont {Dhavamani},
  \citenamefont {Arnold},\ and\ \citenamefont {Zanni}}]{son2022}%
  \BibitemOpen
  \bibfield  {author} {\bibinfo {author} {\bibfnamefont {M.}~\bibnamefont
  {Son}}, \bibinfo {author} {\bibfnamefont {Z.~T.}\ \bibnamefont {Armstrong}},
  \bibinfo {author} {\bibfnamefont {R.~T.}\ \bibnamefont {Allen}}, \bibinfo
  {author} {\bibfnamefont {A.}~\bibnamefont {Dhavamani}}, \bibinfo {author}
  {\bibfnamefont {M.~S.}\ \bibnamefont {Arnold}},\ and\ \bibinfo {author}
  {\bibfnamefont {M.~T.}\ \bibnamefont {Zanni}},\ }\bibfield  {title} {\enquote
  {\bibinfo {title} {Energy cascades in donor-acceptor exciton-polaritons
  observed by ultrafast two-dimensional white-light spectroscopy},}\
  }\href@noop {} {\bibfield  {journal} {\bibinfo  {journal} {Nat.\ Comm.}\
  }\textbf {\bibinfo {volume} {13}},\ \bibinfo {pages} {7305} (\bibinfo {year}
  {2022})}\BibitemShut {NoStop}%
\bibitem [{\citenamefont {Yu}\ \emph {et~al.}(2009)\citenamefont {Yu},
  \citenamefont {Xiong}, \citenamefont {Chen}, \citenamefont {Wang},
  \citenamefont {Xiao},\ and\ \citenamefont {Zhang}}]{yu2009}%
  \BibitemOpen
  \bibfield  {author} {\bibinfo {author} {\bibfnamefont {X.}~\bibnamefont
  {Yu}}, \bibinfo {author} {\bibfnamefont {D.}~\bibnamefont {Xiong}}, \bibinfo
  {author} {\bibfnamefont {H.}~\bibnamefont {Chen}}, \bibinfo {author}
  {\bibfnamefont {P.}~\bibnamefont {Wang}}, \bibinfo {author} {\bibfnamefont
  {M.}~\bibnamefont {Xiao}},\ and\ \bibinfo {author} {\bibfnamefont
  {J.}~\bibnamefont {Zhang}},\ }\bibfield  {title} {\enquote {\bibinfo {title}
  {Multi-normal-mode splitting of a cavity in the presence of atoms: {A} step
  towards the superstrong-coupling regime},}\ }\href@noop {} {\bibfield
  {journal} {\bibinfo  {journal} {Phys.\ Rev.\ A}\ }\textbf {\bibinfo {volume}
  {79}},\ \bibinfo {pages} {061803} (\bibinfo {year} {2009})}\BibitemShut
  {NoStop}%
\bibitem [{\citenamefont {Sundaresan}\ \emph {et~al.}(2015)\citenamefont
  {Sundaresan}, \citenamefont {Liu}, \citenamefont {Sadri}, \citenamefont
  {Sz{\H{o}}cs}, \citenamefont {Underwood}, \citenamefont {Malekakhlagh},
  \citenamefont {T{\"u}reci},\ and\ \citenamefont {Houck}}]{sundaresan2015}%
  \BibitemOpen
  \bibfield  {author} {\bibinfo {author} {\bibfnamefont {N.~M.}\ \bibnamefont
  {Sundaresan}}, \bibinfo {author} {\bibfnamefont {Y.}~\bibnamefont {Liu}},
  \bibinfo {author} {\bibfnamefont {D.}~\bibnamefont {Sadri}}, \bibinfo
  {author} {\bibfnamefont {L.~J.}\ \bibnamefont {Sz{\H{o}}cs}}, \bibinfo
  {author} {\bibfnamefont {D.~L.}\ \bibnamefont {Underwood}}, \bibinfo {author}
  {\bibfnamefont {M.}~\bibnamefont {Malekakhlagh}}, \bibinfo {author}
  {\bibfnamefont {H.~E.}\ \bibnamefont {T{\"u}reci}},\ and\ \bibinfo {author}
  {\bibfnamefont {A.~A.}\ \bibnamefont {Houck}},\ }\bibfield  {title} {\enquote
  {\bibinfo {title} {Beyond strong coupling in a multimode cavity},}\
  }\href@noop {} {\bibfield  {journal} {\bibinfo  {journal} {Phys.\ Rev.\ X}\
  }\textbf {\bibinfo {volume} {5}},\ \bibinfo {pages} {021035} (\bibinfo {year}
  {2015})}\BibitemShut {NoStop}%
\bibitem [{\citenamefont {George}\ \emph {et~al.}(2016)\citenamefont {George},
  \citenamefont {Chervy}, \citenamefont {Shalabney}, \citenamefont {Devaux},
  \citenamefont {Hiura}, \citenamefont {Genet},\ and\ \citenamefont
  {Ebbesen}}]{george2016}%
  \BibitemOpen
  \bibfield  {author} {\bibinfo {author} {\bibfnamefont {J.}~\bibnamefont
  {George}}, \bibinfo {author} {\bibfnamefont {T.}~\bibnamefont {Chervy}},
  \bibinfo {author} {\bibfnamefont {A.}~\bibnamefont {Shalabney}}, \bibinfo
  {author} {\bibfnamefont {E.}~\bibnamefont {Devaux}}, \bibinfo {author}
  {\bibfnamefont {H.}~\bibnamefont {Hiura}}, \bibinfo {author} {\bibfnamefont
  {C.}~\bibnamefont {Genet}},\ and\ \bibinfo {author} {\bibfnamefont {T.~W.}\
  \bibnamefont {Ebbesen}},\ }\bibfield  {title} {\enquote {\bibinfo {title}
  {Multiple {R}abi splittings under ultrastrong vibrational coupling},}\
  }\href@noop {} {\bibfield  {journal} {\bibinfo  {journal} {Phys.\ Rev.\
  Lett.}\ }\textbf {\bibinfo {volume} {117}},\ \bibinfo {pages} {153601}
  (\bibinfo {year} {2016})}\BibitemShut {NoStop}%
\bibitem [{\citenamefont {Hertzog}\ and\ \citenamefont
  {B{\"o}rjesson}(2020)}]{hertzog2020}%
  \BibitemOpen
  \bibfield  {author} {\bibinfo {author} {\bibfnamefont {M.}~\bibnamefont
  {Hertzog}}\ and\ \bibinfo {author} {\bibfnamefont {K.}~\bibnamefont
  {B{\"o}rjesson}},\ }\bibfield  {title} {\enquote {\bibinfo {title} {The
  effect of coupling mode in the vibrational strong coupling regime},}\
  }\href@noop {} {\bibfield  {journal} {\bibinfo  {journal} {ChemPhotoChem}\
  }\textbf {\bibinfo {volume} {4}},\ \bibinfo {pages} {612} (\bibinfo {year}
  {2020})}\BibitemShut {NoStop}%
\bibitem [{\citenamefont {Wright}, \citenamefont {Nelson},\ and\ \citenamefont
  {Weichman}(2023{\natexlab{b}})}]{wright2023_2}%
  \BibitemOpen
  \bibfield  {author} {\bibinfo {author} {\bibfnamefont {A.~D.}\ \bibnamefont
  {Wright}}, \bibinfo {author} {\bibfnamefont {J.~C.}\ \bibnamefont {Nelson}},\
  and\ \bibinfo {author} {\bibfnamefont {M.~L.}\ \bibnamefont {Weichman}},\
  }\bibfield  {title} {\enquote {\bibinfo {title} {A versatile platform for
  gas-phase molecular polaritonics},}\ }\href@noop {} {\bibfield  {journal}
  {\bibinfo  {journal} {J.\ Chem.\ Phys.}\ }\textbf {\bibinfo {volume} {159}},\
  \bibinfo {pages} {164202} (\bibinfo {year} {2023}{\natexlab{b}})}\BibitemShut
  {NoStop}%
\bibitem [{\citenamefont {Liu}, \citenamefont {Menon},\ and\ \citenamefont
  {Sfeir}(2021)}]{liu2021}%
  \BibitemOpen
  \bibfield  {author} {\bibinfo {author} {\bibfnamefont {B.}~\bibnamefont
  {Liu}}, \bibinfo {author} {\bibfnamefont {V.~M.}\ \bibnamefont {Menon}},\
  and\ \bibinfo {author} {\bibfnamefont {M.~Y.}\ \bibnamefont {Sfeir}},\
  }\bibfield  {title} {\enquote {\bibinfo {title} {Ultrafast thermal
  modification of strong coupling in an organic microcavity},}\ }\href@noop {}
  {\bibfield  {journal} {\bibinfo  {journal} {APL Photonics}\ }\textbf
  {\bibinfo {volume} {6}},\ \bibinfo {pages} {016103} (\bibinfo {year}
  {2021})}\BibitemShut {NoStop}%
\bibitem [{\citenamefont {Reber}, \citenamefont {Chen},\ and\ \citenamefont
  {Allison}(2016)}]{reber2016}%
  \BibitemOpen
  \bibfield  {author} {\bibinfo {author} {\bibfnamefont {M.~A.}\ \bibnamefont
  {Reber}}, \bibinfo {author} {\bibfnamefont {Y.}~\bibnamefont {Chen}},\ and\
  \bibinfo {author} {\bibfnamefont {T.~K.}\ \bibnamefont {Allison}},\
  }\bibfield  {title} {\enquote {\bibinfo {title} {Cavity-enhanced ultrafast
  spectroscopy: {U}ltrafast meets ultrasensitive},}\ }\href@noop {} {\bibfield
  {journal} {\bibinfo  {journal} {Optica}\ }\textbf {\bibinfo {volume} {3}},\
  \bibinfo {pages} {311} (\bibinfo {year} {2016})}\BibitemShut {NoStop}%
\bibitem [{\citenamefont {Reitz}, \citenamefont {Koner},\ and\ \citenamefont
  {Yuen-Zhou}(2024)}]{reitz2024}%
  \BibitemOpen
  \bibfield  {author} {\bibinfo {author} {\bibfnamefont {M.}~\bibnamefont
  {Reitz}}, \bibinfo {author} {\bibfnamefont {A.}~\bibnamefont {Koner}},\ and\
  \bibinfo {author} {\bibfnamefont {J.}~\bibnamefont {Yuen-Zhou}},\ }\bibfield
  {title} {\enquote {\bibinfo {title} {Nonlinear semiclassical spectroscopy of
  ultrafast molecular polariton dynamics},}\ }\href@noop {} {\bibfield
  {journal} {\bibinfo  {journal} {arXiv:2410.16630}\ } (\bibinfo {year}
  {2024})}\BibitemShut {NoStop}%
\bibitem [{\citenamefont {Grafton}\ \emph {et~al.}(2021)\citenamefont
  {Grafton}, \citenamefont {Dunkelberger}, \citenamefont {Simpkins},
  \citenamefont {Triana}, \citenamefont {Hern{\'a}ndez}, \citenamefont
  {Herrera},\ and\ \citenamefont {Owrutsky}}]{grafton2021}%
  \BibitemOpen
  \bibfield  {author} {\bibinfo {author} {\bibfnamefont {A.~B.}\ \bibnamefont
  {Grafton}}, \bibinfo {author} {\bibfnamefont {A.~D.}\ \bibnamefont
  {Dunkelberger}}, \bibinfo {author} {\bibfnamefont {B.~S.}\ \bibnamefont
  {Simpkins}}, \bibinfo {author} {\bibfnamefont {J.~F.}\ \bibnamefont
  {Triana}}, \bibinfo {author} {\bibfnamefont {F.~J.}\ \bibnamefont
  {Hern{\'a}ndez}}, \bibinfo {author} {\bibfnamefont {F.}~\bibnamefont
  {Herrera}},\ and\ \bibinfo {author} {\bibfnamefont {J.~C.}\ \bibnamefont
  {Owrutsky}},\ }\bibfield  {title} {\enquote {\bibinfo {title} {Excited-state
  vibration-polariton transitions and dynamics in nitroprusside},}\ }\href@noop
  {} {\bibfield  {journal} {\bibinfo  {journal} {Nat.\ Comm.}\ }\textbf
  {\bibinfo {volume} {12}},\ \bibinfo {pages} {214} (\bibinfo {year}
  {2021})}\BibitemShut {NoStop}%
\bibitem [{\citenamefont {Scholes}(2020)}]{scholes2020}%
  \BibitemOpen
  \bibfield  {author} {\bibinfo {author} {\bibfnamefont {G.~D.}\ \bibnamefont
  {Scholes}},\ }\bibfield  {title} {\enquote {\bibinfo {title} {Polaritons and
  excitons: {H}amiltonian design for enhanced coherence},}\ }\href@noop {}
  {\bibfield  {journal} {\bibinfo  {journal} {Proc.\ R.\ Soc.\ A}\ }\textbf
  {\bibinfo {volume} {476}},\ \bibinfo {pages} {20200278} (\bibinfo {year}
  {2020})}\BibitemShut {NoStop}%
\bibitem [{\citenamefont {Du}\ and\ \citenamefont {Yuen-Zhou}(2022)}]{du2022}%
  \BibitemOpen
  \bibfield  {author} {\bibinfo {author} {\bibfnamefont {M.}~\bibnamefont
  {Du}}\ and\ \bibinfo {author} {\bibfnamefont {J.}~\bibnamefont {Yuen-Zhou}},\
  }\bibfield  {title} {\enquote {\bibinfo {title} {Catalysis by dark states in
  vibropolaritonic chemistry},}\ }\href@noop {} {\bibfield  {journal} {\bibinfo
   {journal} {Phys.\ Rev.\ Lett.}\ }\textbf {\bibinfo {volume} {128}},\
  \bibinfo {pages} {096001} (\bibinfo {year} {2022})}\BibitemShut {NoStop}%
\bibitem [{\citenamefont {Liu}, \citenamefont {Yin},\ and\ \citenamefont
  {Xiong}(2025)}]{liu2025}%
  \BibitemOpen
  \bibfield  {author} {\bibinfo {author} {\bibfnamefont {T.}~\bibnamefont
  {Liu}}, \bibinfo {author} {\bibfnamefont {G.}~\bibnamefont {Yin}},\ and\
  \bibinfo {author} {\bibfnamefont {W.}~\bibnamefont {Xiong}},\ }\bibfield
  {title} {\enquote {\bibinfo {title} {Unlocking delocalization: {H}ow much
  coupling strength is required to overcome energy disorder in molecular
  polaritons?}}\ }\href@noop {} {\bibfield  {journal} {\bibinfo  {journal}
  {Chem.\ Sci.}\ }\textbf {\bibinfo {volume} {16}},\ \bibinfo {pages} {4676}
  (\bibinfo {year} {2025})}\BibitemShut {NoStop}%
\bibitem [{\citenamefont {Pang}\ \emph {et~al.}(2020)\citenamefont {Pang},
  \citenamefont {Thomas}, \citenamefont {Nagarajan}, \citenamefont {Vergauwe},
  \citenamefont {Joseph}, \citenamefont {Patrahau}, \citenamefont {Wang},
  \citenamefont {Genet},\ and\ \citenamefont {Ebbesen}}]{pang2020}%
  \BibitemOpen
  \bibfield  {author} {\bibinfo {author} {\bibfnamefont {Y.}~\bibnamefont
  {Pang}}, \bibinfo {author} {\bibfnamefont {A.}~\bibnamefont {Thomas}},
  \bibinfo {author} {\bibfnamefont {K.}~\bibnamefont {Nagarajan}}, \bibinfo
  {author} {\bibfnamefont {R.~M.}\ \bibnamefont {Vergauwe}}, \bibinfo {author}
  {\bibfnamefont {K.}~\bibnamefont {Joseph}}, \bibinfo {author} {\bibfnamefont
  {B.}~\bibnamefont {Patrahau}}, \bibinfo {author} {\bibfnamefont
  {K.}~\bibnamefont {Wang}}, \bibinfo {author} {\bibfnamefont {C.}~\bibnamefont
  {Genet}},\ and\ \bibinfo {author} {\bibfnamefont {T.~W.}\ \bibnamefont
  {Ebbesen}},\ }\bibfield  {title} {\enquote {\bibinfo {title} {On the role of
  symmetry in vibrational strong coupling: {T}he case of charge-transfer
  complexation},}\ }\href@noop {} {\bibfield  {journal} {\bibinfo  {journal}
  {Angew.\ Chem.\ Int.\ Ed.}\ }\textbf {\bibinfo {volume} {59}},\ \bibinfo
  {pages} {10436} (\bibinfo {year} {2020})}\BibitemShut {NoStop}%
\bibitem [{\citenamefont {Ahn}\ and\ \citenamefont {Simpkins}(2020)}]{ahn2020}%
  \BibitemOpen
  \bibfield  {author} {\bibinfo {author} {\bibfnamefont {W.}~\bibnamefont
  {Ahn}}\ and\ \bibinfo {author} {\bibfnamefont {B.}~\bibnamefont {Simpkins}},\
  }\bibfield  {title} {\enquote {\bibinfo {title} {Raman scattering under
  strong vibration-cavity coupling},}\ }\href@noop {} {\bibfield  {journal}
  {\bibinfo  {journal} {J.\ Phys.\ Chem.\ C}\ }\textbf {\bibinfo {volume}
  {125}},\ \bibinfo {pages} {830} (\bibinfo {year} {2020})}\BibitemShut
  {NoStop}%
\bibitem [{\citenamefont {Nelson}\ and\ \citenamefont
  {Weichman}(2024)}]{nelson24}%
  \BibitemOpen
  \bibfield  {author} {\bibinfo {author} {\bibfnamefont {J.~C.}\ \bibnamefont
  {Nelson}}\ and\ \bibinfo {author} {\bibfnamefont {M.~L.}\ \bibnamefont
  {Weichman}},\ }\bibfield  {title} {\enquote {\bibinfo {title} {More than just
  smoke and mirrors: {G}as-phase polaritons for optical control of
  chemistry},}\ }\href@noop {} {\bibfield  {journal} {\bibinfo  {journal} {J.\
  Chem.\ Phys.}\ }\textbf {\bibinfo {volume} {161}},\ \bibinfo {pages} {074304}
  (\bibinfo {year} {2024})}\BibitemShut {NoStop}%
\bibitem [{\citenamefont {Wang}\ \emph {et~al.}(2014)\citenamefont {Wang},
  \citenamefont {Vasa}, \citenamefont {Pomraenke}, \citenamefont {Vogelgesang},
  \citenamefont {De~Sio}, \citenamefont {Sommer}, \citenamefont {Maiuri},
  \citenamefont {Manzoni}, \citenamefont {Cerullo},\ and\ \citenamefont
  {Lienau}}]{wang2014}%
  \BibitemOpen
  \bibfield  {author} {\bibinfo {author} {\bibfnamefont {W.}~\bibnamefont
  {Wang}}, \bibinfo {author} {\bibfnamefont {P.}~\bibnamefont {Vasa}}, \bibinfo
  {author} {\bibfnamefont {R.}~\bibnamefont {Pomraenke}}, \bibinfo {author}
  {\bibfnamefont {R.}~\bibnamefont {Vogelgesang}}, \bibinfo {author}
  {\bibfnamefont {A.}~\bibnamefont {De~Sio}}, \bibinfo {author} {\bibfnamefont
  {E.}~\bibnamefont {Sommer}}, \bibinfo {author} {\bibfnamefont
  {M.}~\bibnamefont {Maiuri}}, \bibinfo {author} {\bibfnamefont
  {C.}~\bibnamefont {Manzoni}}, \bibinfo {author} {\bibfnamefont
  {G.}~\bibnamefont {Cerullo}},\ and\ \bibinfo {author} {\bibfnamefont
  {C.}~\bibnamefont {Lienau}},\ }\bibfield  {title} {\enquote {\bibinfo {title}
  {Interplay between strong coupling and radiative damping of excitons and
  surface plasmon polaritons in hybrid nanostructures},}\ }\href@noop {}
  {\bibfield  {journal} {\bibinfo  {journal} {{ACS} {N}ano}\ }\textbf {\bibinfo
  {volume} {8}},\ \bibinfo {pages} {1056} (\bibinfo {year} {2014})}\BibitemShut
  {NoStop}%
\bibitem [{\citenamefont {Brawley}\ \emph {et~al.}(2025)\citenamefont
  {Brawley}, \citenamefont {Pannir-Sivajothi}, \citenamefont {Yim},
  \citenamefont {Poh}, \citenamefont {Yuen-Zhou},\ and\ \citenamefont
  {Sheldon}}]{brawley2025}%
  \BibitemOpen
  \bibfield  {author} {\bibinfo {author} {\bibfnamefont {Z.~T.}\ \bibnamefont
  {Brawley}}, \bibinfo {author} {\bibfnamefont {S.}~\bibnamefont
  {Pannir-Sivajothi}}, \bibinfo {author} {\bibfnamefont {J.~E.}\ \bibnamefont
  {Yim}}, \bibinfo {author} {\bibfnamefont {Y.~R.}\ \bibnamefont {Poh}},
  \bibinfo {author} {\bibfnamefont {J.}~\bibnamefont {Yuen-Zhou}},\ and\
  \bibinfo {author} {\bibfnamefont {M.}~\bibnamefont {Sheldon}},\ }\bibfield
  {title} {\enquote {\bibinfo {title} {Vibrational weak and strong coupling
  modify a chemical reaction via cavity-mediated radiative energy transfer},}\
  }\href@noop {} {\bibfield  {journal} {\bibinfo  {journal} {Nat.\ Chem.}\
  }\textbf {\bibinfo {volume} {17}},\ \bibinfo {pages} {439} (\bibinfo {year}
  {2025})}\BibitemShut {NoStop}%
\bibitem [{\citenamefont {Schwartz}\ and\ \citenamefont
  {Hutchison}(2025)}]{schwartz2025}%
  \BibitemOpen
  \bibfield  {author} {\bibinfo {author} {\bibfnamefont {T.}~\bibnamefont
  {Schwartz}}\ and\ \bibinfo {author} {\bibfnamefont {J.~A.}\ \bibnamefont
  {Hutchison}},\ }\bibfield  {title} {\enquote {\bibinfo {title} {Comment on
  `{N}on-polaritonic effects in cavity-modified photochemistry': {O}n the
  importance of experimental details},}\ }\href@noop {} {\bibfield  {journal}
  {\bibinfo  {journal} {Adv.\ Mater.\ 2404602}\ } (\bibinfo {year}
  {2025})}\BibitemShut {NoStop}%
\bibitem [{\citenamefont {Thomas}\ and\ \citenamefont
  {Barnes}(2025)}]{thomas2025}%
  \BibitemOpen
  \bibfield  {author} {\bibinfo {author} {\bibfnamefont {P.~A.}\ \bibnamefont
  {Thomas}}\ and\ \bibinfo {author} {\bibfnamefont {W.~L.}\ \bibnamefont
  {Barnes}},\ }\bibfield  {title} {\enquote {\bibinfo {title} {Response to
  {C}omment on `{N}on-polaritonic effects in cavity-modified photochemistry:
  {O}n the importance of experimental details'},}\ }\href@noop {} {\bibfield
  {journal} {\bibinfo  {journal} {Adv.\ Mater.\ 2501509}\ } (\bibinfo {year}
  {2025})}\BibitemShut {NoStop}%
\bibitem [{\citenamefont {Barachati}, \citenamefont {De~Liberato},\ and\
  \citenamefont {K{\'e}na-Cohen}(2015)}]{barachati2015}%
  \BibitemOpen
  \bibfield  {author} {\bibinfo {author} {\bibfnamefont {F.}~\bibnamefont
  {Barachati}}, \bibinfo {author} {\bibfnamefont {S.}~\bibnamefont
  {De~Liberato}},\ and\ \bibinfo {author} {\bibfnamefont {S.}~\bibnamefont
  {K{\'e}na-Cohen}},\ }\bibfield  {title} {\enquote {\bibinfo {title}
  {Generation of {R}abi-frequency radiation using exciton-polaritons},}\
  }\href@noop {} {\bibfield  {journal} {\bibinfo  {journal} {Phy.\ Rev.\ A}\
  }\textbf {\bibinfo {volume} {92}},\ \bibinfo {pages} {033828} (\bibinfo
  {year} {2015})}\BibitemShut {NoStop}%
\bibitem [{\citenamefont {Dunkelberger}\ \emph {et~al.}(2019)\citenamefont
  {Dunkelberger}, \citenamefont {Grafton}, \citenamefont {Vurgaftman},
  \citenamefont {Soykal}, \citenamefont {Reinecke}, \citenamefont {Davidson},
  \citenamefont {Simpkins},\ and\ \citenamefont {Owrutsky}}]{dunkelberger2019}%
  \BibitemOpen
  \bibfield  {author} {\bibinfo {author} {\bibfnamefont {A.~D.}\ \bibnamefont
  {Dunkelberger}}, \bibinfo {author} {\bibfnamefont {A.~B.}\ \bibnamefont
  {Grafton}}, \bibinfo {author} {\bibfnamefont {I.}~\bibnamefont {Vurgaftman}},
  \bibinfo {author} {\bibfnamefont {O.~O.}\ \bibnamefont {Soykal}}, \bibinfo
  {author} {\bibfnamefont {T.~L.}\ \bibnamefont {Reinecke}}, \bibinfo {author}
  {\bibfnamefont {R.~B.}\ \bibnamefont {Davidson}}, \bibinfo {author}
  {\bibfnamefont {B.~S.}\ \bibnamefont {Simpkins}},\ and\ \bibinfo {author}
  {\bibfnamefont {J.~C.}\ \bibnamefont {Owrutsky}},\ }\bibfield  {title}
  {\enquote {\bibinfo {title} {Saturable absorption in solution-phase and
  cavity-coupled tungsten hexacarbonyl},}\ }\href@noop {} {\bibfield  {journal}
  {\bibinfo  {journal} {ACS Photonics}\ }\textbf {\bibinfo {volume} {6}},\
  \bibinfo {pages} {2719} (\bibinfo {year} {2019})}\BibitemShut {NoStop}%
\bibitem [{\citenamefont {Lee}\ \emph {et~al.}(2024{\natexlab{b}})\citenamefont
  {Lee}, \citenamefont {Lee}, \citenamefont {Choi}, \citenamefont {Choi},\ and\
  \citenamefont {Gong}}]{lee2024}%
  \BibitemOpen
  \bibfield  {author} {\bibinfo {author} {\bibfnamefont {S.~W.}\ \bibnamefont
  {Lee}}, \bibinfo {author} {\bibfnamefont {J.~S.}\ \bibnamefont {Lee}},
  \bibinfo {author} {\bibfnamefont {W.~H.}\ \bibnamefont {Choi}}, \bibinfo
  {author} {\bibfnamefont {D.}~\bibnamefont {Choi}},\ and\ \bibinfo {author}
  {\bibfnamefont {S.-H.}\ \bibnamefont {Gong}},\ }\bibfield  {title} {\enquote
  {\bibinfo {title} {Ultra-compact exciton polariton modulator based on van der
  {W}aals semiconductors},}\ }\href@noop {} {\bibfield  {journal} {\bibinfo
  {journal} {Nat.\ Comm.}\ }\textbf {\bibinfo {volume} {15}},\ \bibinfo {pages}
  {2331} (\bibinfo {year} {2024}{\natexlab{b}})}\BibitemShut {NoStop}%
\end{thebibliography}%

\end{document}